\documentclass[%
 preprint,
groupedaddress,
 amsmath,amssymb,prd,
 aps
]{revtex4-1}

\usepackage{graphicx}% Include figure files
\usepackage[caption=false]{subfig}
\usepackage{dcolumn}% Align table columns on decimal point
\usepackage{bm}% bold math
\usepackage{xcolor,graphicx}

\newcommand{\MK}[1]{\color{violet}{[{\bf MK}: #1] }\color{black}}

\newcommand{\ZZ}[1]{}
\newcommand{\const}{\textrm{const}}

\newcommand{\UD}[2]{\ensuremath{^{#1}_{\phantom{#1} #2}}}

\newcommand{\calO}{\ensuremath{\mathcal{O}}}

\newcommand{\calR}{{\ensuremath{\cal R}}}

\newcommand{\dd}{\ensuremath{\textrm{d}}}

\newcommand{\mnras}{MNRAS}

\newcommand{\beq}{\begin{equation}}
\newcommand{\eeq}{\end{equation}}
\newcommand{\bea}{\begin{eqnarray}}
\newcommand{\eea}{\end{eqnarray}}
\newcommand{\bit}{\begin{itemize}}
\newcommand{\eit}{\end{itemize}}
\newcommand{\bfi}{\begin{figure}}
\newcommand{\efi}{\end{figure}}
\newcommand{\bfic}{\begin{figure*}}
\newcommand{\efic}{\end{figure*}}
\newcommand{\bce}{\begin{center}}
\newcommand{\ece}{\end{center}}
\newcommand{\bt}{\begin{table}}
\newcommand{\et}{\end{table}}
\newcommand{\btb}{\begin{tabular}}
\newcommand{\etb}{\end{tabular}}

\newcommand{\bfL}{\ensuremath{\mathbf{L}}}
\newcommand{\bfU}{\ensuremath{\mathbf{U}}}
\newcommand{\bfX}{\ensuremath{\mathbf{X}}}
\newcommand{\bfW}{\ensuremath{\mathbf{W}}}

\newcommand{\calA}{\ensuremath{\mathcal{A}}}
\newcommand{\calP}{\ensuremath{\mathcal{P}}}

\newcommand{\calE}{\ensuremath{\mathcal{E}}}

\newtheorem{theorem}{Theorem}[section]
\newtheorem{lemma}[theorem]{Lemma}

\newcommand{\qed}{\nobreak \ifvmode \relax \else
      \ifdim\lastskip<1.5em \hskip-\lastskip
      \hskip1.5em plus0em minus0.5em \fi \nobreak
      \vrule height0.75em width0.5em depth0.25em\fi}

\usepackage{bbold}
\usepackage{amsmath}
\usepackage{empheq}
\usepackage{amssymb}
\usepackage{comment}
\usepackage{bm}

\newcommand{\wxx}{W_{XX}}
\newcommand{\wxl}{W_{XL}}
\newcommand{\wlx}{W_{LX}}
\newcommand{\wll}{W_{LL}}
\newcommand{\mw}{\mathbf{W}}

\newcommand{\bmu}{\bar{\mu}}
\newcommand{\bnu}{\bar{\nu}}
\newcommand{\bal}{\bar{\alpha}}
\newcommand{\bbet}{\bar{\beta}}

\newcommand{\lz}{\lambda_{\mathcal{O}}}
\newcommand{\lep}{\lambda_{\mathcal{E}}}

\newcommand{\wxlin}{W^{-1}_{XL}}
\newcommand{\wxlinzer}{W^{-1\ (0)}_{XL\ }}
\newcommand{\wxlion}{W^{-1\ (1)}_{XL}}

\newcommand{\uoo}{U_{\mathcal{O}\mathcal{O}}}
\newcommand{\uoe}{U_{\mathcal{O}\mathcal{E}}}
\newcommand{\ueo}{U_{\mathcal{E}\mathcal{O}}}
\newcommand{\uee}{U_{\mathcal{E}\mathcal{E}}}

\newcommand{\rll}{R^{\bmu}_{\ \bal\bbet\bnu}(\lambda)l^{\bal}l^{\bbet}}

\newcommand{\uu}{\mathbf{U}}
\newcommand{\dl}{\Delta\lambda}

\begin{document}

\title{Weighing the spacetime along the line of sight using times of arrival of electromagnetic signals}% Force line breaks with \\

\author{Miko\l{}aj Korzy\'n{}ski}
\affiliation{%
 Center for Theoretical Physics, Polish Academy of Sciences, Warsaw, Poland
}
\email{korzynski@cft.edu.pl}
 \author{Jan Mi\'s{}kiewicz}
\affiliation{
 Center for Theoretical Physics, Polish Academy of Sciences, Warsaw, Poland
}
\affiliation{Faculty of Physics, University of Warsaw, Poland}
 
 \author{Julius Serbenta}
 \affiliation{
 Center for Theoretical Physics, Polish Academy of Sciences, Warsaw, Poland
}

\date{\today}% It is always \today, today,

\begin{abstract}
We present a new method of measuring the mass density along the line of sight, based on  precise measurements of the variations of the times of arrival (TOA's) of   electromagnetic signals   propagating between two distant regions of spacetime. The TOA variations are measured between a number of slightly displaced pairs of points from the two regions. These  variations are due to the nonrelativistic geometric effects (R{\o}mer delays and finite distance effects) as well as the gravitational effects in the light propagation (gravitational  ray bending and Shapiro delays). We show that from a sufficiently broad sample of TOA measurements we can determine two scalars quantifying the impact of the spacetime curvature on the light propagation, directly related to the first two moments of the mass density distribution along the line of sight. The values of the scalars  are independent of the angular positions or the states of motion of the
 two clock ensembles  we use for the measurement and  free from any influence of  masses  off the line of sight. These properties can make the mass density measurements  very robust.
 The downside of the method is the need for extremely precise signal timing.
\end{abstract}

\pacs{Valid PACS appear here}% PACS, the Physics and Astronomy
\maketitle

%\tableofcontents

\section{Introduction}
In this paper we develop a method of tomographic measurement of mass density of matter along the line of sight using variations of times of arrival (TOA's) of pulses of electromagnetic radiation between pairs of points from two small, distant regions connected by a null geodesic. The result may be seen as a continuation of the research program initiated in \cite{Grasso:2018mei, Korzynski:2017nas}  on determining the spacetime geometry, or -- more precisely -- the spacetime curvature, directly from precise optical or astrometric measurements, but this time focusing on a different observable, namely the precise time when a sharp discontinuity of the electromagnetic field reaches a receiver.

The TOA's of  electromagnetic signals are among the most important observables in general relativity. They have been studied in many fields of relativity and astronomy: in pulsar and binary pulsar timing \cite{Damour1992, Lorimer2008, Dahal2020},  
time-delay cosmography using strong lenses in cosmology \cite{Refsdahl1964,  Suyu2017}, experiments with atomic clocks and clock ensembles in gravitational fields \cite{Bauch}, relativistic geodesy 
\cite{puetzfeldbooke},
navigation on Earth \cite{Bahder2003, Ashby2003} or measurements of Shapiro delays 
from massive bodies \cite{shapiro:1964}. They also played an important role in the early days of special relativity
as the main observable in the Einstein's radar method of time and distance measurements.  The general problem of TOA's in a curved spacetime is rather difficult; it has been considered by Synge \cite{SyngeBook},
later by Teyssandier, Le Poncin-Lafitte and Linet \cite{Teyssandier-book, Teyssandier:2008, LinetTeyssandier}. 

In this work we propose a differential measurement in which we compare the TOA's between many pairs of distant of points, with the points contained in two fixed distant regions. The points in each region are  displaced with respect to each other in both space and time.  The measurement can be performed with the help of two  ensembles (or groups) of co-moving synchronized clocks, each located in one of the two distant regions. We assume that the clocks are equipped with transmitters and receivers of  electromagnetic radiation and are able to emit pulses of radiation at prescribed moments. They can also  measure precisely the moments of reception of signals emitted by other clocks, recognizing at the same time the origin of each received signal. The clocks within each ensemble are in free fall and at rest with respect to each other. We assume that  they are positioned fairly close to each other  and in a prescribed manner within the ensembles, while the ensembles themselves are located  far apart, see Fig.~\ref{fig:ideaofmeasurement}.
\bfi
\includegraphics[width=0.8\textwidth]{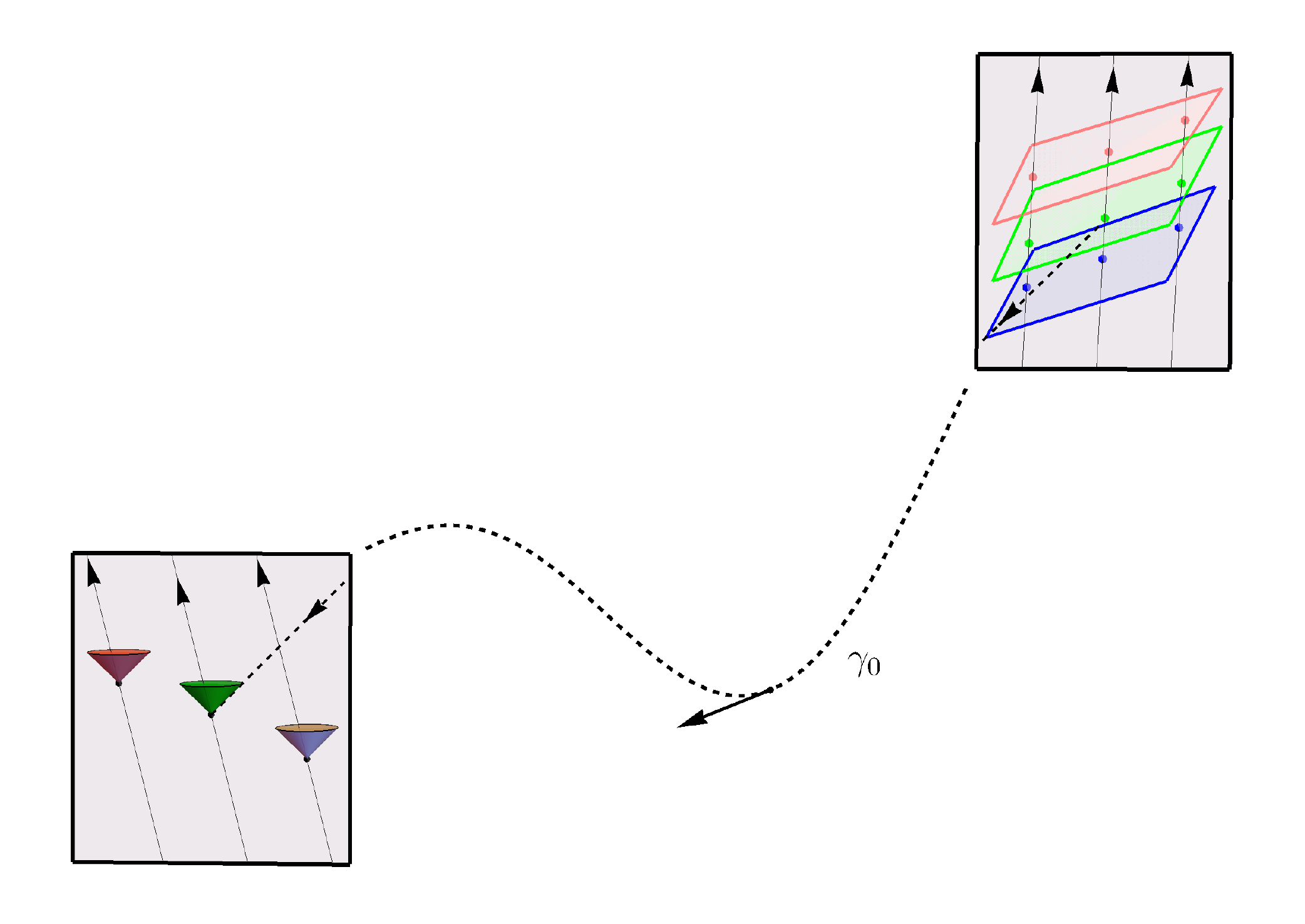}
\caption{The synchronized, comoving clocks in the emitters' ensemble (left) emit pulses of radiation in all directions at prescribed moments defined by their local time. The pulses travel then
a long distance through the curved spacetime and finally reach the receivers' ensemble (right), where each of the synchronized and comoving clocks records the time of arrival of each signal. The regions 
of emission and reception are connected by the fiducial null geodesic $\gamma_0$.}
\label{fig:ideaofmeasurement}
\efi

 Each of the two ensembles plays a different role in the measurement. The clocks in the emitters' ensemble emit signals in the form of pulses according to a fixed measurement protocol. Each signal should contain
  a time marker, defining the emission moment as sharply as possible, for example by the means of a discontinuity in a component of the electromagnetic field.
 The
  clocks in the receivers' group register the moments of arrival of the markers, see Fig.~\ref{fig:ideaofmeasurement}. We assume that the receivers can also recognize
the origin of each signal, meaning  the emitting clock and the moment of emission \footnote{This point is important, because in certain cases two or more signals may arrive almost at  the same moment, leading to ambiguities in the data interpretation.
Distinguishing of the incoming signals can be achieved, for example, by including a data stamp  containing an identifier encoded  in the signal  (for example, see the method used to transmit the P-code in the 
Global Positioning System \cite{Bahder2003}), or by using different frequencies for each signal according to a prescribed sequence. }.
 This way the system is able to record  the TOA's between \emph{all} emitter-receiver pairs and for various moments of emission. In this paper we provide  a method of combining all these data and extracting information about the spacetime curvature and the matter content along the line of sight, assuming that we know precisely the placement and the motions of the emitters and receivers as measured in their respective local inertial frames.

The two ensembles resemble  two clock-based gravitational compasses as discussed in \cite{PhysRevD.102.044027, PhysRevD.98.024032, PhysRevD.93.044073}. Recall that the clock-based gravitational compass is
a curvature-measuring device using differences in ticking rates of precise clocks. It generalises the notion of
the gravitational compass introduced by Szekeres \cite{Szekeres1965} and  developed  also by other authors \cite{PhysRevD.34.1018, Bazanski1, Bazanski2, Pirani1956}.  The  setup we discuss may be also be considered  a special case of an extended gravitational clock compass in the terminology of \cite{PhysRevD.102.044027, PhysRevD.98.024032, PhysRevD.93.044073}, consisting of two distinct groups of clocks. Unlike the former, it is only  extended along a single null geodesic, and therefore it is only able to measure some of the curvature components. On the other hand, this extension is so large that we need to allow  the curvature to vary along the way.

 Direct measurements of the spacetime curvature in GR are usually quite challenging for a simple reason: with the exception of cosmology, black hole theory or strong gravitational lensing, the scale of the curvature is usually very large in comparison to all other time and length scales involved. Thus the curvature corrections to observables are usually tiny  and  need to be distinguished from many other effects. However, the measurement presented here is designed in such a way that potentially much larger contributions to the TOA's  due to the kinematical effects (states of motion of the two ensembles), attitude effects (angular position or attitude of both ensembles with respect to each other or to the line of sight) and influence of nearby masses off the line of sight (Shapiro delays from nearby masses) cancel out completely. 
 Effectively it is only the gravity of the mass density located along the  null geodesic linking the two regions that contributes to the results. This implies a strong resistance of the measurement to perturbations by external masses or due to the effects of unknown velocities or  misalignment
 of the two clock ensembles. This property  makes the measurement  very robust despite very small magnitude of the expected signal.
  Moreover, thanks to the advancements in the atomic clock techniques, the time measurements are now among the most precise 
 types of measurements possible today,  and therefore the TOA's  can in principle be measured with high precision.
 
 The quantities determined by the measurement are the \emph{distance slip} $\mu$, introduced in \textcite{Grasso:2018mei, korzynskivilla}, and a closely related quantity $\nu$. Recall that the dimensionless scalar $\mu$ measures the difference between the parallax 
 effects and the magnification between two points along a null geodesic, suitably averaged over all possible baseline orientations. In a flat spacetime the effect of the transverse displacement of the receiver on the 
 apparent position of a distant luminous source (i.e. the trigonometric parallax) is identical
 to the effect of the same displacement of the source  in the opposite direction, leading to $\mu=0$. In a curved spacetime this is no longer the case, and,  as a result, the two methods of 
 distance measurement to a single body, by angular size and by parallax, may yield different numbers, resulting in nonvanishing $\mu$. We showed in \cite{Grasso:2018mei} that the measurement of $\mu$ is selectively sensitive to the curvature along the line of sight and that for short distances it is sensitive to an integral of a single component of the stress-energy tensor. 
 The same applies to $\nu$ and therefore the method we present may be seen as a tricky differential measurement of the Shapiro delays, sensitive only to the signal originating from the mass density along the line of sight.
 
 In the meantime, as an important side result, we develop further the bi-local approach to light propagation in the curved spacetime, introduced first in \cite{Grasso:2018mei}. 
In this framework the TOA's, just like other observables, are expressed as functions of the curvature tensor along the line of sight, describing the influence of spacetime geometry on light propagation, as well as the momentary positions and motions of both sources and receivers, as described in their locally flat coordinates.  The dependence of the TOA's on the spacetime geometry is clearly separated from the dependence on the momentary positions and motions of the clocks. We then consider the emitters and the receivers as contained in two distant, free-falling Einstein elevators. The direct influence of gravity is undetectable within each elevator within the timescale of the experiment. However, light propagating over long distances from one elevator to the other does feel the impact of the curved spacetime
and, as we show in this paper, this impact can be detected.

\subsection{Geometric description and mathematical apparatus}
The description of light propagation using bi-local operators is quite different from the more standard approach in which we use an approximation (for example post-Newtonian,
post-Minkowskian or linearized gravity), calculate the potentials or metric components from the mass distribution and then trace the perturbed null geodesics as well as the perturbed worldlines of the emitters and receivers of the pulses. 
In contrast, the bi-local approach works in any spacetime and does not require an explicit decomposition the metric into a global flat background plus perturbations. It shares many features with
the time transfer function formalism of Teyssandier, Le Poncin-Lafitte and Linet \cite{Teyssandier:2008, Teyssandier-book, LinetTeyssandier}, including the use of the Synge's world function, but it also differs in a number of important ways.
Firstly, our formalism does not aim to be global: it provides expressions for the variations of the TOA's in the form of a Taylor expansion valid in the immediate neighbourhood of a given pair of points, rather than the exact or approximate value of the TOA's between any emitter-receiver pair in a given spacetime. Consequently, it does not need to make use of a global time coordinate as a reference.

 Following \cite{Grasso:2018mei},  we proceed here by
deriving a number of exact geometric relations between the spacetime geometry, the kinematical quantities describing the momentary positions and states of motion of the emitters and receivers in their local inertial frames, and the TOA's they measure.   The whole framework is therefore formulated in a coordinate-independent, geometric way. 

 The dependence of the TOA's on the spacetime geometry enters only via 2 covariantly defined tensors and 1 bitensor  encoding the impact of the spacetime on the  wavefronts, just like in 
 \cite{Grasso:2018mei}. These linear operators  are defined as functionals 
of the  components the Riemann curvature tensor along the line of sight, independently from any tetrads, frames or other structures. This is again in contrast to \cite{Teyssandier:2008, Teyssandier-book, LinetTeyssandier}, where the dependence of TOA's on the spacetime geometry enters directly via the components of the metric expressed  in a particular coordinate system, related to the post-Newtonian or post-Minkowskian approximation. As we will see, this last feature is crucial 
for deriving the key result of this paper, namely the method for extracting the bare curvature effects from the variations
of the TOA's.

Since the definitions and the measurements of the observables in question require no external structures like global coordinate systems (for example the Solar System barycentric coordinates), large-scale non-rotating reference frames, Killing vectors etc. we avoid  any problems of the astrometric data reduction in the data interpretation \cite{Klioner_2003}.
\subsection{The basic idea of the measurement} \label{sec:basic}
We consider the exact moment of arrival of a signal sent from a point in the emitters' region $N_\calE$ as measured by a clock performing the measurement in
a local inertial frame in the receivers' region $N_\calO$. We fix the receivers' reference frame by fixing their common 4-velocity $u_\calO^\mu$ and thus also the corresponding coordinate time. The TOA depends now on the spatial position and the moment of emission in $N_\calE$ and the spatial position of the receiver in $N_\calO$. 
We consider two reference points $\calE \in N_\calE$ and $\calO \in N_\calO$ such that a signal from $\calE$ reaches $\calO$. We can then describe the positions and the emission moment by the displacement vectors from $\calO$ (three spatial components only) and $\calE$ (4-dimensional spacetime vectors). We can then expand the TOA up to the next-to-leading, second order in the 
displacements. At the leading, linear order we just see the R{\o}mer delays, i.e.  the dependence of the TOA's on the position of the emitter and the receiver along the line of sight, and the
frequency/time transfer effects (the redshift or blueshift of frequencies and difference between the receiver's proper time lapse and the observed emitter's time lapse), see Fig.~\ref{fig:TOAs}, upper panel.  These effects do not depend explicitly on the spacetime curvature and can be described using just special relativity (see \textcite{2014A&A...570A..62B} for a discussion in the context of precise astrometry). However, as we will see,
 the  quadratic term in the displacements contains curvature corrections on top of the standard, distance-dependent effects  present also in a flat spacetime (Fig.~\ref{fig:TOAs}, lower panel). The idea is to recognize and evaluate these curvature corrections to the ``flat'' finite distance effects. 
 
 \bfi
 \centering
    \subfloat{\includegraphics[width=0.2\linewidth]{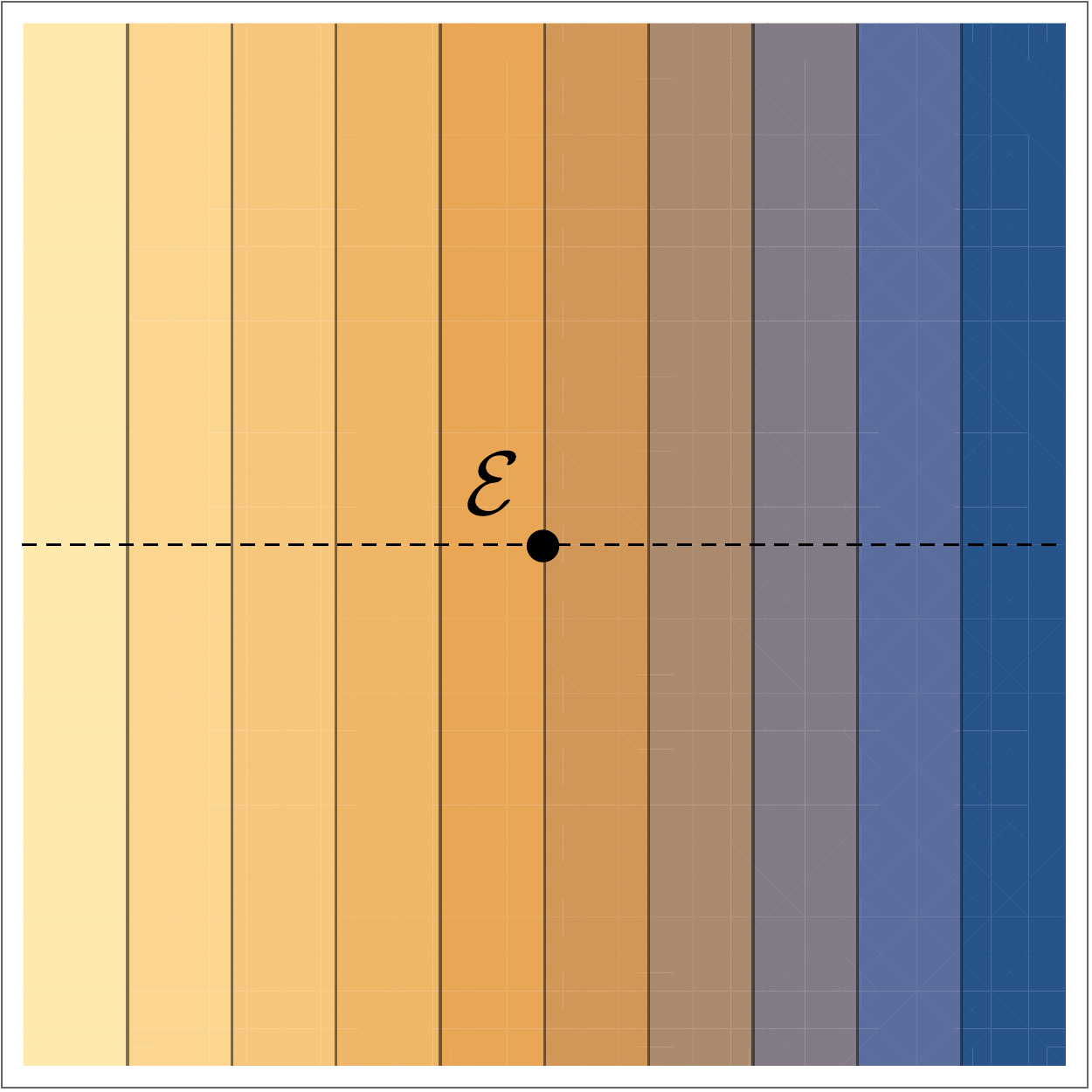} } \qquad
    \subfloat{\includegraphics[width=0.2\linewidth]{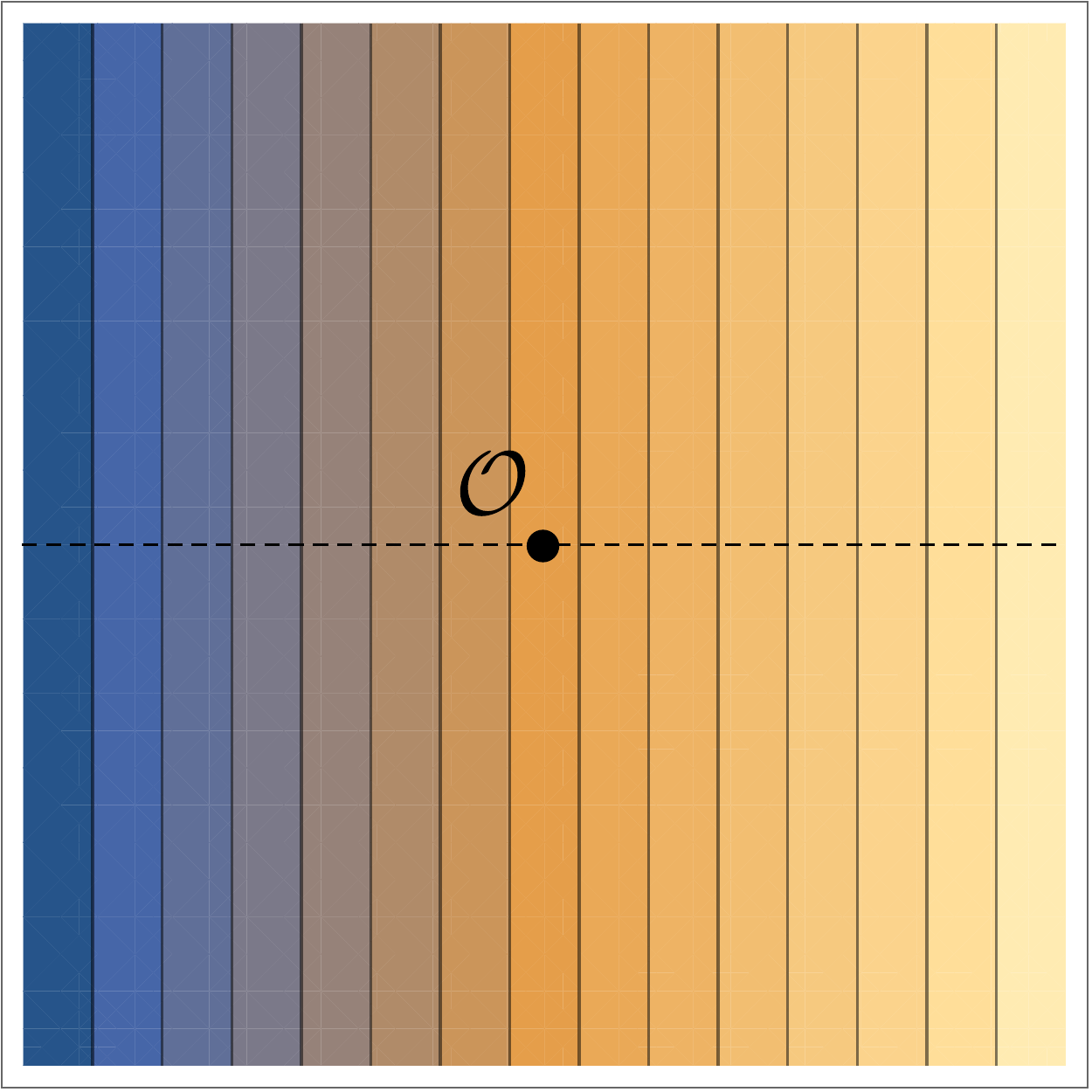} } \\ \vspace{1cm}
    \subfloat{\includegraphics[width=0.2\linewidth]{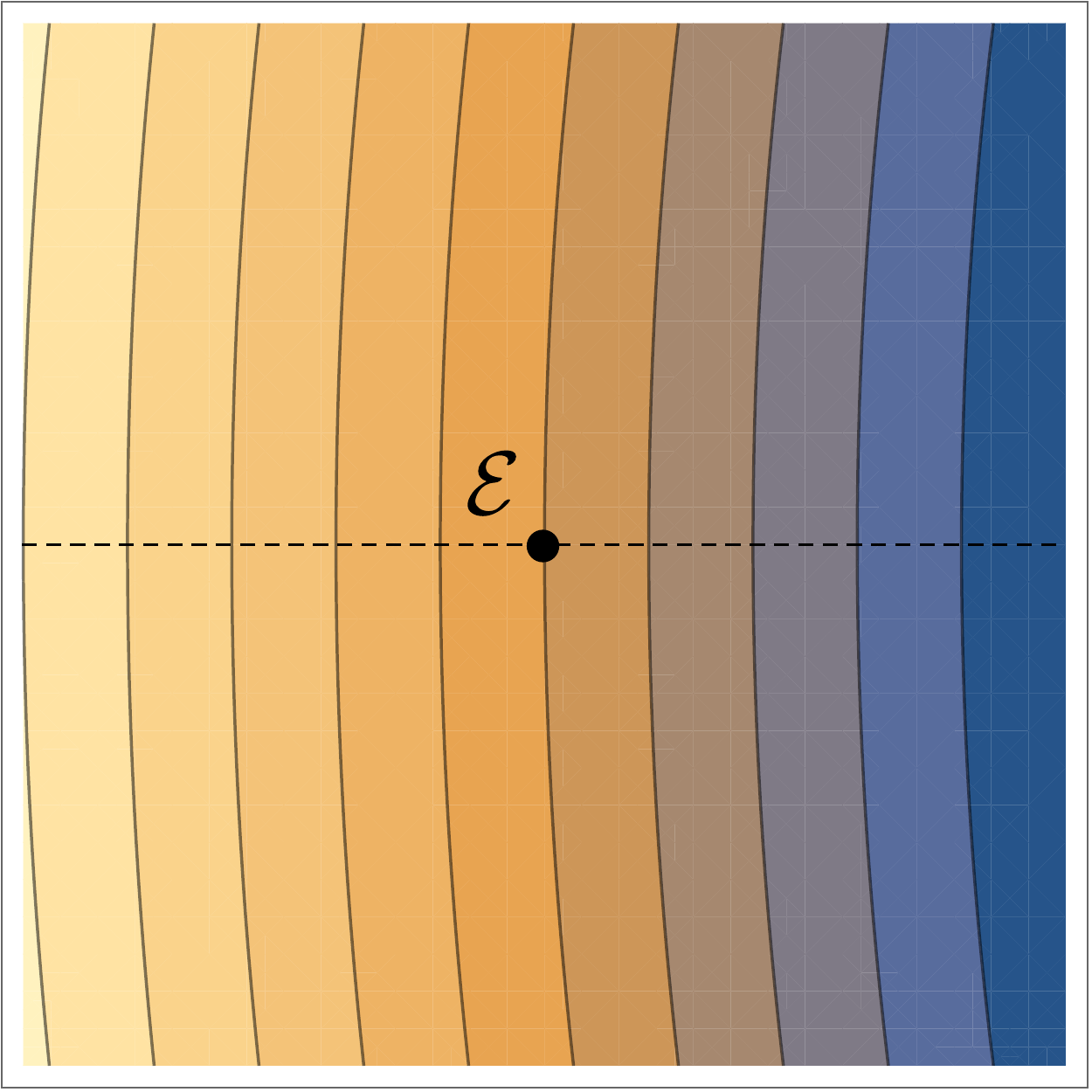} } \qquad
    \subfloat{\includegraphics[width=0.2\linewidth]{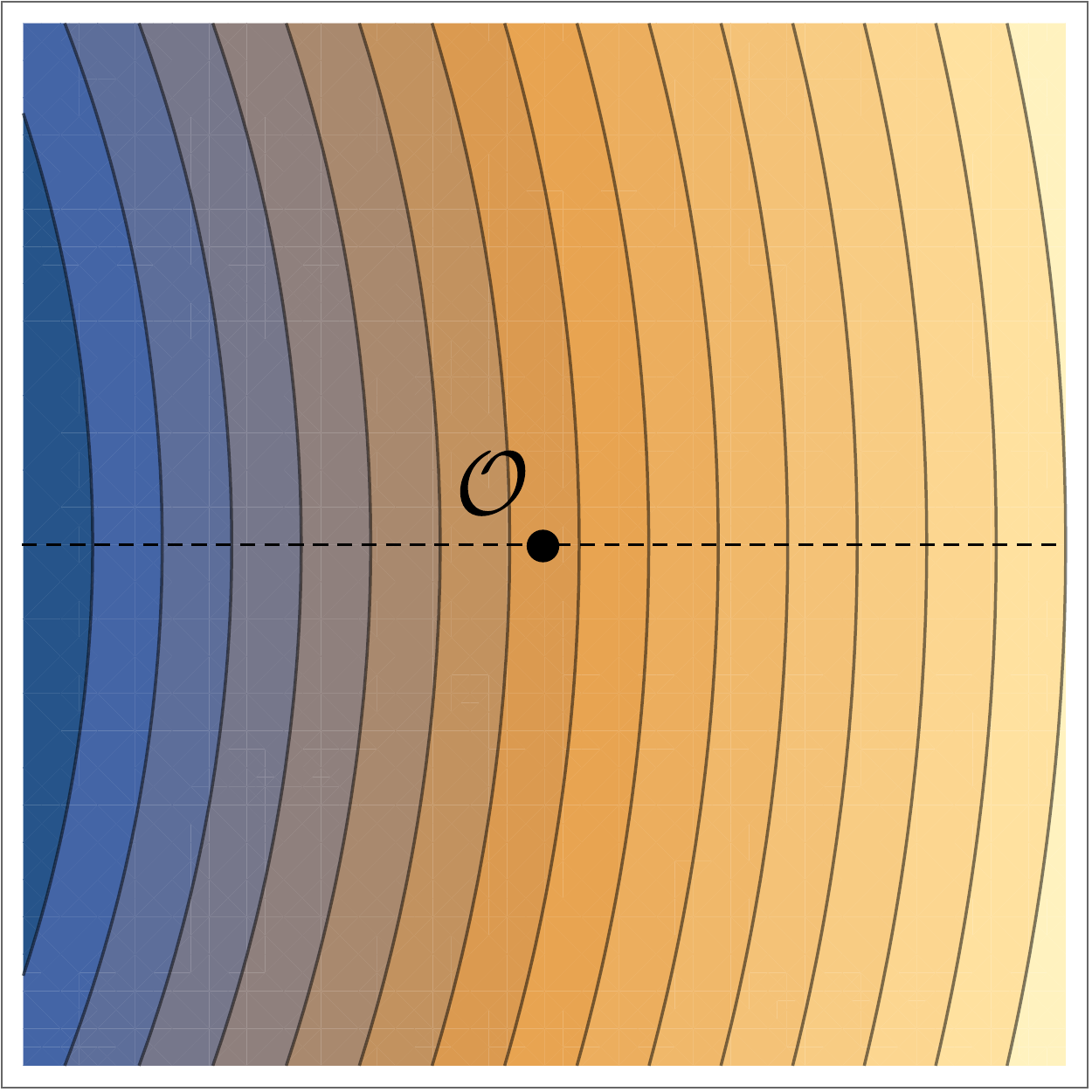} } 
\caption{Upper panel, left: The variations of the TOA's of light signals, as registered by the receiver in his or her proper time, plotted for a fixed observation point $\calO$, but variable emission point. Blue colour denotes an earlier TOA, while yellow a later TOA. For sufficiently small displacements we see only the the leading order, linear effects in displacements. Upper panel, right: the same TOA's, but for a fixed emission  point $\calE$ and 
variable observation points.  Note that in both cases in the linear order we only see the dependence on 
the displacements along the line of sight (R{\o}mer delays). Due to special relativistic effects the dependence on displacements does not  need to be exactly the same on both ends. Lower panel, left: Again TOA's for a fixed observation point $\calO$ and displaced emission points. For sufficiently large displacements we begin to see the quadratic order effects in displacements, in the form of transverse delays. Lower panel, right: The TOA's for a fixed emission point $\calO$ and displaced observation points, together with the second order effects. In a general spacetime the magnitude of the second order effects may be different at the two endpoints.}
\label{fig:TOAs}
\efi

 The quadratic term in question encodes the subleading, transverse effects in the TOA variations as well as the cross effects of displacements on both ends. They are usually much smaller than the linear effects and are responsible for, among other things, the curved shape of the light cones and the wave fronts.
  The TOA's depend also  on the receivers' and emitters' rest frames via the standard special relativity effects, but we prove here that the two quantities we measure are Lorentz-invariant with respect to both the emitters' and receivers' frames.

The basic idea of the measurement has  a simple geometric interpretation: 
the TOA's between two regions can be geometrically described by a subset of the product manifold $M\times M$ constructed from the spacetime $M$. We consider the  locus of pairs of spacetime events which can be connected by a null geodesic. Under fairly mild assumptions this locus is (locally) an embedded hypersurface in $M\times M$ and we will call it  the \emph{local surface of communication} (LSC). Its shape in the vicinity of a given point $(\calO,\calE)$ 
can be approximated by its second order tangent surface by the means of a Taylor series. The second order term in this approximation, related to the extrinsic curvature of the LSC as an embedded hypersurface in $M\times M$, contains spacetime curvature corrections and, as we show, can be related to the Riemann tensor along the line of sight. The measurement itself amounts to sampling the LSC over a large number of nearby emission/reception event pairs, with the help of clocks equipped with electromagnetic radiation transmitters and receivers.  Note that by definition each emission-reception pair we register yields a point in $M \times M$ lying on the LSC. If the sampling is sufficiently broad, the 
shape of the  LSC in the second order Taylor expansion can be recovered completely from the data and the extrinsic curvature tensor can be obtained componentwise. Finally, the two bi-local scalars $\mu$ and $\nu$ measuring the spacetime curvature are
calculated from the extrinsic curvature and the first order term.

We point out here that the TOA's of mechanical waves in elastic media and their variations have been studied extensively in seismology. In fact, the formalism developed in this paper may be also regarded as the general relativistic counterpart of the  second order approximation for travel times in the seismic ray theory  \cite{cerveny_2001, farra2002, cerveny2012,  waheed2013}.

\subsection{Assumptions and limitations of the approach}

We assume the validity of the geometric optics approximation throughout the work. This means that the electromagnetic wave's wavelength is assumed to be much smaller than all other length scales involved and the
radiation intensity small enough as not to produce any significant contribution to the stress-energy tensor. With these assumptions we can consider the electromagnetic radiation as 
propagating along null geodesics of a fixed metric  \cite{PhysRev.166.1263, PhysRev.166.1272}. Consequently, we can assume that a pulse of radiation originating from single emission event propagates along this event's future light cone.

The assumptions regarding the  two  distant regions are similar to those in \cite{Grasso:2018mei}: we assume that both regions we consider are sufficiently small in comparison with the 
spacetime curvature scale that the light propagation between them can be approximated using the first order geodesic deviation equation. As we will see in this paper, this is 
equivalent to assuming that the Synge's world function can be approximated by its Taylor series truncated at the second order. In a more physical language, this means that we demand 
the curvature to be constant in the transverse directions across the long, thin cylinder  connecting both regions. It follows that we need to assume  the matter density and the tidal effects
to be constant in the transverse directions as well.

We assume that each process of emission and reception of a signal happens at one single event and without instrumental delays. We also disregard any non-gravitational effects of refraction or pulse dispersion due to the presence of a medium, for example
ionized hydrogen, along the line of sight. This kind of medium produces additional frequency-dependent delays related to the electron column density along the line of sight \cite{Lorimer2008}.

\subsection{Structure of the paper}
The paper is organized as follows. In Section \ref{sec:geometry} we describe in detail the geometry of the problem, introduce the product manifold and discuss all types of coordinate systems we 
use in the rest of the paper. Then, in Section \ref{sec:shape},  we introduce the mathematical machinery of the paper: we  derive  relation between the shape of the local surface of communication 
and the spacetime curvature, first generally and then in the small curvature limit, derive exact expressions for the TOA and finally discuss the inverse problem of reconstructing the shape of the local surface of communication from TOA measurements.

 In Section \ref{sec:two} we introduce the two scalar observables measuring the curvature corrections and prove their properties. 
Section \ref{sec:expansioninsmall} contains the derivation of the order-of-magnitude estimates of various TOA effects in terms of the characteristic scales of the problem. Additionally, we 
relate the two observables to the first two moments of the mass distribution along the line of sight. We also summarize the most important results of the first five sections.
Section \ref{sec:protocol} contains an example of the measurement protocol.  We conclude the paper with a summary and conclusions. The Appendix contains the details and the derivations of more elaborate technical results of the paper.

\subsection{Notation and conventions} \label{sec:notation}

In this work we use many different types of indices, related to various types of  geometric objects and bases (tetrads), and running over different sets of integers. We will now summarize our index conventions.

The Greek indices $\mu,\nu,\dots$ run from to 0 to 3 and denote geometric objects expressed in various coordinate bases. We borrow the notation from \cite{Poisson2011} for bi-local (or two-point) geometric objects, in which
primed Greek indices $\mu',\nu',\dots$ refer to the tangent space at the observation point $\calO$, while the unprimed ones refer to the emission point $\calE$. 
The Latin indices $i',j',k',\dots$ and $i,j,k,\dots$ run from 1 to 3, again referring to $\calO$ and $\calE$ respectively. They will be used for the spatial components of vectors and other objects.
We also use  indices with an overline $\bar{\mu}, \bar{\nu}, \dots$ for geometric objects expressed in a parallel propagated tetrad along the fiducial null geodesic $\gamma_0$, independently of the point along $\gamma_0$. 
The Latin capitals, both with overline ($\bar{A}, \bar{B}, \dots$) or without ($A, B, \dots$ or $A,' B', \dots$), run from 1 to 2 and will be used for the transverse spatial components, orthogonal to the line of sight.

Finally, we also introduce notation for 8-dimensional vectors  in the direct sum of the tangent spaces at $\calO$ and $\calE$, i.e. in $ T_\calO M \oplus T_\calE M$, as well as tensors on this space. The geometric objects themselves will be denoted by boldface capitals, i.e. ${\bf X}, {\bf Y}, \dots$. Moreover, all index numbers in this case will be 
denoted by the boldface font;  the indices run by convention from $\bf 0$ to $\bf 7$. The first four components (i.e. $\bf0$-$\bf 3$) will refer to the $\mu'$ components in $T_\calO M$, while the latter four ($\bf 4$-$\bf 7$) to the $\mu$ components from $T_\calE M$. We will also use the boldface latin letters $\bf i, \bf j, \dots$ to denote the 7 components ranging from $\bf 1$ to $\bf 7$, i.e. omitting the $\bf 0$ component.

For the sake simplicity we will assume that $c=1$ throughout the paper, i.e. we are using distance units to express time.

\section{Geometrical preliminaries} \label{sec:geometry}

\subsection{Geometric setup}

\bfi
\includegraphics[width=0.5\textwidth]{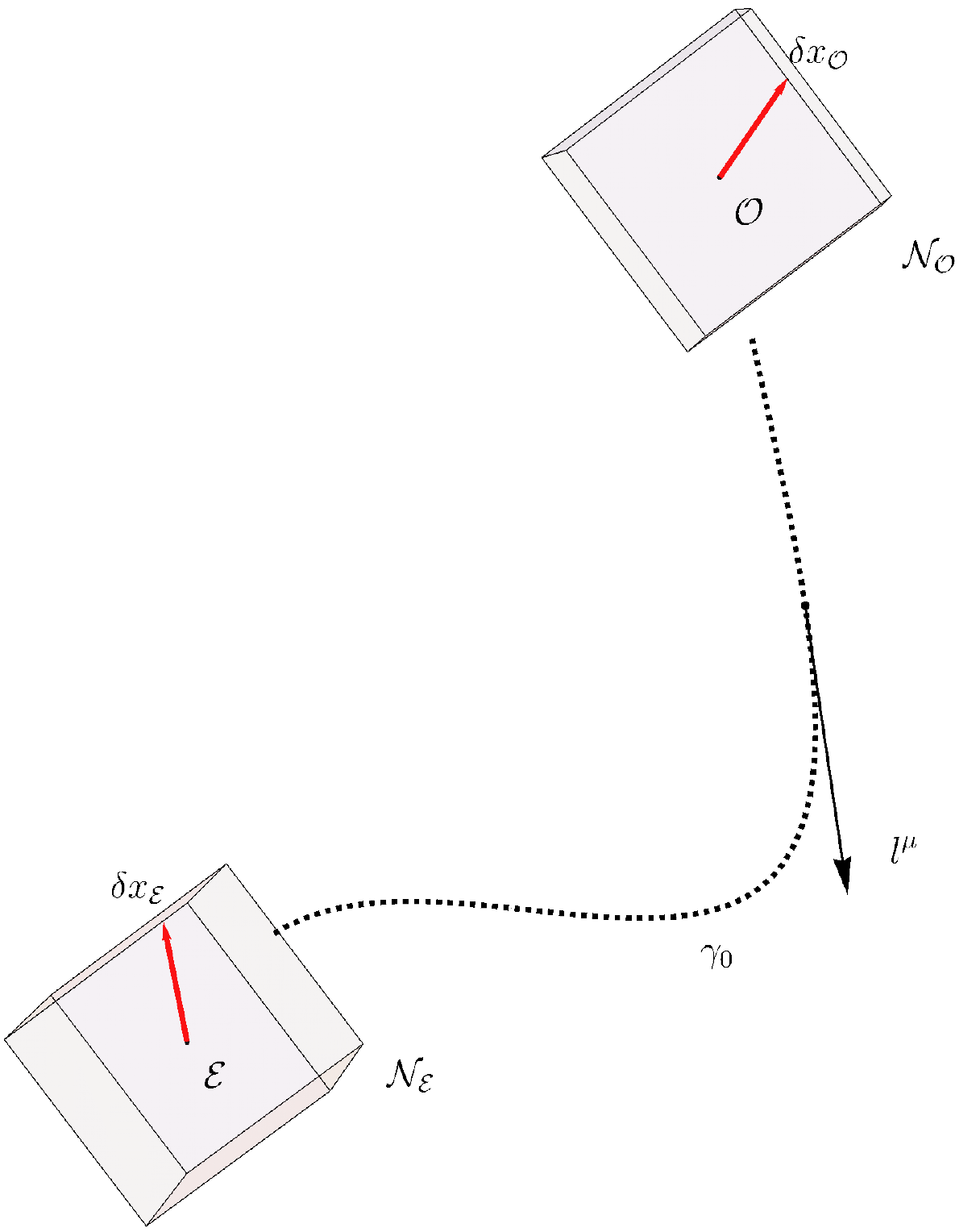}
\caption{Geometric setup of the paper, with two locally flat regions $N_\calO$ and $N_\calE$, connected by a null geodesic $\gamma_0$ passing through $\calO$ and $\calE$. The displacement vectors $\delta x_\calO$ and $\delta x_\calE$ identify points in both regions. The tangent vector to $\gamma_0$ is denoted by $l^\mu$. }
\label{fig:Boxes}
\efi

The geometric setup is similar to that of \cite{Grasso:2018mei}, see Fig.~\ref{fig:Boxes}:  let $M$ be the spacetime of dimension $4$, equipped with a smooth metric $g$ of signature $(-,+,+,+)$. 
We consider two points $\calO$ and $\calE$ such that an electromagnetic signal emitted at $\calE$ can be received at $\calO$. In other words, we assume that $\calO$ lies
on the future light cone centered at $\calE$. In this case $\calO$ and $\calE$ can be connected by a null geodesic $\gamma_0$. We also consider two small regions $N_\calO \subset M$ and $N_\calE \subset M$ around points $\calO$ and $\calE$ respectively, extending in both space and time. Under the assumptions above we may expect that signals from other points (events) in $N_\calE$ can be received at other points in $N_\calO$. 

We assume both $N_\calE$ and $N_\calO$ to be
sufficiently small to be effectively flat. Namely, if $\calR$ denotes the spacetime curvature radius scale and $L$ the size of both regions, then we assume that $L^2/\calR^2$, a dimensionless quantity scaling the curvature corrections within both regions, is negligibly small. In this case we may simply identify points in $N_\calO$ with vectors in
the tangent space $T_\calO M$ with the help of the exponential map and the spacetime metric $g$ with the flat metric on the tangent space; the same construction works for $N_\calE$.
With this identification the points $\calO$ and $\calE$ will serve as reference for other points in $N_\calO$ and $N_\calE$ and vectors in $T_\calO M$ and $T_\calE M$, denoted $\delta x_\calO$ and $\delta x_\calE$, as displacement vectors representing points. 
 The null geodesic $\gamma_0$ connecting $\calO$ and $\calE$ will be referred to as the \emph{line of sight} (LOS) or the \emph{fiducial null geodesic} and will be used as reference for other geodesics connecting points in $N_\calO$ and $N_\calE$. 
 We assume that $\gamma_0$ is parametrized affinely by the parameter $\lambda$, although we make no assumptions regarding the normalization of this parametrization. In order to be consistent with \cite{Grasso:2018mei, Korzynski:2017nas} we will assume that the parametrization runs backward in time, i.e. from $\calO$ to $\calE$. The null tangent vector to $\gamma_0$ will be denoted by $l$.
 
 In a general spacetime it is possible that more than one geodesic connects a pair of points from $N_\calO$ and $N_\calE$. In the context of geometrical optics this means that we may expect  multiple imaging and, consequently, many TOA's of a single signal from a given event. The TOA's becomes then multivalued for a given receiver.
 For the sake of simplicity we will assume throughout this paper that both $N_\calE$ and $N_\calO$ lie in the normal convex neighbourhood of $\calE$. Under this assumption the geodesics 
 connecting such pairs are unique and no multiple TOA's are possible. This assumption also allows to define a single-valued  world function, an important object in this work.
 
 We point out, however, that the results of this paper should also hold if multiple imaging is present, but $N_\calO$ and $N_\calE$ do not contain conjugate points, i.e. they are away from caustics, and they are both  sufficiently small. In this case
 we simply need to limit the geodesics considered to those contained in a sufficiently narrow tube around a single connecting geodesic $\gamma_0$ \cite{Grasso:2018mei}, and this way limit our interest to a particular single image of the distant region $N_\calE$ on the celestial sphere. However, in this paper we will not consider this situation in detail.

 We assume that the light propagating between $\calE$ and $\calO$ is affected by the spacetime curvature, but this effect can be efficiently described in terms of the 
 Riemann tensor along $\gamma_0$. In other words, the effects of higher derivatives of the curvature along the LOS are negligible. 
 It follows that the behaviour of geodesics connecting $N_\calO$ and $N_\calE$ can be very well approximated using the first order geodesic deviation equation (GDE). 
 This is again true provided that $N_\calO$ and $N_\calE$ are small enough \cite{Grasso:2018mei}. 
 As we will see, this assumption has an important effect on the TOA's between the two regions. Namely, it means that we can approximate the variations of the  TOA's using the second order
 Taylor expansion.

\subsection{The product manifold. } The geometric construction behind the direct curvature measurements from the TOA variations is much simpler to explain if we 
perform it at the level of  the product manifold $M \times M$ instead of the spacetime $M$ itself. 
Consider then the Cartesian product $M\times M$, consisting of pairs $(p,q)$ of points in $M$. It is also a smooth  manifold and its tangent vectors can be identified with \emph{pairs} of vectors in $M$ at two, usually distinct, points. Namely, for ${\bf X} \in T_{(p,q)} (M\times M)$ we have
${\bf X} \equiv (X_1, X_2)$, where $X_1 \in T_p M$ and $X_2 \in T_q M$. More formally, at every $(p,q)$ we have a natural isomorphism $T_{(p,q)} (M\times M) \cong T_pM \oplus T_q M$ between the tangent space to the product manifold and a direct sum of the tangent spaces to the spacetime.

Moreover, $M \times M$ as a manifold can be equipped with a smooth metric tensor $\bf h$ constructed from $g$. 
There are many ways to define it, but we propose here the following one:  for ${\bf X, Y} \in T_pM \oplus T_q M $ we define ${\bf h}({\bf X},{\bf Y}) = g_p(X_1, Y_1) - g_q(X_2,Y_2)$, where
$X_1, Y_1 \in T_p M$ and $X_2, Y_2 \in T_qM$ denote vectors from the decomposition  of $\bf X$ and $\bf Y$ respectively, while $g_p$ and $g_q$ denote the spacetime metric $g$ at $p$ and $q$.
The reader may check that this metric is non-degenerate and of split signature (4,4), i.e. with equal number of positive and negative signs in the diagonal form.

The advantage of the product space construction is that now the bi-local scalars or tensors on $M$, defined for pairs of point, can be identified with strictly local functions or tensors in $M \times M$. Moreover, as we will see, the problem of TOA variations has a particularly simple and elegant geometric formulation in the language of the product manifold.

\subsection{Local surface of communication.}
Consider now the neighbourhood of  the point $(\calO,\calE) \in M\times M$, given by $N_\calO \times N_\calE \subset M\times M$. The signals propagating between both regions define
the set $\Sigma \subset N_\calO \times N_\calE$ of pairs of points which can be connected by a null geodesics. Physically they correspond to pairs of points $(x',x)$ such that the 
signal emitted from $x$ can be received at $x'$. Note that by assumption we always have $(\calO,\calE) \in \Sigma$. 

In a general situation $\Sigma$ may have a complicated geometric structure:  the wave fronts, which constitute simply the sections of $\Sigma$ though surfaces $x_\calE^\mu = \const, x_\calO^{0'}=\const$, typically develop folds, cusps and other types of singularities \cite{low:1998, perlick-lrr}. However,  in the absence of strong lensing we may expect it to form an embedded submanifold of codimension 1 passing through $(\calO,\calE)$, at least locally within $N_\calO\times N_\calE$. 
We will therefore refer to $\Sigma$ as the \emph{local surface of communication} (LSC),  the word ``local'' meaning simply that we restrict our considerations to a small neighbourhood of $(\calO,\calE)$.

The physical interpretations of $\Sigma$ all follow from the fact that 
$\Sigma$ governs the shapes of the past light cones centered at points in $N_\calO$, as registered in  $N_\calE$, and the shapes of the future light cones centered at points
in $N_\calE$, as measured in $N_\calO$. More precisely,  the sections of $\Sigma$ with the surfaces $x_\calO^{\mu'} = \const$ yield the past light cones in $N_\calE$ with vertex at chosen
 $x_\calO^{\mu'}$, while the sections with $x_\calE^\mu = \const$ give light cones in $N_\calO$ with vertex at $x_\calE^\mu$.  The shapes of the light cones on the other hand govern the times of arrival as well as the shapes of the wavefronts. The latter means that it also controls the direction of light propagation. 
Later in this paper we will show that the  shape of $\Sigma$ near $(\calO,\calE)$ is directly related to the spacetime curvature tensor along $\gamma_0$.

\subsection{Coordinate systems and orthonormal tetrads. } We will use a number of distinct types of coordinate systems thoughout the paper. We will list them below 
from the most general to the most specific. 

The most general type is any coordinates in $M$ covering both $N_\calO$ and $N_\calE$. However, we will more often use the \emph{locally flat}  (LF) coordinates, i.e. any coordinates covering $N_\calO$ and $N_\calE$ such that the Christoffel symbols vanish at $\calO$ and $\calE$:
\bea
 \Gamma\UD{\mu}{\nu\alpha}(\calE) = \Gamma\UD{\mu'}{\nu'\alpha'}(\calO) = 0, \nonumber
\eea
or, equivalently, that the first derivatives of the metric components vanish at $\calO$ and $\calE$. 
 A special type of coordinates of this kind can be obtained from a pair of orthonormal tetrads, one
in $T_\calO M$ and the other in $T_\calE M$. Namely, given two such orthonormal tetrads $(u_\calO,e_i)$, $(u_\calE, f_j)$, consistent with the spacetime orientation and with both $u_\calO$ and $u_\calE$ timelike, future-pointing,
we can introduce locally flat coordinates at $N_\calO$ and $N_\calE$ as a pair of Riemann normal coordinates defined by the two tetrads: 
 we define the point $p\in N_\calO$ corresponding to (sufficiently small) coordinates $(y^\mu) \in \mathbb{R}^4$ as $\exp(y^0\,u_\calO+y^i\,e_i)$, with $\exp$ denoting the exponential map at $\calO$ (analogous construction works in $N_\calE$).
 We will call such coordinates the \emph{orthonormal locally flat} (OLF) coordinates.
 
  Finally, given a single orthonormal tetrad $(u, e_i)$ at $\calO$, properly oriented with future pointing, timelike $ u$, we can parallel-propagate it along $\gamma_0$.
This provides us with a parallel-propagated tetrad at every point along $\gamma_0$, useful for describing geometric objects along the line of sight, as well as a corresponding orthonormal tetrad at $\calE$. Repeating the same construction of the Riemann normal coordinates  yields \emph{parallel-propagated locally flat} (PLF) coordinates covering  $N_\calO$ and $N_\calE$ as a special case of OLF coordinates. For any of these coordinate systems we will use the corresponding coordinate tetrads at $T_\calO M$ and $T_\calE M$ for decomposing the tensors and the bi-tensors into components.

Given two pairs of tetrads used for defining the OLF coordinates, say $\left((u_\calO,e_i), (u_\calE, f_j)\right)$  and  $((\tilde u_\calO,\tilde e_i),(\tilde u_\calE, \tilde f_j))$, we can
always transform the first pair into the second one by a pair of proper, orthochronal Lorentz transforms acting on the two tetrads. This way we see that the OLF coordinates are defined uniquely up to the action of a pair of special, orthochronal Lorentz groups. On the other hand, the PLF coordinates are defined uniquely up to the action of a single proper, orthochronal Lorentz group. 

As a special type of oriented and time-oriented orthonormal tetrads along $\gamma_0$ we will  consider the \emph{adapted tetrads}, i.e. those for which the third spatial vector is aligned along the null tangent vector
$l^\mu$. For a tetrad $(u,e_i)$ this simply means that $l^\mu = Q( - u^\mu + e_3^\mu)$ for some tetrad-dependent $Q > 0$. Adapted tetrads will be denoted by $(u, e_A, e_3)$ and we will call the two spatial vectors $e_A^\mu$, orthogonal to $l^\mu$,  the \emph{transverse vectors}, forming a Sachs basis \cite{perlick-lrr}. Adapted tetrads can be transformed  into other adapted tetrads by the elements of the stabilizer of the null direction of $l^\mu$, a 4-dimensional subgroup of the proper orthochronal Lorentz group defined by the condition $\Lambda\UD{\alpha}{\beta}\,l^\beta = C\,l^\alpha$ for some $C > 0$, with $\Lambda\UD{\alpha}{\beta}$ denoting the group element. %\MK{Cite something!!! Weinberg??? MTB???? $\textrm{Sim}(2)$ group}. %The action of this group is transitive. 
Any oriented and time-oriented orthonormal tetrad can be transformed into an adapted tetrad by a spatial rotation aligning $e_3^\mu$ with the spatial projection of  $l^\mu$. 

Let us  stress here that we assume  that each of our  coordinate systems and tetrads is  properly oriented and the tetrads are also time-oriented. This is an important restriction, as it makes
the spatial volume 3-form unique and well-defined. We will make use of this fact in Section~\ref{sec:extracting}.

Finally, any coordinate system $(\xi^\mu)$ covering $N_\calO$ and $N_\calE$ provides automatically a coordinate system $(\xi^\mu, \xi^{\mu'})$  on $N_\calO \times N_\calE$. Moreover, it is easy to see that any of the locally flat coordinates at $\calO$ and $\calE$, i.e. LF, OLF or PLF coordinates, produce this way a locally flat coordinate system at the point $(\calO,\calE) \in N_\calO \times N_\calE$ with respect to the metric $\bf h$.

\section{The curvature tensor and the shape of the local surface of communication} \label{sec:shape}

The goal of this section is to derive the first  key  result of this paper, i.e. the relation between the shape of the LSC and the Riemann tensor along the LOS. The derivation uses the Synge's world function for its intermediate steps and therefore we begin by a short review of its definition and properties. 

The  world function is a powerful tool for tracing geodesics between pairs of points \cite{SyngeBook, Poisson2011} and determining which pairs of points can be connected by null geodesics \cite{LinetTeyssandier,Teyssandier:2008, Teyssandier-book}. It also helps to express the relations between geodesics in curved spacetime in a covariant way and to solve the geodesic deviation equations of the first and higher orders \cite{Vines:2014oba}. It has also found its applications in the problem of navigation in a curved spacetime using electromagnetic signals, as in the Global Positioning System (GPS) \cite{Bahder2003, BahderBook}.
We will now briefly remind its most important properties, for a more detailed treatment and full derivations see \cite{SyngeBook, Poisson2011}. In this work we follow the notation and conventions of \cite{Poisson2011}.

The world function $\sigma$ is defined on pairs of points from $M$. Let $x' \in M$ be one point (called the \emph{base point}) and $x \in M$ (the \emph{field point}) be another one belonging to the normal convex neighbourhood of $x'$. In this case there is a unique, affine parametrized geodesic segment $\gamma(\lambda)$
through these two points. We define 
\bea
\sigma(x,x') = \frac{\Delta \lambda}{2}\,\int_{\lambda_{x'}}^{\lambda_x} g_{\mu\nu}\, \frac{\dd x^{\mu}}{\dd \lambda} \, \frac{\dd x^\nu}{\dd \lambda}   \,\dd \lambda, \nonumber
\eea
where the integration is performed along $\gamma(\lambda)$, $\lambda_x$ is the value of the affine parameter $\lambda$ corresponding to $x$, $\lambda_{x'}$ to $x'$ and $\Delta \lambda = \lambda_x - \lambda_{x'}$ is the affine distance between the two endpoints.

The partial derivatives of $\sigma$ wrt the components $x'$ and $x$, taken in any coordinate system, are proportional to the components of the tangent vectors to $\gamma$ at $x$ and $x'$ with lowered indices:
\bea
\sigma_{,\nu} &=& \Delta \lambda\,\left.\frac{\dd x^\mu}{\dd \lambda}\right|_{\lambda=\lambda_x} \,g_{\mu\nu} \label{eq:dsigma1}\\
\sigma_{,\nu'} &=& -\Delta \lambda\,\left.\frac{\dd x^{\mu'}}{\dd \lambda}\right|_{\lambda=\lambda_{x'}} \,g_{\mu'\nu'}. \label{eq:dsigma2}
\eea
Here, following \cite{Poisson2011}, the primed indices refer to the differentiation wrt $x'$ and the unprimed to $x$.  Note that in any coordinates $\sigma_{,\nu} = \sigma_{;\nu}$ and $\sigma_{,\nu'} = \sigma_{;\nu'}$, because the
covariant and the partial first derivatives of a 2-point function always coincide.
$\sigma$ obeys also the identity $2\sigma = \sigma^{;\alpha}\,\sigma_{;\alpha} = \sigma^{;\alpha'}\,\sigma_{;\alpha'}$. Differentiating it covariantly
wrt to $x$ and $x'$ yields the following relations:
\bea
\sigma_{,\nu'} &=& \sigma_{;\mu'\nu'}\,\sigma^{,\mu'} = \sigma_{;\mu\nu'}\,\sigma^{,\mu} \label{eq:ddsigmadsigma1}\\
\sigma_{,\nu}&=& \sigma_{;\mu'\nu}\,\sigma^{,\mu'} = \sigma_{;\mu\nu}\,\sigma^{,\mu} \label{eq:ddsigmadsigma2}
\eea
Moreover, we always have $\sigma_{;\mu\nu} = \sigma_{;\nu\mu}$, $\sigma_{;\mu'\nu'} = \sigma_{;\nu'\mu'}$, $\sigma_{;\mu\nu'} = \sigma_{;\nu'\mu}$.

The applicability of the world function to the problem of signal propagation follows from the observation that $\sigma(x,x')=0$ iff $x$ and $x'$ are linked by a null geodesic \cite{LinetTeyssandier}.
This means that the local surface of communication between the neighbourhoods of $x$ and $x'$ may be identified with the zero level set of $\sigma$, i.e.
$\Sigma = \left\{ (p',q) \in U' \times U  \big| \sigma( p',q) = 0 \right\}$, where $U$ and $U'$ are sufficiently small neighbourhoods of $x$ and $x'$.

%The world function is defined as a function of two points on $M$.
%Introduce the world function \cite{SyngeBook,  Poisson2011}. It can be used for determining the points which can be connected by a null geodesic \cite{LinetTeyssandier,Teyssandier:2008, Teyssandier-book}.
%Namely, its level set corresponding to 0 can be identified with the surface of communication: $\Sigma = \left\{ (p',q) \in N_\calO\times N_\calE \big| \sigma(q, p') = 0 \right\}$.
%Difference in the approach with \cite{Teyssandier:2008, Teyssandier-book}: we do not use any particular coordinate system, fully covariant description. Spacetime geometry
%influence via the curvature along the LOS.

%We begin by considering the 2nd order Taylor expansion of $\sigma$ in a general coordinate system.  From \cite{Poisson2011} we have
%\bea
%\sigma(x,x') = \frac{1}{2}\,\int_{0}^{1} g_{\mu\nu}\, \frac{\dd x^\mu}{\dd \tilde\lambda} \, \frac{\dd x^\nu}{\dd \tilde\lambda} \,\dd \tilde\lambda,
%\eea
%where $\tilde\lambda$ is the affine parametrization of the connecting geodesic $\gamma$ such that $x'$ corresponds to $\tilde\lambda = 0$ and $x$ to $\tilde\lambda = 1$. In this paper we prefer  not to impose 
%any  affine parametrization on our geodesics. We therefore rewrite the formula above in arbitrary affine parametrization $\lambda$:

%\ZZ{Z61 p. 66}, also \cite{Poisson2011}. 

Our goal is to understand the TOA's and the shape of $\Sigma$ in the neighbourhood of $\calE$ and $\calO$. 
We will begin by considering the 2nd order Taylor expansion of $\sigma$ near this pair of points. Let  $x_\calE$ and $x_\calO$ be the coordinates of the points $\calE$ and $\calO$ in a general coordinate system and let $\delta x_\calO$ and $\delta x_\calE$ be small displacements expressed in the same coordinates.  As noted before, for points contained in $N_\calO$ and $N_\calE$  we may
identify $\delta x_\calO$ and $\delta x_\calE$  with tangent vectors at $T_\calO M$ and $T_\calE M$. By convention $\calO$ will play the role of the base point $x'$, while $\calE$ will be identified with the 
field point $x$. In this setup we have the following Taylor expansion:
\bea
\sigma(x_\calE + \delta x_\calE, x_\calO + \delta x_\calO) &=& \sigma_{,\mu'} \,\delta x_\calO^{\mu'} + \sigma_{,\mu} \,\delta x_\calE^{\mu} + \frac{1}{2}\,\sigma_{,\mu'\nu'}\,\delta x_\calO^{\mu'}\,\delta x_\calO^{\nu'} + \sigma_{,\mu'\nu}\,\delta x_\calO^{\mu'}\,\delta x_\calE^\nu \nonumber \\ 
&+& \frac{1}{2}\,\sigma_{,\mu\nu}\,\delta x_\calE^{\mu}\,
\delta x_\calE^{\nu}  + O(\delta x^3), \nonumber
\eea
the higher order terms being of the order of 3 and up in the displacements. Since $\sigma(x_\calE, x_\calO) = 0$ (recall that by assumption $\calE$ and $\calO$ are connected by the fiducial \emph{null} geodesic $\gamma_0$) we have no free term in this expansion.
The partial derivatives of $\sigma$ are evaluated at $(x_\calE, x_\calO)$. Now, while the formula in this form works in any coordinate system, in LF coordinates it takes a particularly useful form:  namely, in this case we have $\sigma_{;\mu\nu} = \sigma_{,\mu\nu}$, $\sigma_{;\mu'\nu'} = \sigma_{,\mu\nu}$. On top of that, we always have
$\sigma_{;\mu\nu'}= \sigma_{,\mu\nu'}$, $\sigma_{;\mu'\nu}= \sigma_{,\mu'\nu}$, $\sigma_{;\mu} = \sigma_{,\mu}$ and $\sigma_{;\mu'}=\sigma_{,\mu'}$, so we can rewrite the Taylor expansion using the \emph{covariant} second derivatives of bitensors only:
\bea
\sigma(x_\calE + \delta x_\calE, x_\calO + \delta x_\calO) &=& \sigma_{;\mu'} \,\delta x_\calO^{\mu'} + \sigma_{;\mu} \,\delta x_\calE^{\mu} + \frac{1}{2}\,\sigma_{;\mu'\nu'}\,\delta x_\calO^{\mu'}\,\delta x_\calO^{\nu'} + \sigma_{;\mu'\nu}\,\delta x_\calO^{\mu'}\,\delta x_\calE^\nu \nonumber\\ 
&+& \frac{1}{2}\,\sigma_{;\mu\nu}\,\delta x_\calE^{\mu}\,
\delta x_\calE^{\nu}  + O(\delta x^3) \label{eq:sigmaexpansion2}
\eea
Therefore in LF coordinates the expansion contains only the covariant derivatives of $\sigma$, which turn out to have a special geometric meaning.  We will explore this fact in the rest of this section.

Define now the 1-by-8 matrix
\bea
 \bfL =  - \frac{1}{\Delta \lambda} \left(\begin{array}{ll} \sigma_{,\mu'} & \sigma_{,\mu} \end{array} \right),
 \label{eq:Lll}
\eea
which constitutes the matrix representation the rescaled gradient of $\sigma$ as a function on $M\times M$ at $(\calO, \calE)$, i.e. the linear mapping $\bfL\colon T_\calO M\oplus T_\calE M \to \mathbb{R}$. From the properties of the first derivatives of the world function (\ref{eq:dsigma1})-(\ref{eq:dsigma2}) we can easily prove that it is in fact composed of the components tangent vectors to $\gamma$, calculated with the  parametrization $\lambda$ at the two endpoints, and
with lowered indices:
\bea
 \bfL =  \left(\begin{array}{ll} l_{\calO\,\mu'} & -l_{\calE\,\mu} \end{array} \right).
\eea
We also  define the 8-by-8 matrix
\bea
 \bfU = -\frac{1}{\Delta\lambda}\,\left( \begin{array}{ll} \sigma_{;\mu'\nu'} & \sigma_{;\mu'\nu} \\ \sigma_{;\mu\nu'} & \sigma_{;\mu\nu}  \end{array} \right)  \label{eq:Udef},
\eea
i.e.  the rescaled Hessian of  $\sigma$. It is the matrix representation of a symmetric bi-linear mapping $\bfU\colon (T_\calO M\oplus T_\calE M) \times  (T_\calO M\oplus T_\calE M) \to \mathbb{R}$.  As a matrix it decomposes naturally into four 4-by-4 matrices in the block decomposition:
\bea
 \bfU = \left( \begin{array}{ll} U_{\calO\calO} & U_{\calO\calE} \\ U_{\calE\calO} & U_{\calE\calE} \end{array} \right) , \nonumber
\eea
with $U_{\calO\calO\,\mu'\nu'}$ and $U_{\calE\calE\,\mu\nu}$ being tensors  at $\calO$ and $\calE$ respectively and $U_{\calO\calE\,\mu'\nu}$ and $U_{\calE\calO\,\mu\nu'}$ being bitensors.

Define now the 8-dimensional vector in $T_\calO M \oplus T_\calE M$, composed of the two displacement vectors \footnote{Note that in the definitions of $M\times M$, $\bfL$ and $\bfX$ we have swapped  the order of coordinates at $\calO$ and $\calE$
in comparison to the definition of the world function: we first take the four components related to the reception region $N_\calO$, and then the four components related to the emission region $N_\calE$, the opposite of the definition of the world function. Hopefully this should not lead to confusion.}
\bea
\bfX = \left(\begin{array}{l} \delta x_{\calO}^{\mu'} \\ \delta x_{\calE}^{\mu} \end{array} \right) \nonumber.
\eea
With these definitions and with the help of (\ref{eq:sigmaexpansion2}) we can re-write the equation $\sigma(x',x)=0$ for the surface of communication $\Sigma$ in LF coordinates as an equation in $T_\calO M \oplus T_\calE M$:
\bea
\bfL (\bfX) + \frac{1}{2} \bfU (\bfX, \bfX) + O(\bfX^3)= 0 \label{eq:sigmaexpansion}
\eea
If we neglect the third and higher order terms we obtain an equation of a 7-dimensional quadric in an 8-dimensional vector space. Equation (\ref{eq:sigmaexpansion}) has no free term, which means that the quadric approximating $\Sigma$ must pass through the origin $\bfX = 0$, i.e. point $(\calO,\calE)$. This is again  a consequence of
our initial assumption that $\calO$ and $\calE$ can be linked by the fiducial \emph{null} geodesic $\gamma_0$. The quadric equation (\ref{eq:sigmaexpansion}) is not the normal form, but, as we show in Appendix \ref{app:normalform}, can be transformed to the normal form. We also discuss there the signature of the quadratic form $\bf U$.

We point out that in this approach $\bfX$, $\bfL$ and $\bfU$ can always be  interpreted as  8-dimensional geometric objects, living on $M\times M$, and with the summation implied in the equations performed over all 8 components of vectors in $T_\calO M \oplus T_\calE M$. However,  we can also divide the   components of $\bfX$  into the 4 components corresponding to $T_\calO M$ and the 4 components corresponding $T_\calE M $, and  sum over them separately. This way the 1-form $\bf L$ decomposes into two 1-forms $l_\calO$ and $l_\calE$ as in (\ref{eq:Lll}) and the 2-form $\bf U$ decomposes  into four 2-forms as in (\ref{eq:Udef}).
While  the first representation is more concise and simpler from the geometric point of view, the second one is more closely related to the spacetime itself and  has therefore a more straightforward physical interpretation. In the rest of the paper we will freely
switch between these two equivalent representations of equations and geometric objects.

\subsection{Physical interpretation and algebraic properties of $\bfL$ and $\bfU$} \label{sec:physinter}

We will now briefly summarize  the algebraic properties and the physical interpretation of the co-vector $\bfL$ and the quadratic form $\bfU$.
We note first that both objects are given up to a common rescaling. Namely, the transformation $\bfL \to  C\cdot \bfL$,  $\bfU \to  C\cdot \bfU$
with $C >0 $ amounts to 
an affine reparametrization of the fiducial null geodesic $\lambda \to C^{-1}\cdot \lambda$ and therefore leaves all physical quantities invariant. Therefore, the global scaling of both objects plays the role of a gauge degree of freedom
\cite{Korzynski:2017nas, Grasso:2018mei}. 

$\bfL$ is always null with respect to the metric $\bf h$, i.e. ${\bf h}^{-1}(\bfL,\bfL) = 0$. Moreover, both of its constituent co-vectors are null with respect to the spacetime metric,
i.e.  $l_\calO^{\mu'}\,l_{\calO\,\mu'} = l_\calE^{\mu}\,l_{\calE\,\mu}  = 0$. Since $l_\calO^{\mu'}$ and $l_\calE^{\mu}$ are the tangent vectors to the fiducial null geodesic $\gamma_0$ at $\calO$ and $\calE$, $\bfL$  describes the way events at and near $\calE$ appear to an observer at $\calO$. Namely,  
$l_\calO^{\mu'}$ defines the apparent position on the sky of an object at $\calE$, as it is registered by observers at $\calO$ and, by extension, the approximate position
of every light signal from $N_\calE$ as seen by observers at $N_\calO$ \cite{Grasso:2018mei}. $l_\calE^{\mu}$ on the other hand is related to the viewing angle, i.e. the direction from which observers at $N_\calO$ observe the events at $N_\calE$. They are also both related to the redshift, or frequency transfer, between a frame defined by a normalized, future-pointing timelike vector $u_\calE^\mu$ at $\calE$ and a frame 
defined by another such vector at $u_\calO^{\mu'}$, via the standard relation $1 + z =\frac{u_\calE^{\mu}\,l_{\calE\,\mu}}{u_\calO^{\mu'}\,l_{\calO\,\mu'}}$ for the redshift $z$ \cite{LinetTeyssandier, Grasso:2018mei}.

From the definition (\ref{eq:Udef})  it is straightforward to see that $\bfU$ is a symmetric matrix, i.e. 
\bea
 \bfU^T = \bfU, \label{eq:Usymmetric}
\eea
or, equivalently, that $U_{\calO\calO\alpha'\beta'} = U_{\calO\calO\beta'\alpha'} $, $U_{\calE\calE\alpha\beta} = U_{\calE\calE\beta\alpha}$, 
$U_{\calE\calO\alpha\beta'} = U_{\calO\calE\beta'\alpha} $.
Additional algebraic relations between $\bfU$ and $\bfL$ follow from the properties for the world function (\ref{eq:ddsigmadsigma1})-(\ref{eq:ddsigmadsigma2}). Combining them with (\ref{eq:Usymmetric})
yields four relations between the submatrices of $\bfU$ and the tangent vectors $l_\calO$, $l_\calE$: 
\bea
 U_{\calO\calO\,\alpha'\beta'}\,l_\calO^{\beta'} &=& U_{\calO\calO\,\beta'\alpha'}\,l_\calO^{\beta'} = -\frac{1}{\Delta\lambda}\,l_{\calO\,\alpha'} \label{eq:Uoowithl} \\
 U_{\calE\calO\,\alpha\beta'}\,l_\calO^{\beta'} &=& U_{\calO\calE\,\beta'\alpha}\,l_\calO^{\beta'}  = \frac{1}{\Delta\lambda}\,l_{\calE\,\alpha} \label{eq:Ueowithl}  \\
 U_{\calO\calE\,\alpha'\beta}\,l_\calE^{\beta} &=& U_{\calE\calO\,\beta\alpha'}\,l_\calE^{\beta}  = \frac{1}{\Delta\lambda}\,l_{\calO\,\alpha'} \label{eq:Uoewithl} \\
 U_{\calE\calE\,\alpha\beta}\,l_\calE^{\beta} &=& U_{\calE\calE\,\beta\alpha}\,l_\calE^{\beta}  = -\frac{1}{\Delta\lambda}\,l_{\calE\,\alpha} \label{eq:Ueewithl} 
\eea
\ZZ{Z62 p.56}

The bilocal operator $\bfU$ has a simple geometric interpretation: it relates the variations of the endpoints of a geodesic to the variations of the tangent vectors at the endpoints, calculated at linear order
around $\gamma_0$.
Let $\delta x_\calO^{\mu'}$ and $\delta x_\calE^{\nu}$ be the endpoints variations, expressed in general coordinates, and let $\delta l_{\calO\,\mu'}$ and $\delta l_{\calE\,\nu}$ be the variations of the components of the
tangent vectors. We introduce the covariant variations $\Delta l_{\calO\,\mu'} = \delta l_{\calO\,\mu'} - \Gamma\UD{\nu'}{\mu'\sigma'}(\calO)\,l_{\calO\,\nu'}\,\delta x_\calO^{\sigma'}$ and
$\Delta l_{\calE\,\mu} = \delta l_{\calE\,\mu} - \Gamma\UD{\nu}{\mu\sigma}(\calE)\,l_{\calE\,\nu}\,\delta x_\calE^{\sigma}$;  with this notation we have
 \bea
 \Delta l_{\calO\,\mu'} &=& U_{\calO\calO\,\mu'\nu'} \,\delta x_\calO^{\nu'} +  U_{\calO\calE\,\mu'\nu}\,\delta x_\calE^{\nu} \label{eq:Udef1}\\
- \Delta l_{\calE\,\mu} &=& U_{\calE\calO\,\mu\nu'} \,\delta x_\calO^{\nu'} +  U_{\calE\calE\,\mu\nu}\,\delta x_\calE^{\nu}, \label{eq:Udef2}
\eea
see Appendix \ref{app:Uvariations} for the proof. 

The basic principle of the measurement we discuss in this paper in based on the fact that $\bfU$ can be expressed as a functional of the curvature tensor along the LOS. 
The relation we present here is indirect: $\bfU$ is related by a non-linear transform to another geometrical object describing the
propagation of light between $N_\calO$ and $N_\calE$, the bilocal geodesic operator, which in turn is related to the curvature as a solution of a simple ODE. We will explain this point in detail. 

Recall first the bilocal geodesic operator (BGO) $\bfW \colon T_\calO M\oplus T_\calO M \to T_\calE M\oplus T_\calE M$, relating the perturbed initial data for a geodesic near $\calO$ to the perturbed final data near $\calE$, discussed extensively in \cite{Grasso:2018mei}. In the context of null geodesics it may be 
seen as a general relativistic generalization of the ray bundle transfer matrix from non-relativistic optics  \cite{Uzun_2020}. $\bfW$ can represented by 4 bitensors
\bea
{\bfW} &=& \left(\begin{matrix} {W_{XX}}\UD{\mu}{\nu'} & {W_{XL}}\UD{\mu}{\nu'} \\ {W_{LX}}\UD{\mu}{\nu'}  & {W_{LL}}\UD{\mu}{\nu'}  \end{matrix} \right), \nonumber
\eea
each mapping vectors form $T_\calO M$ to $T_\calE M$. By definition it
links the displacement and the direction deviation vectors of a perturbed geodesic (not necessary null) at $\lambda = \lambda_\calO$ 
to the displacement and the direction deviation vectors at $\lambda = \lambda_\calE$, at the linear order:
\bea 
\delta x_\calE^\mu &=& W_{XX}{}\UD{\mu}{\nu'}\,\delta x_\calO^{\nu'} + W_{XL}{}\UD{\mu}{\nu'}\,\Delta l_\calO^{\nu'} \label{eq:Wdef1}\\
 \Delta l_\calE^\mu &=& W_{LX}{}\UD{\mu}{\nu'}\,\delta x_\calO^{\nu'} + W_{LL}{}\UD{\mu}{\nu'}\,\Delta l_\calO^{\nu'}  \label{eq:Wdef2},
\eea
with $\Delta l_\calO^{\mu'} = \delta l_\calO^{\mu'}  + \Gamma\UD{\mu'}{\nu'\sigma'}(\calO)\,l_{\calO}^{\nu'}\,\delta x_\calO^{\sigma'}$ and
$\Delta l_\calE^{\mu} = \delta l_\calE^{\mu}  + \Gamma\UD{\mu}{\nu\sigma}(\calE)\,l_{\calE}^{\nu}\,\delta x_\calE^{\sigma}$.
The equations above constitute the relativistic counterpart of the decomposition of the ray bundle transfer matrix into the $ABCD$ blocks in the non-relativistic optics \cite{Uzun_2020} .

One can now show that $\bf U$ between $\calO$ and $\calE$ is related to $\bf W$ between the same pair of points by a nonlinear matrix transform,
derived in Appendix \ref{app:UfromW}. In the block matrix form with indices suppressed it reads
%\bea
% \bfU = \left(  \begin{matrix}
%  -\eta\,W_{XL}^{-1} \,W_{XX} && \eta\,W_{XL}^{-1} \\ \eta\left(W_{LL}\,W_{XL}^{-1}\,W_{XX} - W_{LX}\right) && -\eta\,W_{LL}\,W_{XL}^{-1}
%\end{matrix}   \right). \label{eq:UfromW}
%\eea
\bea
 \bfU = \left(  \begin{matrix}
  -g_{\calO}\,W_{XL}^{-1} \,W_{XX} && g_\calO\,W_{XL}^{-1} \\ g_\calE\left(W_{LL}\,W_{XL}^{-1}\,W_{XX} - W_{LX}\right) && -g_\calE\,W_{LL}\,W_{XL}^{-1}
\end{matrix}   \right). \label{eq:UfromW}
\eea
\ZZ{Z62 p. 34}
Here  $g$ stands for the lower-index metric tensor in the coordinate frame, taken at $\calO$ or $\calE$.  It lowers the first index in each submatrix, consistently with the definition of $\bfU$ above. The transformation (\ref{eq:UfromW}) works in any coordinate system and any tetrads.
 Moreover, since $U_{\calE\calO} = U_{\calO\calE}^T$ there is actually no need to evaluate the complicated, lower-left expression for $U_{\calE\calO}$, it is enough to evaluate the upper-right one for $U_{\calO\calE}$ and transpose.
Note that the transformation works as long as $W_{XL}$ is invertible. With this assumption we can also derive the inverse transform:
\bea\label{eq:WfromU}
 \bfW = \left( \begin{matrix} 
 -\,\uoe^{-1}\uoo && \uoe^{-1}\,g_\calO \\
 g_\calE^{-1}\,(\uee\uoe^{-1}\uoo - \ueo) && -g_\calE^{-1}\,\uee\uoe^{-1}\,g_\calO
 \end{matrix}         \right),
\eea
%\bea\label{eq:WfromU}
 %\bfW = \left( \begin{matrix} 
 %-\,\uoe^{-1}\uoo && \uoe^{-1}\,\eta \\
 %\eta^{-1}\,(\uee\uoe^{-1}\uoo - \ueo) && -\eta^{-1}\,\uee\uoe^{-1}\,\eta
 %\end{matrix}         \right).
%\eea
with $g^{-1}$ denoting the upper-index metric.
We may therefore regard $\bfU$ and $\bfW$ as two equivalent methods of encoding the geometric information about the behaviour of perturbed geodesics passing near
the two endpoints  $\calE$ and $\calO$. $\bfW$ corresponds to the initial value problem, parametrized by the initial data 
$(\delta x_\calO^{\mu'},\Delta l_\calO^{\mu'})$ according to (\ref{eq:Wdef1})-(\ref{eq:Wdef2}), while $\bfU$ governs the boundary problem parametrized by $(\delta x_\calO^{\mu'},\delta x_\calE^\mu)$ via (\ref{eq:Udef1})-(\ref{eq:Udef2}).

The relation between the pair of matrices $W_{XX}$, $W_{XL}$, known as Jacobi propagators \cite{DeWittBrehme, Vines:2014oba, Dixon2}, and  the second derivatives of the world function, proportional to $\bfU$, is an old result dating back to Dixon \cite{Dixon2}, see also \cite{Vines:2014oba}. It 
 has found its applications in the study of equations of motions of massive objects \cite{Poisson2011}. Here we have presented this relation in a  complete form, including all four constituent
bitensors of $\bfW$, and applied it in the context of geometric optics and perturbed light rays.

The relation of $\bfW$ to the curvature is best described in a parallel propagated tetrad along $\gamma_0$ and the corresponding  PLF coordinates. In this case $\bfW$ is simply the resolvent matrix of the first order geodesic deviation equation (GDE) with $\lambda = \lambda_\calO$ as the initial data point \cite{Grasso:2018mei, Uzun_2020}. Namely, for a solution $\xi(\lambda)$ of the GDE 
in a parallel propagated tetrad
\bea
\ddot \xi^{\bar \mu}(\lambda) - R\UD{\bar\mu}{\bar{\alpha}\bar{\beta}\bar{\nu}}(\lambda)\,l^{\bar{\alpha}}\,l^{\bar{\beta}}\,\xi^{\bar \nu}(\lambda) = 0 \nonumber
\eea
with the initial data $\xi(\lambda_\calO)^{\bar\mu} = \delta x_\calO^{\bar\mu}$, $\dot \xi(\lambda_\calO)^{\bar\mu} = \Delta l_\calO^{\bar\mu}$
we have the following relation:
\bea
 \left( \begin{matrix} \xi^{\bar\mu}(\lambda) \\ \dot \xi^{\bar\nu}(\lambda) \end{matrix} \right) = \bfW(\lambda)  \left( \begin{matrix} \delta x_\calO^{\bar\mu} \\ 
 \Delta l_\calO^{\bar\nu} \end{matrix} \right), \nonumber
\eea
with $\bfW(\lambda)$ expressed as an 8-by-8 matrix in the parallel-propagated tetrad. Therefore, we can obtain $\bfW$ between $\calO$ and $\calE$ as a solution of the resolvent matrix ODE in the parallel-propagated tetrad, with the value taken at $\lambda=\lambda_\calE$. This ODE and the initial condition read
\bea
 \dot \bfW(\lambda) &=& \left(\left( \begin{matrix}
 0 && I_4 \\
 0 && 0 \end{matrix} \right) + \left(\begin{matrix}
 0 && 0 \\
 R_{ll}(\lambda)  && 0 \end{matrix}  \right)\right) \bfW(\lambda)  \label{eq:ODEW}\\
 \bfW(\lambda_\calO) &=& I_8
\eea
where $I_n$ is the $n$-dimensional unit matrix and $R_{ll}(\lambda)$ is the optical tidal matrix $R\UD{\bar\mu}{\bar{\alpha}\bar{\beta}\bar{\nu}}(\lambda)\,l^{\bar{\alpha}}\,l^{\bar{\beta}}$ in the PP frame.
\ZZ{(Z61 p. 50.)} 
Note that the equations above are formally identical to the equations for the time evolution operator in quantum mechanics, with $\lambda$ playing the role of the time and with the ``Hamiltonian'' consisting of  a $\lambda$-independent ``free evolution'' term  and 
the $\lambda$-dependent  ``perturbation'' given by appropriate components of the Riemann tensor. The ``free Hamiltonian'' fully governs $\bfW$ (and thus also $\bfU$) in a flat spacetime and it is therefore responsible for the finite-distance optical effects mentioned in Section \ref{sec:basic}. The second term on the other hand generates the curvature corrections present in non-flat spacetimes. This way  we obtain a formal separation of the optical effects into the finite-distance effects, present also in the Minkowski space, and the ``pure" curvature effects.

\begin{comment}
\MK{Necessary paragraph?} Summarizing, $\bfU$ and $\bfL$ govern the TOA's measured between the regions $N_\calE$ and $N_\calO$ via equation (\ref{eq:sigmaexpansion}). $\bfL$ governs the dominating effects related to the placement of the emitter and the receiver along the line of sight, while $\bfU$ encodes the second order effects, usually much smaller, including the transverse variations of the TOA's.  $\bfU$ can be related to the 
spacetime geometry along the LOS indirectly, via relations (\ref{eq:UfromW}) and (\ref{eq:ODEW}). We can distinguish here the finite distance effects, present also in a flat spacetime and related to the distance between the two regions, 
and the curvature effects, related to the Riemann tensor along the LOS. 
\end{comment}

\subsection{Small curvature limit} \label{sec:smallcurvature}

In the absence of strong lensing  we may expect the curvature terms appearing in (\ref{eq:ODEW})  to be small in comparison to the rest.  In Section \ref{sec:expansioninsmall} we 
will make this statement more precise by deriving more precise estimates, but here  we simply   assume that we may treat the curvature terms as small corrections superimposed on top of the finite distance terms. 
The standard procedure in this case is to apply the  perturbative expansion in powers of  the Riemann tensor \cite{PhysRevD.83.083007, PhysRevD.97.084010, PhysRevD.99.084044}. We will follow this approach and truncate the expansion at the first order, effectively linearizing the dependence on the curvature in all relations above. 

The details of the calculations are laid out in  the Appendix: first, the derivation of $\bfW$ at first order from (\ref{eq:ODEW})  is described in  Appendix \ref{app:UW1},
and then the linearization of (\ref{eq:UfromW}) around a ``flat'' solution, leading to the analogous expansion in powers of curvature for $\bfU$, is described in Appendix \ref{app:1PTR}.
As expected,  the final expression for $\bf U$ contains a finite distance term and a curvature correction.  In a parallel transported tetrad and PLF coordinates it reads
\bea
 \bfU = \bfU^{(0)} + \bfU^{(1)}
\eea
with
\bea
 \bfU^{(0)} =  \frac{1}{\Delta\lambda}\left(\begin{matrix} -\eta_{\bar{\mu}\bar\nu} && \eta_{\bar{\mu}\bar\nu} \\ \eta_{\bar{\mu}\bar\nu} && -\eta_{\bar{\mu}\bar\nu} \end{matrix} \right),
 \label{eq:UDform}
\eea
and
\bea
 \bfU^{(1)} =  -\int_{\lambda_\calO}^{\lambda_\calE}   \left(\begin{array}{l|l}  R_{\bar{\mu}\bar\alpha\bar\beta\bar\nu}(\lambda)\,l^{\bar \alpha}\,l^{\bar \beta} \, \frac{(\lambda_\calE - \lambda)^2}{\Delta\lambda^2}\,&
  R_{\bar{\mu}\bar\alpha\bar\beta\bar\nu}(\lambda)\,l^{\bar\alpha}\,l^{\bar\beta}\, \frac{(\lambda - \lambda_\calO)(\lambda_\calE - \lambda)}{\Delta\lambda^2} \\ \hline R_{\bar{\mu}\bar\alpha\bar\beta\bar\nu}(\lambda)\,l^{\bar{\alpha}}\,l^{\bar{\beta}}\,\frac{(\lambda - \lambda_\calO)(\lambda_\calE - \lambda)}{\Delta\lambda^2}  & R_{\bar{\mu}\bar\alpha\bar\beta\bar\nu}(\lambda)\,l^{\bar{\alpha}}\,l^{\bar{\beta}}\,\frac{(\lambda - \lambda_\calO)^2}{\Delta\lambda^2} \end{array} \right) \dd\lambda, \label{eq:URintegral}
\eea
The leading order term $\bfU^{(0)}$ gives the finite distance effects, inversely proportional to the affine distance $\Delta\lambda$ between $\calE$ and $\calO$, while $\bfU^{(1)}$ governs the leading order curvature corrections, proportional to the integrals of the tidal deformation tensor with  kernels of the form of second order polynomials in the affine parameter $\lambda$.
\ZZ{Z61 p.58-62}

\subsection{Times of arrival of pulsed signals} \label{sec:TOAs}

We now use the results derived above to approximate the shape of $\Sigma$ in LF coordinates near $(\calO,\calE)$ by its second order tangent.
 This approximation automatically yields an approximation for the TOA's of pulsed electromagnetic signals between
displaced points.

We begin by solving perturbatively (\ref{eq:sigmaexpansion}) near 
${\bfX} = 0$,
treating $\bfU$ as small. Formally we write $\bfX = \bfX_{(0)} + \bfX_{(1)}$ and obtain the relations:
\bea
\mathbf{L}\left(\mathbf{X}_{(0)}\right) &=& 0 \label{eq:Xperturb1}, \\
\mathbf{L}\left(\mathbf{X}_{(1)}\right) &=& -\frac{1}{2}\mathbf{U}\left(\mathbf{X}_{(0)},\mathbf{X}_{(0)}\right). \label{eq:Xperturb2}
\eea
At the leading, linear order we simply obtain the equation of the tangent space to $\Sigma$, given by $\mathbf{L}\left(\mathbf{X}_{(0)}\right) = 0$. Expressed back in the spacetime notation
this is equivalent to the condition for the time of arrival of the form
\bea
l_{\calO\,\mu'}\,\delta x_\calO^{\mu'} = l_{\calE\,\mu}\,\delta x_\calE^{\mu} \nonumber
\eea
in LF coordinates. This condition is covariant, i.e. it maintains the same form in any coordinate system. Its physical significance has been discussed  in \cite{Grasso:2018mei}, where it is called the \emph{flat lightcones approximation} (FLA).  The FLA takes into account the effects of the delays along the LOS (or R{\o}mer delays), the frequency shift between the rest frames of the emitters and the receivers (of whatever physical origin), but not
the transverse delay effects. In order to evaluate them we need to go to the sub-leading, quadratic order, given by (\ref{eq:Xperturb2}).

Note that from (\ref{eq:Xperturb1}) the leading term $\bfX_{(0)}$ is orthogonal to the normal vector $\bfL$, or $\bfX_{(0)} \in \bfL^{\perp}$, where 
$\bfL^\perp$ denotes the 7-dimensional normal  subspace to $\bfL$. 
Therefore, (\ref{eq:Xperturb2}) depends effectively only on the operator $\bfU$ restricted 
to the subspace $\bfL^{\perp}$. We will denote this restriction as $\bfU^{\perp} = \bfU \big|_{\bfL^{\perp}}$; the equations now take the form of
\bea
\mathbf{L}\left(\mathbf{X}_{(0)}\right) &=& 0 \label{eq:Xperturb3} \\
\mathbf{L}\left(\mathbf{X}_{(1)}\right) &=& -\frac{1}{2}\mathbf{U}^{\perp}\left(\mathbf{X}_{(0)},\mathbf{X}_{(0)}\right) \label{eq:Xperturb4}.
\eea

In order to describe the TOA conditions in detail we introduce an OLF coordinate system, constructed from a pair of tetrads $(u_\calO,e_i)$, $(u_\calE, f_j)$. $u_\calO$ is chosen here to be aligned with the 4-velocities
of the receivers,  $u_\calE$ with the 4-velocities of the emitters, while the remaining spacelike vectors are so far arbitrary. The rest of the equations in this section are expressed in these coordinates.
We introduce the (tetrad-dependent) gauge condition
\bea
 l_{\calO 0'} \equiv {\bf L}_{\bf 0} = 1, \label{eq:gauge}
 \eea
 which fixes  the normalization of $\bfL$ and $\bfU$.
 
 In our measurement the time of arrival $\delta x_\calO^{0'}$ plays a different role than other components of $\bf X$. Namely, we may treat the spatial positions of the
 receivers $\delta x_\calO^{i'}$, the spatial positions of the emitters $\delta x_\calE^i$ and the moments of emission $\delta x_\calE^0$ as controlled by the experimentator. The time of arrival of
 the signals on the other hand is an uncontrolled, measured quantity. Therefore in the subsequent analysis we will treat $\delta x_\calO^{0'}$ as a function of the other seven components, which in turn play the role of the independent variables:
\bea
\bfX^{\bf 0} \equiv \delta x_\calO^{0'} \equiv \tau(\delta x_\calO^{i'},\delta x_\calE^{\mu}) \equiv \tau({\bfX}^{\bf a}), \nonumber
\eea 
with index ${\bf a}$ running from $\bm 1$ to $\bm 7$. The functional dependence is defined implicitly by equation (\ref{eq:sigmaexpansion}), or, more generally, by the vanishing condition for the world function $0=\sigma(x_\calE^\mu + \delta x_\calE^\mu, x_\calO^{0'} + \tau( \delta x_\calO^{i'}, \delta x_\calE^\mu), x_\calO^{i'} + \delta x_\calO^{i'}) $. 
   This way we represent $\Sigma$ locally as a graph of a function $\bfX^{\bf 0} = \tau(\bfX^{\bf a})$ in $T_\calO M\oplus T_\calE M$. In the following calculations we will expand $\tau(\bfX^{\bf a})$ in terms of the other 7 components $\bfX^{\bf a}$ up to the second order, consistently with the expansions (\ref{eq:Xperturb3})--(\ref{eq:Xperturb4}) and (\ref{eq:sigmaexpansion}).
 The expansion can be obtained directly from (\ref{eq:sigmaexpansion}):
\bea
\mathbf{X}^{\bf 0} &=& -\bfL_{\bf a}\,\bfX^{\bf a} - \frac{1}{2}\,{\bf Q}_{\bf a b}\,\bfX^{\bf a}\,\bfX^{\bf b} + O\left(\left({\bf X}^{\bf a}\right)^3\right)\label{eq:TOAfromQX}
\eea
with ${\bf Q}_{\bf a\bf b}$ being a symmetric square matrix of dimension 7 given by
\bea
{\bf Q}_{\bf a b }= \bfU_{\bf a b}  - 2 \bfU_{\bf 0 \left(a\right.} \,\bfL_{\left. \bf b \right)} + \bfU_{\bf 0 0} \, \bfL_{\bf a}\,\bfL_{\bf b}. \nonumber
\eea
Higher order terms on the right hand side are of the order of $O\left(\left(\bf X^{\bf a}\right)^3\right)$ and will be neglected from now on. Returning to the standard notation on $M$ this can be written as
\bea
\delta x_\calO^{0'} &=& -l_{\calO\,i'}\,\delta x_\calO^{i'} + l_{\calE\,\mu}\,\delta x_\calE^\mu - \frac{1}{2}\left(Q_{\calO\calO\,i'j'}\,\delta x_\calO^{i'}\,\delta x_\calO^{j'} 
+ 2Q_{\calO\calE\,i'\mu}\,\delta x_\calO^{i'} \,\delta x_\calE^{\mu}\right.\nonumber\\
&+& \left. Q_{\calO\calO\,\mu\nu}\,\delta x_\calE^{\mu}\,\delta x_\calE^{\nu}\right), \label{eq:TOAfromdeltax}
\eea
 with $Q_{\calO\calO\,i'j'}$, $Q_{\calE\calE\,\mu\nu}$, $Q_{\calO\calE\,i'\mu}$ being the submatrices of $\bf Q_{\bf a \bf b}$:
\bea
Q_{\calO\calO\,i'j'} &=& U_{\calO\calO\,i'j'} - 2U_{\calO\calO\,0'(i'}\,l_{\calO\,j')} + U_{\calO\calO\,0'0'}\,l_{\calO\,i'} \,l_{\calO\,j'} \label{eq:Qoo}\\
Q_{\calE\calE\,\mu\nu} &=& U_{\calE\calE\,\mu\nu} - 2U_{\calO\calE\,0'(\mu}\,l_{\calE\,\nu)} + U_{\calO\calO\,0'0'}\,l_{\calE\,\mu} \,l_{\calE\,\nu} \label{eq:Qee} \\
Q_{\calO\calE\,i'\mu} &=& U_{\calO\calE\,i'\mu} - U_{\calO\calE\,0'\mu}\,l_{\calO\,i'} - U_{\calO\calO\,0'i'}\,l_{\calE\,\mu} + U_{\calO\calO\,0'0'}\,l_{\calO\,i'} \,l_{\calE\,\mu}. \label{eq:Qoe}
\eea 
\ZZ{Z62 p. 31, 32, 57}   Taken together they form ${\bf Q}_{\bf a \bf b}$ according to
\bea
 {\bf Q}_{\bf a \bf b} = \left(\begin{array}{l|l} Q_{\calO\calO} & Q_{\calO\calE} \\ \hline  Q_{\calE\calO} & Q_{\calE\calE} \end{array}\right), \label{eq:Qqq}
\eea
(note that the submatrices here are of unequal shape and size and that $Q_{\calE\calO} = Q_{\calO\calE}^T$). In the terminology of \cite{Teyssandier-book,Teyssandier:2008} equation (\ref{eq:TOAfromdeltax}) provides the Taylor expansion (up to a constant) for
the reception time transfer function ${\cal T}_r$ expressed in locally flat coordinates at $\calO$ and $\calE$.

The 7-dimensional  sub-vectors $\bf X^{\bf a}$ contain  now the  degrees of freedom under control of the experimentator, i.e. the positions of emitters and receivers wrt their local inertial frames, plus the moment of emission in the emitter's frame.
 Together with $\bfX^{\bf 0}$ they form an
8-dimensional vector 
in $T_\calO M \oplus T_\calE M$. The
matrix $\bf Q$ has a simple algebraic interpretation as a method to express $\bf U^\perp$ using the components $\bf a = 1\dots 7$. Namely,
let $E_{\bm \mu}$ denote the 8 basis vectors $(u_\calO, e_i, u_\calE, f_j)$ in $T_\calO M\oplus T_\calE M$. 
Then we  have
\bea
{\bf Q}_{\bf a\bf b} = \mathbf{U}\left( E_{\bf a} - \mathbf{L}_{\bf a}\,E_{\bf 0}, E_{\bf b} - \mathbf{L}_{\bf b}\,E_{\bf 0}\right), \nonumber
\eea
Since the combinations $E_{\bf a} - \bfL_{\bf a}\,E_{\bf 0}$ are all orthogonal to $\bfL$, the components above
can be calculated from $\bf U^{\perp}$ only.

\paragraph{Remark. }We can obtain a deeper geometric insight into the problem if we consider $\Sigma$ as a surface embedded isometrically in $(M\times M, \bf h)$, see
Fig.~\ref{fig:surfaces}. The reader may check that in this interpretation 
$\bf L$ is simply  the normal vector to $\Sigma$ at $(\calO,\calE)$, while $\bf U^\perp$ can be identified with its covariant derivative in the tangent directions, i.e. the extrinsic curvature (or the second fundamental form), of $\Sigma$ at $(\calO,\calE)$.
The vector $\bf L$ is null wrt to the metric $\bf h$ and therefore cannot be normalized: both $\bfL$ and $\bf U^{\perp}$ are thus given only up to rescalings, unless we fix them with a condition
of type (\ref{eq:gauge}).
In this picture the fundamental equations (\ref{eq:TOAfromQX}) and (\ref{eq:TOAfromdeltax}) for the TOA's may be interpreted as resulting simply from approximating $\Sigma$ by its second order tangent, defined  by
the null normal and extrinsic curvature, in locally flat coordinates in $M\times M$. In particular, $\bf Q_{\bf a \bf b}$ represents in this case the extrinsic curvature at $(\calO,\calE)$ expressed in the coordinate system on $\Sigma$ given by
the 7 components $\bf X^{\bf a}$ restricted to $\Sigma$, see again Fig.~\ref{fig:surfaces}. In Appendix~\ref{app:normalform} we discuss in more detail the algebraic properties of  $\bf U^\perp$.

\bfi
\includegraphics[width=0.8\textwidth]{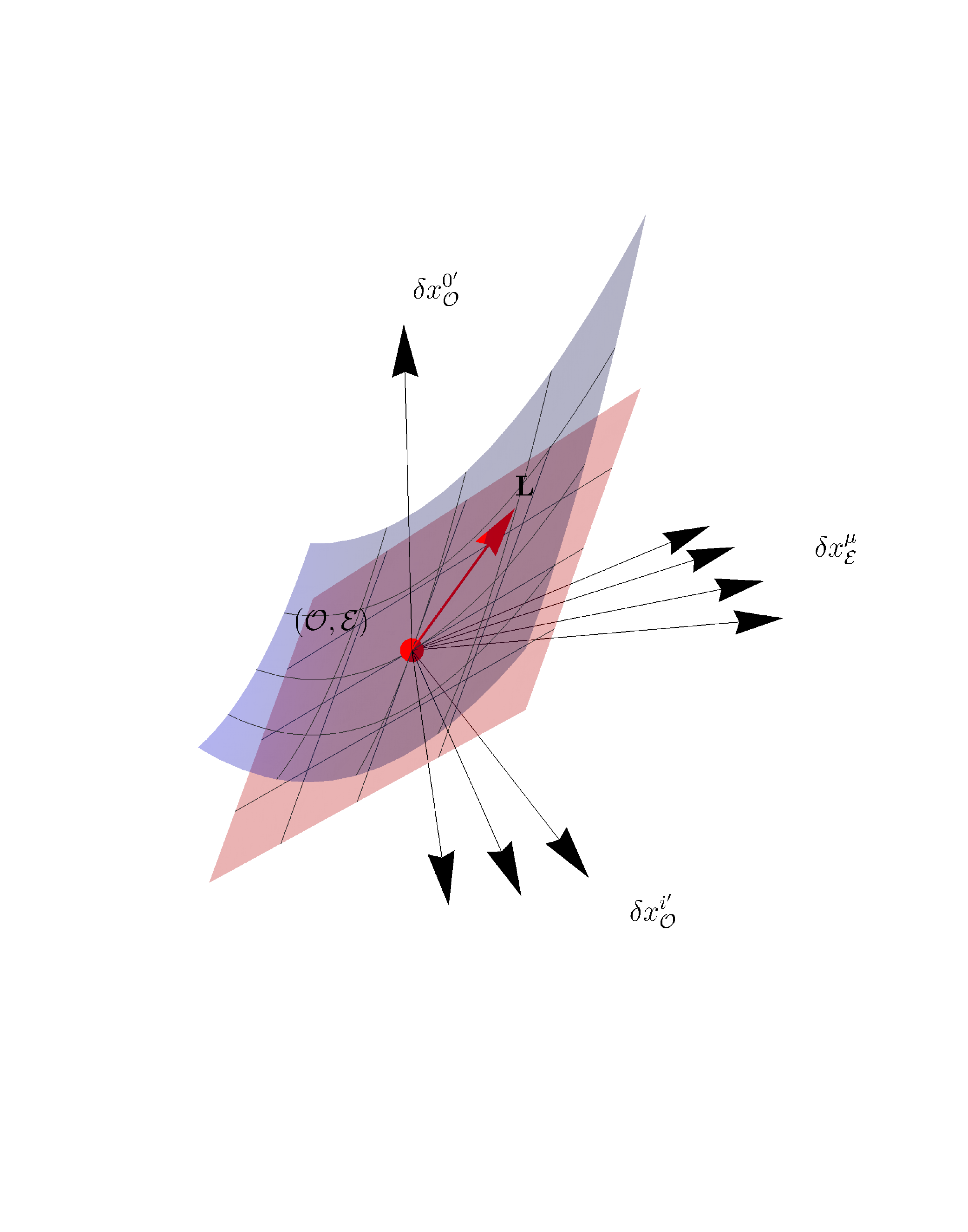}
\caption{Approximating of $\Sigma$ by its first and second order tangents hypersurfaces in $M\times M$. The first order tangent is parametrized 
just by the normal vector $\bf L$. The second order tangent requires also the extrinsic curvature $\bf Q$. The surface may be then represented by the graph with $\delta x_\calO^{0'}$ as the dependent variable.}
\label{fig:surfaces}
\efi

\begin{comment}
\MK{Repetition?} Summarizing, the covector $\bf L$ controls the line-of-sight effects in  the TOA's between $N_\calE$ and $N_\calO$, i.e.  the R{\o}mer delays and the redshift  effects (the Doppler  effect etc.). It
does not govern the transverse delays though. The matrix $\bf Q_{\bf a \bf b}$ on the other hand
determines the higher order effects, including the transverse delays of  the TOA's. It is made of components of $\bf U^\perp$, itself related to the curvature along the LOS via (\ref{eq:UfromW}) and (\ref{eq:ODEW}).
It contains the finite distance effects mixed with (usually much smaller) curvature corrections, see (\ref{eq:UDform})-(\ref{eq:URintegral}).
\end{comment}

\subsection{The inverse problem:  $\bfL$ and $\bfU^\perp$ from the measurements of TOA's. } \label{sec:inverse}

Equations (\ref{eq:TOAfromQX}) and (\ref{eq:TOAfromdeltax}) provide also a method for the inverse problem, i.e. the problem of determining the components of $\bfL$ and $\bf Q$ from measurements. As our input dataset  we may consider a finite  sample
of pairs of events connected by null geodesics. Assume thus that we have obtained by direct measurements a sample of pairs of points $\delta x_{\calE\,(i)}^\mu$, 
$\delta x_{\calO\,(i)}^{\mu'}$, $i = 1,\dots,n$, in $N_\calE$ and $N_\calO$, such that signals from the former event reach the latter. The pairs form a set of 8-vectors
 $\bfX_{(i)}$. We can now write down equation
(\ref{eq:TOAfromQX}) for each $\bfX_{(i)}$, expressed in a system of OLF coordinates, and treat the resulting system of $n$ equations as a linear system  for the components of $\bf L$ and $\bf Q$, with the components of $\bfX_{(i)}$ now considered known. 
Since the  number of independent coordinates of $\bfL$ and $\bf Q$, subject to (\ref{eq:gauge}), is $7+28=35$, and since the problem is linear, we expect that with at least 35 independent measurements we may determine this way all components of $\bf L$ and $\bf Q$.\footnote{In practice the number of measurements can be made lower because the components of $\bfL$ and $\bfU$ are not independent, see (\ref{eq:Uoowithl})-(\ref{eq:Ueewithl}). However,  we do not make any use of this possibility in this paper.}

Geometrically this procedure is equivalent to determining the shape of the LSC $\Sigma$ in the 2nd order Taylor approximation  by probing it at a finite number of points near the origin. In this approximation $\Sigma$ must be a quadric passing through the origin, therefore its shape in the end parametrized by 35 numbers. The inverse problem becomes then the question of finding a unique quadric of type (\ref{eq:TOAfromdeltax}) through a set of known points
in dimension 8. 

Note that the feasibility of this procedure crucially depends on the linear independence of the resulting equations for $\bf Q_{\bf a b}$ and $\bfL_{\bf a}$. This obviously imposes additional restrictions on
the choice of the points over which we probe the TOA's. The detailed analysis of these restrictions is an algebraic geometry problem in  dimension 8 and it
 is beyond the scope of this article. We just note here that  the independence condition for the equations is equivalent to a nondegeneracy condition for a 35-by-35 matrix: namely, the determinant made  of 
the components and products of components of the 7-vectors $\bfX_{(i)}^{\bf a}$ should not vanish. We may therefore expect the independence to hold generically, i.e. always except a measure-zero set of 
degenerate configurations of sampling points. However, in this paper we do not attempt to provide a rigorous proof;   in Section \ref{sec:protocol} we simply present a particular choice of sampling points (i.e. a choice of the spatial positions of the emitters and the times of emission as well as the spatial positions of the receivers) for which we may prove that the inverse problem can always be
solved. Nevertheless, many other sampling strategies should be possible here. 

  We also note here that the  reconstruction of  $\bfU$, defining the shape of the surface of communication via the quadric equation (\ref{eq:sigmaexpansion}), from the TOA measurements near $\calO$ and $\calE$, is necessary incomplete: as long as
the approximation (\ref{eq:TOAfromQX}) is valid, only the components of $\bfU^\perp$ can be reconstructed, while  the rest is hidden in the unobservable, higher order effects in $\bfX^{\bf a}$. However, as we will see in the next section, this is not really a problem, because $\bfU^\perp$ contains all the information about the matter density along the line of sight we need.

\section{Two scalar quantities measuring the  curvature imprint} \label{sec:two}

The operator $\bfU^\perp$ consists of the finite distance part and much smaller curvature corrections.  In order to evaluate the curvature imprint we need to find a way to measure the latter independently of the former. It should preferably be done in  a covariant, tetrad-independent manner. In this section we present two functions of $\bfU^\perp$ and $\bfL$ which measure the curvature impact in a tetrad-invariant way.

Let    $(u_\calO, f_A, f_3)$  and $(u_\calE, g_A, g_3)$ be a pair of adapted tetrads at $\calO$ and $\calE$ respectively. %\MK{Change to $(u_\calO, f_A, f_3)$  and $(u_\calE, k_A, k_3)$, $g$ is also used for the metric tensor, $e$ for a general ON frame.} 
In the previous papers \cite{Grasso:2018mei, korzynskivilla} we have introduced the distance slip, i.e. a scalar, dimensionless quantity, defined as
\bea 
 \mu = 1 - \frac{\det \Pi\UD{A'}{B'}}{\det M\UD{A'}{B}} . \nonumber
\eea
Here $\Pi\UD{A'}{B'}$ is the parallax matrix, describing the linear relation between the perpendicular displacement of the 
observer $\delta x_\calO^{B'}$ and the apparent displacement of the image $\delta \theta^{A'} = (u_\calO^{\mu'}\,l_{\calO\,\mu'})^{-1}\,\Delta l_\calO^{A'}$ of an object at $\calE$ on the observer's celestial sphere, expressed in radians: 
\bea
 \delta \theta^{A'} = -\Pi\UD{A'}{B'}\,\delta x_\calO^{B'}. \nonumber
\eea
$M\UD{A'}{B}$ on the other hand 
 is the magnification matrix, relating the perpendicular displacement of the emitter $\delta x_\calE^B$ to the apparent displacement of the image $\delta \theta^{A'}$ at linear order:
 \bea
 \delta \theta^{A'} = M\UD{A'}{B}\,\delta x_\calE^B. \nonumber
 \eea
 $\mu$ measures the spacetime curvature along the LOS by comparing two methods of distance determination to a single object: by parallax and by the angular size. It vanishes in a flat spacetime, or whenever the Riemann tensor along the LOS vanishes \cite{Grasso:2018mei, korzynskivilla}.
From (\ref{eq:Udef1}) we can prove that this quantity can be calculated directly from $\bfU$ via the ratio of the determinants of its two transverse submatrices:
\bea
 \mu = 1 - \frac{\det U_{\calO\calO\,A'B'}}{\det U_{\calO\calE\,A'B}} = 1 - \frac{\det U_{\calO\calO}{}\UD{A'}{B'}}{\det U_{\calO\calE}{}\UD{A'}{B}}, \label{eq:muviaU}
\eea
We can also define its counterpart with the roles of $\calO$ and $\calE$ reversed:
\bea
 \nu = 1 - \frac{\det U_{\calE\calE\,AB}}{\det U_{\calO\calE\,A'B}}  \label{eq:nuviaU}
\eea
It is also a dimensionless scalar and its physical interpretation is  similar to that of $\mu$: it measures the difference of the variations of the viewing angle wrt to displacements at $\calO$ and at $\calE$,
and $\nu = 0$ in a flat spacetime. Both quantities are insensitive to the rescalings of $\bfU$ and $\bfL$ mentioned in Section \ref{sec:physinter}.

The definitions of  $\mu$ and $\nu$ between events $\calO$ and $\calE$ can be reformulated in terms of the cross-sectional area of infinitesimal bundles of null geodesics along $\gamma_0$, parallel at $\calO$ and $\calE$ respectively. The language of bundles of null geodesics and its evolution along a null geodesic is probably more familiar to most relativists than the bilocal formulation (see for example \textcite{perlick-lrr}), so we present this formulation in Appendix~\ref{app:cross}. As a side result we also derive alternative expressions for $\mu$ and $\nu$ in terms of ODE's involving the shear and expansion of an appropriate bundle.

\paragraph{Independence from the adapted tetrads.} While the definitions above require introducing two adapted tetrads at $\calO$ and $\calE$, both these quantities are in fact independent of the choice of these tetrads. More formally, the expressions (\ref{eq:muviaU}) and (\ref{eq:nuviaU}) define 
two functions of the components of $\bfU$ expressed in adapted tetrads: $\mu \equiv \mu(U_{\calO\calO\,A'B'}, U_{\calO\calE\,A'B})$, $\nu \equiv \nu(U_{\calE\calE\,A'B'}, U_{\calO\calE\,A'B})$. On the other hand, we may also regard them as functions of the  ``abstract'', tetrad-independent linear mappings $\bfU$, $\bfL$, and a pair of adapted ON tetrads $(u_\calO, f_A, f_3)$,
$(u_\calE, g_A, g_3)$ in which we have expressed their components:
\bea
  \mu &\equiv & \mu\left(\mathbf{U},\mathbf{L},\left(u_\calO, f_A, f_3\right), \left(u_\calE, g_A, g_3\right)\right) \label{eq:mufromG2}  \\
  \nu &\equiv & \nu\left(\mathbf{U},\mathbf{L},\left(u_\calO, f_A, f_3\right), \left(u_\calE, g_A, g_3\right)\right). \label{eq:nufromG2} 
 \eea
In the Appendix \ref{app:proof} we show explicitly that for any other pair of adapted ON tetrads $(\tilde u_\calO, \tilde f_A, \tilde f_3)$ and $(\tilde u_\calE, \tilde g_A, \tilde g_3)$
the values of both $\mu$ and $\nu$ are the same. We do it by showing that the transverse submatrices of $U_{\calO\calO}$, $U_{\calO\calE}$ and $U_{\calE\calE}$ transform
very simply under the change of the adapted ON tetrads, i.e. via  two-dimensional rotations of the transverse components. These transformations do not affect the values of the determinants 
in  (\ref{eq:muviaU}) and (\ref{eq:nuviaU}). This way we prove that $\mu$ and $\nu$ are adapted tetrad-independent functions of the geometric objects $\bfU$ and $\bfL$ only:
\bea
  \mu &\equiv & \mu\left(\mathbf{U},\mathbf{L}\right)  \label{eq:muinvariance} \\
  \nu &\equiv & \nu\left(\mathbf{U},\mathbf{L}\right).  \label{eq:nuinvariance} 
 \eea

\subsection{Extracting $\mu$ and $\nu$ from  $\bf Q$.} \label{sec:extracting}
 
We have defined $\mu$ and $\nu$ as a function of  $\bfL$ and $\bfU$. However, we can show that it is only the components of $\bfU^\perp$ which contribute, i.e.
 \bea
  \mu &\equiv &   \mu\left(\mathbf{U}^\perp,\mathbf{L}\right) \label{eq:muofUperpL}\\
  \nu &\equiv &  \nu\left(\mathbf{U}^\perp,\mathbf{L}\right)  \label{eq:nuofUperpL}.
 \eea
Define first the four transverse 8-vectors labelled by $\calO$, $\calE$ and $A=1,2$:
\bea
{\bf Z}_{\calO\,A} &=& \left(\begin{array}{l} f_A^{\mu'} \\ 0 \end{array}\right)  \nonumber\\
{\bf Z}_{\calE\,A} &=& \left(\begin{array}{l} 0 \\ g_A^\mu \end{array}\right). \nonumber
\eea 
The reader may check that all four are orthogonal to $\bfL$. Now, we have obviously $U_{\calO\calO\,AB} = \bfU\left({\bf Z}_{\calO\,A},{\bf Z}_{\calO\,B}\right)$,
$U_{\calE\calO\,AB} = \bfU\left({\bf Z}_{\calE\,A},{\bf Z}_{\calO\,B}\right)$, $U_{\calE\calE\,AB} = \bfU\left({\bf Z}_{\calE\,A},{\bf Z}_{\calE\,B}\right)$. But since all transverse 8-vectors  are
orthogonal to $\bfL$, we simply have $U_{\calO\calO\,AB} = \bfU^\perp\left({\bf Z}_{\calO\,A},{\bf Z}_{\calO\,B}\right)$,
$U_{\calE\calO\,AB} = \bfU^\perp\left({\bf Z}_{\calE\,A},{\bf Z}_{\calO\,B}\right)$, $U_{\calE\calE\,AB} = \bfU^\perp\left({\bf Z}_{\calE\,A},{\bf Z}_{\calE\,B}\right)$, i.e. we only need 
$\bfU^\perp$ to calculate them. It follows that we can calculate $\mu$ and $\nu$ via (\ref{eq:muviaU}) and (\ref{eq:nuviaU}) just from $\bfU^\perp$. This in turn means that
it should be possible to express both scalars by the components of $\bf Q_{\bf a b}$ expressed in a pair of adapted tetrads.

During the measurement we cannot expect the spatial vectors of the tetrads we use for defining the positions and directions to be aligned precisely  along the line of sight. Therefore, we need to relax the requirement for the ON tetrads to be adapted. We will simply assume
 that after the measurement we are given ${\bf Q}_{\bf a b}$ in an \emph{arbitrary} pair of ON (and properly oriented) tetrads $(u_\calO, e_i)$ and $(u_\calE, f_j)$. Note that after the measurement of the components of $\bfL$ we can simply rotate both tetrads
to make $e_3^{\mu'}$ and $f_3^\mu$ aligned with $l_\calO^{\mu'}$ and $l_\calE^{\mu}$ and later work in a pair of adapted tetrads. The reader may check that in this case
the transverse components of  $U_{\calO\calO}$ and $U_{\calO\calE}$ and $U_{\calE\calE}$ coincide with the corresponding 
components of $Q_{\calO\calO}$, $Q_{\calO\calE}$, $Q_{\calE\calE}$ (see equations (\ref{eq:Qoo})-(\ref{eq:Qee})), so we can directly substitute $Q_{\calO\calO\,A'B'}$, $Q_{\calO\calE\,A'B}$ and $Q_{\calE\calE\,AB}$
to (\ref{eq:muviaU})-(\ref{eq:nuviaU}).   However, the tetrad adjustment can be avoided if we introduce more general, rotationally invariant definitions of $\mu$, $\nu$.

Given the decomposition of $\bf Q_{\bf a b}$ in not necessary adapted $(u_\calO, e_i)$, $(u_\calE, f_j)$ we define
\bea
 \mu &=& 1 - \frac{Q_{\calO\calO\,i'j'}\,Q_{\calO\calO\,k'l'} \, \epsilon_\calO^{i'k'}\,\epsilon_\calO^{j'l'}}{Q_{\calO\calE\,i'j}\,Q_{\calO\calE\,k'l} \, \epsilon_\calO^{i'k'}\,\epsilon_\calE^{jl}} \label{eq:mufromQepsilon} \\
 \nu &=& 1 - \frac{Q_{\calE\calE\,ij}\,Q_{\calE\calE\,kl} \, \epsilon_\calE^{ik}\,\epsilon_\calE^{jl}}{Q_{\calO\calE\,i'j}\,Q_{\calO\calE\,k'l} \, \epsilon_\calO^{i'k'}\,\epsilon_\calE^{jl}}, \label{eq:nufromQepsilon} 
\eea
with the two antisymmetric, spatial 2-tensors
\bea
  \epsilon_\calO^{i'j'} &=& \frac{1}{l_{\calO\,0'}} \,\varepsilon^{i'j'k'}_\calO\,l_{\calO\,k'} \label{eq:epsilonO}\\
  \epsilon_\calE^{ij} &=& \frac{1}{l_{\calE\,0}}\,\varepsilon^{ijk}_\calE\,l_{\calE\,k}. \label{eq:epsilonE}
\eea
$\varepsilon_\calO^{i'j'k'}$ and $\varepsilon_\calE^{ijk}$ denote here the standard, totally antisymmetric spatial 3-tensors, with $\varepsilon_\calO^{1'2'3'} = \varepsilon_\calE^{123} = 1$, i.e. the $SO(3$)-invariant,  upper-index spatial volume forms. 
 The expressions on the right hand sides of (\ref{eq:mufromQepsilon})-(\ref{eq:nufromQepsilon}) are obviously SO(3)-invariant in both $N_\calE$ and $N_\calO$, and thus they are insensitive 
 to spatial rotations of the tetrads at $\calO$ and $\calE$. Thus they can also be applied with non-adapted ON tetrads. On the other hand,  if we do align the
 vectors $f_3$ and $g_3$ with $l_\calO$ and $l_\calE$ respectively, making this way both tetrads adapted, then
 \bea
  \epsilon_\calO^{i'j'} &=& \left\{\begin{array}{rl} 1 & \textrm{for } i'=1',  j'=2' \\ -1 & \textrm{for } i'=2', j'=1' \\ 0 & \textrm{otherwise} \end{array} \right. \nonumber \\
  \epsilon_\calE^{ij} &=& \left\{\begin{array}{rl} 1 & \textrm{for } i=1,  j=2 \\ -1 & \textrm{for } i=2, j=1 \\ 0 & \textrm{otherwise} \end{array} \right. \nonumber
 \eea
It is now easy to see that in equations (\ref{eq:mufromQepsilon})-(\ref{eq:nufromQepsilon}), expressed in adapted tetrads, the contractions of $Q_{\calO\calO}$, $Q_{\calO\calE}$ and $Q_{\calE\calE}$  with $\epsilon_\calO^{i'j'}$ and $\epsilon_\calE^{ij}$ automatically yield their transverse subdeterminants. These are in turn equal to the corresponding transverse subdeterminants of $U_{\calO\calO}$, $U_{\calO\calE}$ and $U_{\calE\calE}$ because of (\ref{eq:Qoo})-(\ref{eq:Qoe}). Therefore, in a pair of adapted ON tetrads we have (\ref{eq:mufromQepsilon}) equal to (\ref{eq:muviaU}) and (\ref{eq:nufromQepsilon}) equal to (\ref{eq:nuviaU}). 

Summarizing, we have proved that both definitions
 (\ref{eq:mufromQepsilon})-(\ref{eq:nufromQepsilon}) and  (\ref{eq:muviaU})-(\ref{eq:nuviaU}) are equivalent, but the equations (\ref{eq:mufromQepsilon})-(\ref{eq:nufromQepsilon})  can be used with \emph{any} pair of ON, oriented and time-oriented tetrads at $\calO$ and $\calE$. In other words,  (\ref{eq:mufromQepsilon})-(\ref{eq:nufromQepsilon}) are fully invariant with respect to proper, orthochronal Lorentz transforms on both geodesic endpoints.
 
\paragraph{Remarks.} In a similar way we may introduce fully Lorentz-invariant versions of (\ref{eq:muviaU}) and (\ref{eq:nuviaU}):
 \bea
 \mu &=& 1 - \frac{U_{\calO\calO\,i'j'}\,U_{\calO\calO\,k'l'} \, \epsilon_\calO^{i'k'}\,\epsilon_\calO^{j'l'}}{U_{\calO\calE\,i'j}\,U_{\calO\calE\,k'l} \, \epsilon_\calO^{i'k'}\,\epsilon_\calE^{jl}} \label{eq:mufromUepsilon} \\
 \nu &=& 1 - \frac{U_{\calE\calE\,ij}\,U_{\calE\calE\,kl} \, \epsilon_\calE^{ik}\,\epsilon_\calE^{jl}}{U_{\calO\calE\,i'j}\,U_{\calO\calE\,k'l} \, \epsilon_\calO^{i'k'}\,\epsilon_\calE^{jl}}. \label{eq:nufromUepsilon} 
\eea
Here both scalars are formally expressed as functions of $\bfU^\perp$ and $\bfL$ as well as a pair of properly oriented ON tetrads:
 \bea
  \mu &\equiv & \mu\left(\mathbf{U}^\perp,\mathbf{L}, \left(u_\calO, e_i\right), \left(u_\calE, f_i\right)\right)  \label{eq:mufromON} \\
  \nu &\equiv & \nu\left(\mathbf{U}^\perp,\mathbf{L}, \left(u_\calO, e_i\right), \left(u_\calE, f_i\right)\right),  \label{eq:nufromON} 
 \eea
 with the tetrad dependence effectively spurious, as in (\ref{eq:muofUperpL})-(\ref{eq:nuofUperpL}).

\subsection{$\mu$ and $\nu$ and the stress-energy tensor}
 From (\ref{eq:URintegral}), (\ref{eq:muviaU}) and (\ref{eq:nuviaU}) respectively we can show that in the leading order of the expansion in the curvature
\bea
 \mu &=& 8\pi\,G \int_{\lambda_\calO}^{\lambda_\calE} T_{ll}(\lambda) \left(\lambda_\calE - \lambda\right)\,\dd\lambda \label{eq:muTintegral} + O\left(\textrm{Riem}^2\right) \\
 \nu &=& 8\pi\,G  \int_{\lambda_\calO}^{\lambda_\calE} T_{ll}(\lambda) \left(\lambda - \lambda_\calO\right)\,\dd\lambda \label{eq:nuTintegral} + O\left(\textrm{Riem}^2\right), 
\eea
where $T_{ll}(\lambda) \equiv  T_{\mu\nu}\,l^{\mu}\,l^{\nu}$. Thus both $\mu$ and $\nu$ effectively depend  on a single component of the stress-energy tensor out of the whole Riemann tensor.
 Equations (\ref{eq:muTintegral})-(\ref{eq:nuTintegral}) are most conveniently derived in a parallel-transported adapted tetrad $(\hat u, \hat e_A, \hat e_3)$, although note  that
both equations are in fact covariant and therefore valid in all possible tetrads or coordinates. 
Since the equation for $\mu$ has already been derived in \cite{Grasso:2018mei}, we present in detail the derivation of equation (\ref{eq:nuTintegral}) for $\nu$. 

 From (\ref{eq:UDform}) and (\ref{eq:URintegral}) we get the first two terms of expansion of the transverse components of $U_{\calE\calE}$ and $U_{\calO\calE}$ in the powers of the Riemann tensor:
\bea
U_{\calE\calE\,\bar{A}\bar{B}} &=& - \Delta\lambda^{-1}\,\delta_{\bar{A}\bar{B}} -  \int_{\lambda_\calO}^{\lambda_\calE} R_{\bar{A}\bar{\mu}\bar{\nu}\bar{B}}(\lambda)\,
l^{\bar{\mu}}\,l^{\bar{\nu}}\,\frac{\left(\lambda - \lambda_\calO\right)^2}{\Delta\lambda^2}\,\dd\lambda + O\left(\textrm{Riem}^2\right)\nonumber \\
U_{\calO\calE\,\bar{A}\bar{B}} &=&  \Delta\lambda^{-1}\,\delta_{\bar{A}\bar{B}} -  \int_{\lambda_\calO}^{\lambda_\calE} R_{\bar{A}\bar{\mu}\bar{\nu}\bar{B}}(\lambda)\,
l^{\bar{\mu}}\,l^{\bar{\nu}}\,\frac{\left(\lambda_\calE - \lambda\right)\left(\lambda-\lambda_\calO\right)}{\Delta\lambda^2}\,\dd\lambda + O\left(\textrm{Riem}^2\right) .\nonumber 
\eea
From this we get the expansion of the determinants
\bea
\det U_{\calE\calE\,\bar{A}\bar{B}} &=& \Delta\lambda^{-2}  + \Delta\lambda^{-1}\, \int_{\lambda_\calO}^{\lambda_\calE} R\UD{\bar{A}}{\bar{\mu}\bar{\nu}\bar{A}}(\lambda)\,
l^{\bar{\mu}}\,l^{\bar{\nu}}\,\frac{\left(\lambda - \lambda_\calO\right)^2}{\Delta\lambda^2}\,\dd\lambda + O\left(\textrm{Riem}^2\right) \nonumber \\
\det U_{\calO\calE\,\bar{A}\bar{B}} &=&  \Delta\lambda^{-2} - \Delta\lambda^{-1}\, \int_{\lambda_\calO}^{\lambda_\calE} R\UD{\bar{A}}{\bar{\mu}\bar{\nu}\bar{A}}(\lambda)\,
l^{\bar{\mu}}\,l^{\bar{\nu}}\,\frac{\left(\lambda_\calE - \lambda\right)\left(\lambda-\lambda_\calO\right)}{\Delta\lambda^2}\,\dd\lambda + O\left(\textrm{Riem}^2\right) .\nonumber 
\eea
In the next step we simplify the integrands. Note that for a null vector $l^{\bar{\mu}}$ the contraction of the Riemann over the two transverse indices is equal to the full contraction over the whole space  \cite{Grasso:2018mei}, 
i.e. $R\UD{\bar A}{\bar \mu \bar \nu \bar A}\,l^{\bar{\mu}}\,l^{\bar{\nu}} = R\UD{\bar \alpha}{\bar \mu \bar \nu \bar \alpha}\,l^{\bar{\mu}}\,l^{\bar{\nu}} = -R_{\bar\mu \bar\nu}\,l^{\bar{\mu}}\,l^{\bar{\nu}} $. Moreover, from the Einstein equations
\bea
R_{\mu\nu} - \frac{1}{2}\,R\,g_{\mu\nu} = 8\pi G\,T_{\mu\nu}-\Lambda\,g_{\mu\nu}\nonumber 
\eea
it is easy to show that $R_{\bar\mu\bar\nu}\,l^{\bar\mu}\,l^{\bar\nu} = 8\pi G\, T_{\bar\mu\bar\nu}\,l^{\bar\mu}\,l^{\bar\nu}$, i.e. only the matter stress-energy tensor contributes to both integrals, while the cosmological constant drops out, thus
\bea
\det U_{\calE\calE\,\bar{A}\bar{B}} &=& \Delta\lambda^{-2}  - \Delta\lambda^{-1}\, 8\pi G\,\int_{\lambda_\calO}^{\lambda_\calE} T_{ll}(\lambda)\,\frac{\left(\lambda - \lambda_\calO\right)^2}{\Delta\lambda^2}\,\dd\lambda + O\left(\textrm{Riem}^2\right) \nonumber  \\
\det U_{\calO\calE\,\bar{A}\bar{B}} &=&  \Delta\lambda^{-2} + \Delta\lambda^{-1}\, 8\pi G\,\int_{\lambda_\calO}^{\lambda_\calE} T_{ll}(\lambda)\,
\frac{\left(\lambda_\calE - \lambda\right)\left(\lambda-\lambda_\calO\right)}{\Delta\lambda^2}\,\dd\lambda + O\left(\textrm{Riem}^2\right) , \nonumber 
\eea
From this  and (\ref{eq:nuviaU}) we get the expansion of $\nu$ in the form of
\bea
\nu &=& \Delta\lambda \,8\pi G\int_{\lambda_\calO}^{\lambda_\calE} T_{ll}(\lambda)\,\left(\frac{\left(\lambda-\lambda_\calO\right)^2}{\Delta\lambda^2} + \frac{\left(\lambda_\calE-\lambda\right)(\lambda - \lambda_\calO)}{\Delta\lambda^2}\right)\,\dd\lambda + O(\textrm{Riem}^2).  \label{eq:muint}%\\
% &=& 8\pi G\int_{\lambda_\calO}^{\lambda_\calE} T_{ll}(\lambda)\,\left(\lambda_\calE - \lambda\right)\,\dd\lambda + O(\textrm{Riem}^2). \label{eq:muint}
\eea
Simplifying the integrand we obtain (\ref{eq:nuTintegral}).
Equation (\ref{eq:muTintegral}) for $\mu$ can be derived the same way from the analogous expansion of $U_{\calO\calO\,\bar{A}\bar{B}}$ in the powers of $R\UD{\mu}{\nu\alpha\beta}$. 
However, note that there is also a short-cut reasoning leading directly  from (\ref{eq:nuTintegral}) to (\ref{eq:muTintegral}): since $\nu$ is simply the same quantity as $\mu$, but with the role of the geodesic endpoints $\calO$ and $\calE$ reversed, we can simply obtain (\ref{eq:muTintegral})   from  (\ref{eq:nuTintegral}) by consistently swapping the $\calO$ and $\calE$ subscripts on the right hand side. 

We also point out that the expansions (\ref{eq:muTintegral})-(\ref{eq:nuTintegral}) in the curvature contain no free term. It follows that $\mu=\nu = 0$ in the absence of curvature.

\section{Expansion in small parameters} \label{sec:expansioninsmall}
 \ZZ{Z62 p. 81-96} 
We will now  compare the magnitude of various terms in equation (\ref{eq:sigmaexpansion}) governing the TOA's in a typical astrophysical situation. We assume that the curvature tensor is of the scale of $\calR^{-2}$, where $\calR$ is the curvature radius scale and $\bfX$ is of the order of  the size of $N_\calO$ and $N_\calE$, denoted by $L$.  For simplicity we pick the affine parameter $\lambda$  normalized in the receivers' frame, i.e. $l_{\calO\,\mu}\,u_{\calO}^\mu = 1$. In this case we can also assume that $\lambda_\calO = 0$ and $\lambda_\calE = D$, where $D$ is the affine distance from $\calO$ to $\calE$ in the receivers' frame. 

We introduce the dimensionless counterparts of the objects appearing in (\ref{eq:sigmaexpansion}), (\ref{eq:UDform}) and (\ref{eq:URintegral}), denoted by tilde:
\bea
 \lambda &=& D\,\widetilde \lambda \\
 \bfX &=& L\,\widetilde \bfX \\
 \bfL &=& \widetilde \bfL \\
 l^{\bar\mu} &=& \widetilde l^{\bar\mu} \\
 R\UD{\bar \mu}{\bar{\nu}\bar{\alpha}\bar{\beta}} &=& \calR^{-2}\,\widetilde R\UD{\bar\mu}{\bar\nu\bar\alpha\bar\beta}  .
\eea
Then (\ref{eq:sigmaexpansion}) can be recast in the following form:
\bea
0 &=&  L\,\widetilde\bfL(\widetilde \bfX) + \frac{1}{2}\,\left(\frac{L^2}{D}\right)\,\widetilde{\bfU}^{(0)}(\widetilde \bfX, \widetilde \bfX) + \frac{1}{2}\,\left(\frac{D\,L^2}{\calR^2}\right)\,\widetilde{\bfU}^{(1)}(\widetilde \bfX, \widetilde \bfX) \label{eq:sigmauptot2dimless},
\eea
with the dimensionless expansion for $\bfU$ given by:
\bea
\widetilde \bfU^{(0)} &=& D\,\bfU^{(0)} =  \left(\begin{array}{rr} -\eta_{\bar{\mu}\bar{\nu}} & \eta_{\bar{\mu}\bar{\nu}} \\ 
\eta_{\bar{\mu}{\bar \nu}} & -\eta_{\bar{\mu}\bar{\nu}} \end{array}\right) \\
\widetilde \bfU^{(1)} &=& \frac{\calR^2}{D}\,\bfU^{(1)} = \int_0^1 \dd \widetilde \lambda \, \widetilde R\UD{\bar \mu}{\bar {\nu}\bar{\alpha}\bar{\beta}}(\widetilde r)\,l^{\bar \nu}\,l^{\bar \alpha}\,P(\widetilde \lambda),
\eea
$P(\widetilde\lambda)$ denoting the appropriate second order polynomials for each submatrix, see Equation (\ref{eq:URintegral}). Dividing the resulting equation by $L$ we obtain:
\bea
0 &=&  \widetilde\bfL(\widetilde \bfX) + \frac{1}{2}\,\left(\frac{L}{D}\right)\,\widetilde{\bfU}^{(0)}(\widetilde \bfX, \widetilde \bfX) + \frac{1}{2}\,\left(\frac{L}{D}\right)\left(\frac{D}{\calR}\right)^2\,\widetilde{\bfU}^{(1)}(\widetilde \bfX, \widetilde \bfX) \label{eq:dimlesscondition}.
\eea
In this form the hierarchy between the three terms is evident. Assuming $L\ll D \ll \calR$ (physically realistic case) we see that the first term, responsible for the LOS effects, is $O(1)$ and thus dominating over the other two. The finite distance corrections term, responsible for the ``flat'' transverse effects in TOA's, scales like the ratio of the size $L$ of the regions $N_\calO$, $N_\calE$ and their mutual distance $D$. Finally, 
the curvature term is a small correction upon the finite distance term, smaller again by the factor $D^2/\calR^2$. We also see that the curvature correction grows with the distance $D$ between the two regions. This is expected, as the signals pick up more and more curvature corrections as they propagate through the curved spacetime. The second term, on the other hand,  decreases as $D$ grows.

We may now solve (\ref{eq:dimlesscondition}) for $\widetilde\bfX^0$, as in Section \ref{sec:TOAs}, this time treating $L/D$ and $D^2/\calR^2$ as legitimate small parameters. We get
the dimensionless version of equation (\ref{eq:TOAfromQX}):
\bea
\widetilde{\bfX}^{\bm 0} &=& -\widetilde{\bfL}_{\bf a}\,{\widetilde\bfX}^{\bf a} - \left(\frac{L}{D}\right)\,\frac{1}{2}\left({\widetilde\bfU}^{(0)}_{\,\bf a b}  - 2 {\widetilde\bfU}^{(0)}_{\,\bf 0 \left(a\right.} \,{\widetilde\bfL}_{\left. \bf b \right)} + {\widetilde\bfU}^{(0)}_{\bf 0 0} \, {\widetilde\bfL}_{\bf a}\,{\widetilde\bfL}_{\bf b}\right)\,{\widetilde \bfX}^{\bf a}\,{\widetilde \bfX}^{\bf b} \nonumber \\
&-&\left(\frac{L}{D}\right)\,\left(\frac{D}{\calR}\right)^2\,\frac{1}{2}\left({\widetilde\bfU}^{(1)}_{\bf a b}  - 2 {\widetilde\bfU}^{(1)}_{\bf 0 \left(a\right.} \,{\widetilde\bfL}_{R\,\left. \bf b \right)} + {\widetilde\bfU}^{(1)}_{\bf 0 0} \, {\widetilde\bfL}_{\bf a}\,{\widetilde\bfL}_{\bf b}\right)\,{\widetilde \bfX}^{\bf a}\,{\widetilde \bfX}^{\bf b} + \textrm{h.o.t.}, \label{eq:X0dimless}
\eea
where the higher order terms include, among other things, terms with higher powers of the small parameters, i.e. $O\left(\left(\frac{L}{D}\right)^2\right)$, $O\left(\left(\frac{D}{\calR}\right)^3\right)$ or their products, which arise when we solve (\ref{eq:dimlesscondition}) for $\widetilde\bfX^{\bm 0}$.
Note that if the curvature corrections are sufficiently small while $L$ is large enough it may happen that 
$\left(\frac{L}{D}\right)^2$ is comparable or larger than $\left(\frac{L}{D}\right)\,\left(\frac{D}{\calR}\right)^2$. In this case we may need to include the second or even higher power terms in $\frac{L}{D}$ in equation
(\ref{eq:X0dimless}) and in the data analysis.  They are of third and higher order in $\widetilde{\bfX}$, i.e. of the form  $\left(\frac{L}{D}\right)^2\,Q^{(3)}_{\bf abc}\,\widetilde{\bfX}^{\bf a}\,\widetilde{\bfX}^{\bf b}\,\widetilde{\bfX}^{\bf c}$ etc. Their presence complicates the inverse problem and thus also the measurement $\mu$ and $\nu$, but the general idea should still remain applicable: with  sufficiently many sampling points $\widetilde \bfX^{\bf a}_{(i)}$ over which we measure the  TOA's  we should be able to obtain exactly also the third or higher order terms in $\widetilde{\bfX}^{\bf a}$ in the Taylor expansion (\ref{eq:X0dimless}), since the 3rd order tangent surface is also parametrized by a finite number of parameters. 

We will not discuss here the inverse problem with a longer Taylor expansions in detail. Let us just note that the number of additional unknown coefficients to be determined in the process is not as large as one might think at first. As an example, note that the term $Q^{(3)}_{\bf abc}$ represents the  higher order corrections to the TOA in a \emph{flat} spacetime, with no curvature corrections. Thus we may expect that this term, just like the one arising from $\bfU^{(0)}$,  has a fairly simple, universal form in a parallel propagated, adapted ON tetrad, as in equation (\ref{eq:UDform}).
In a generic situation this term can be parametrized by $L/D$ and 6 numbers relating the orthonormal tetrads on $\calO$ and $\calE$ we use (this follows from the Lorentz invariance of a flat spacetime). This means that the effective number of new, free parameters we need to include with a 3rd order Taylor expansion is at most 7.

\subsection{Particular case: weak field limit with non-relativistic dust. } \label{sec:weakfield}

Consider a spacetime filled with a static or very slowly varying dust distribution, i.e. $T_{00} = \rho(x^i)$, $T_{0i}$ and $T_{ij}$ negligible. We assume that the the weak field limit of GR holds, i.e. $g_{\mu\nu} = \eta_{\mu\nu} + h_{\mu\nu}$ with $h_{\mu\nu}$ small.  Let the receivers' frame
$u_\calO^\mu$ coincide fairly precisely with the rest frame of the dust. As before, let $\lambda$ denote the affine distance along $\gamma_0$ normalized wrt to $u_\calO^\mu$.

Note that in this case we have $\mu,\nu = O(h)$. It follows that in the leading order in the metric perturbation $h_{\mu\nu}$
we may approximate $\gamma_0$ in the integration by an unperturbed geodesic (i.e.  a null straight line in $\eta_{\mu\nu}$) and the affine parameter by the radial coordinate $r$ centered at $\calO$; corrections to $\mu$ and $\nu$ will be at most $O(h^2)$.
The integrals (\ref{eq:muTintegral}) and (\ref{eq:nuTintegral}) take now the form of
\bea
 \mu &=& 8\pi\,G \int_{0}^{D} \rho(r)\, \left(D - r\right)\,\dd r \label{eq:muTintegralrho} + \textrm{h.o.t.}\\
 \nu &=& 8\pi\,G  \int_{0}^{D} \rho(r) \,r \,\dd r\label{eq:nuTintegralrho} +\textrm{ h.o.t.},
\eea
with the integration performed over a straight radial line.
The higher order terms are either $O(h^2)$ or $O(\beta\cdot h)$, where $\beta = v/c$ is the relative velocity of the receivers and the dust, and we consider them negligible.

The dimensionless parameters $\mu$ and $\nu$ are  not the most convenient tool to describe the mass distribution between $\calO$ and $\calE$, but with the help of the equations above they can be transformed into  quantities with a more clear physical interpretation. We first note that their sum gives the integral of $\rho(r)$, or the 0th moment of the mass distribution along the LOS:
\bea
\mu + \nu = 8\pi G\,D\,\int_0^D \rho(r)\,\dd r, \nonumber
\eea
providing this way the true tomography of the matter distribution $\rho(x^i)$, unlike $\mu$ or $\nu$ themselves, which provide density integrals with linear kernels.
It also follows  that this sum can be translated directly into the average matter density along LOS, defined by
\bea
\langle\rho\rangle = \frac{1}{D}\int_0^{D} \rho(r)\,\dd r. \nonumber 
\eea   
Namely, we have the relation
\bea
\left\langle \rho \right\rangle &=& \frac{\mu + \nu}{8\pi G \, D^2} \label{eq:rhoav1}.
\eea
On top of that, $\mu$ and $\nu$ contain the information about the first moment of the mass distribution along the LOS, related to the position of its center of mass (COM). 
Namely, consider the  dimensionless parameter $\alpha_\textrm{CM}$
\bea
\alpha_\textrm{CM} = \frac{\int_0^{D} \rho(r)\,\frac{r}{D}\,\dd r}{ \int_0^{D} \rho(r)\,\dd r}. \nonumber 
\eea
 It determines the position of the COM along $\gamma_0$: $\alpha_\textrm{CM}$ is equal to 0 for a mass distribution centered at $\calO$, 
1 for a mass distribution centered at $\calE$ and $\frac{1}{2}$ if it has its COM at the midpoint. One can now prove using (\ref{eq:muTintegralrho})-(\ref{eq:nuTintegralrho}) that
\bea
\alpha_\textrm{CM} &=& \frac{\nu}{\mu + \nu} \label{eq:CM}.
\eea
Equations (\ref{eq:rhoav1}) and (\ref{eq:CM}) show how to relate $\mu$, $\nu$ and the distance $D$ between $\calE$ and $\calO$ to physically more interesting quantities: the directional average of the mass distribution and the COM of this distribution.

Note that while equation (\ref{eq:rhoav1}) requires formally the normalized affine distance $D$ between $\calO$ and $\calE$, we may  use instead the angular diameter distance $D_\textrm{ang}$ or the parallax distance $D_\textrm{par}$  in the receivers' frame, much simpler to measure in the astronomical context. These distance measures differ at most by $O(h)$, so 
in the leading order they lead to the same value of $\langle \rho \rangle$.

\subsection{Estimating the signal from a small background matter density} \label{sec:estimate}

We will now  estimate the additional TOA variations caused by a small  mass density along the LOS in the weak field approximation. This estimate may also serve as an indicator of the minimal  precision of pulse timing required for detection of the mass density.  We consider here a simplified scenario in which the spacetime is filled with a tiny, constant background mass density $\rho(x) \equiv \rho$ of pressureless matter. We perform the measurement of $\mu$ and $\nu$ along the line of sight between the two groups of clocks of size $L$ and separated by the distance of $D$. The idea is now to estimate the contribution coming from the constant stress-energy tensor $T_{00} = \rho$, $T_{0i} = T_{ij} = 0$ to the TOA's in the form of (\ref{eq:TOAfromQX}) or, equivalently, of (\ref{eq:X0dimless}), ignoring the Weyl or the cosmological constant terms since their contributions will cancel out anyway.

Since the stress-energy tensor $T_{\mu \nu}$ is of the order of $\rho$, its contribution to the Ricci tensor  is $R_{\mu\nu} \sim 8\pi G \,\rho$. This in turn  corresponds to the curvature radius $\calR^2 = (8\pi G\,\rho)^{-1}$ in the terminology of Section \ref{sec:expansioninsmall}. It follows now from (\ref{eq:X0dimless})  that the pressureless matter contribution to 
the dimensionless TOA $\tilde {\bf X}^{\bm 0}$, contained in the third term on the right hand side, is of the order of $\left(\frac{L}{D}\right)\,\left(\frac{D}{\calR}\right)^2=\left(\frac{L}{D}\right)\,4\pi G\,\rho D^2 = 4\pi G\,\rho D L$. Returning to the dimensional TOA
${\bf X}^{\bm 0} = L\,\tilde {\bf X}^{\bm 0}$ we obtain the estimate ${\bf X}^{\bm 0} \sim  4\pi G\,\rho D L^2$ for the TOA variations induced by the mass density along the LOS. For comparison we include here  the analogous estimate for the finite distance variations, which are of the order of $L^2/D$. 

This estimate can be also reformulated as a simple rule of thumb for the TOA variations. Namely, note that $4\pi \,\rho D L^2$ is simply the total mass $M_{tot}$ contained in a thin 3D cylinder of radius $L$ and height $D$, connecting $N_\calO$ and $N_\calE$ along the spatial dimensions. With this interpretation $4\pi G\,\rho D L^2$  is simply the half of the Schwarzschild radius corresponding to $M_{tot}$.
 The result may therefore be rephrased as follows: the TOA variations caused by the matter along the LOS, expressed in  units of length,  are of the order of the Schwarzschild radius of the total mass contained
 in a spatial cylinder connecting $N_\calO$ and $N_\calE$ whose radius $L$ is equal to the physical size of each of the two clock systems. Note also that the same number may serve as an estimate of the precision of time measurements required to detect the mass density background 
of the order of $\rho$.

\subsection{Summary of the results}

We will pause here to give a short summary of the most important results of Sections \ref{sec:geometry}-\ref{sec:expansioninsmall}. 

\begin{enumerate}
 \item Local variations of the TOA's in the vicinity of two points $\calE$ and $\calO$ can be expanded in a Taylor series consisting of the leading order, linear effects and the quadratic effects in the displacements, see equation (\ref{eq:sigmaexpansion}) or (\ref{eq:TOAfromQX}). The linear effects 
 involve the line of sight delays (R{\o}mer delays) as well as the frequency shifts and are governed   by the components of the 8-vector $\bfL$. The quadratic effects involve, among other things, the transverse variations and are governed by the quadratic form $\bfU^\perp$. They consist of  the finite distance effects, present also in flat spacetimes and related to the distance between the endpoints, as well as the curvature effects, usually much smaller and superimposed on the former.  This approach to the TOA variations has a neat geometric interpretation of approximating the LSC $\Sigma \in M\times M$ by its second order tangent at $(\calO,\calE)$, defined by the normal vector $\bfL$ and the extrinsic curvature $\bfU^\perp$.
\item  $\bfU^\perp$ can be expressed as a functional of the Riemann tensor along the LOS, via an ODE in a parallel-transported tetrad (equations (\ref{eq:UfromW}), (\ref{eq:ODEW})). It can therefore be used to measure the curvature impact
on the propagation of signals.
 \item The linear and quadratic order effects are effectively parametrized by a finite set of numbers, i.e. the independent components of $\bfL$ and $\bfU^\perp$, expressed in a chosen pair of locally flat, orthonormal coordinates. By sampling the TOA's over a sufficient number of evenly distributed points we can determine all of these components and thus reconstruct the shape of the LSC. 
 \item In order to quantify the curvature effects we have defined two scalar, dimensionless functions $\mu$ and $\nu$ of the components of $\bfL$ and $\bfU^\perp$, see equations (\ref{eq:Qoo})-(\ref{eq:Qoe}) and (\ref{eq:mufromQepsilon})-(\ref{eq:nufromQepsilon}). They are both selectively sensitive to the curvature corrections present in $\bfU^\perp$ and insensitive to the finite distance effects, i.e. they both vanish identically in a flat spacetime.
 Moreover, both quantities are invariant with respect to the proper orthochronous Lorentz transformations of the coordinates on both endpoints, see Appendix \ref{app:proof}.
 \item For short distances, at the leading order expansion in the curvature tensor, both $\mu$ and $\nu$ depend only on integrals of the stress-energy tensor (equations (\ref{eq:muTintegral})-(\ref{eq:nuTintegral})), with the contributions from the Weyl tensor (tidal forces) and the cosmological constant vanishing identically. 
 \item Lorentz invariance implies the insensitivity of the measurement to the states of motion and the angular positions (or attitudes) of the two clock ensembles: after  the measurements we may  perform the calculations in the internal ON tetrads of the ensembles, recover the components of $\bfL$ and $\bfU^\perp$ in these tetrads and calculate the two invariants $\mu$ and $\nu$. Their invariance  under passive Lorentz transforms of the coordinates implies thus the invariance wrt to active Lorentz boosts and spatial rotations of the ensembles. On the other hand, the independence from the Weyl tensor implies the insensitivity of the measurement to the influence of masses off the LOS.
 \item The values of $\mu$ and $\nu$ can  be finally related to the zeroth and first moment of the mass density distribution along the LOS (equations (\ref{eq:rhoav1}) and (\ref{eq:CM})), yielding a directional, tomographic measurement of the mass distribution $\rho(x^i)$.
 \end{enumerate}

\section{An example of the measurement protocol} \label{sec:protocol}

As a proof of concept, we will now present a particular method of measurement of $\mu$ and $\nu$ with the help of two ensembles of clocks capable of exchanging electromagnetic signals. We stress that this is not the only
possible measurement protocol based on the geometric principles introduced in this paper. The setup we present is in fact rather wasteful when it comes to  resources: it is neither optimal with respect to the number of clocks involved, nor with respect the total number  of measurements performed, since, as we will see, only $\approx 10\%$ of measured TOA's is finally used in the data processing stage. However, the peculiar geometry of both ensembles we consider here allows for a fairly straightforward reconstruction of the components of $\bf L$ and $\bf Q$ from the TOA's.

Assume we have at our disposal two sets of clocks, 13 receivers
($O_1, \dots, O_{13}$) in $N_\calO$  and and 13 emitters ($E_1, \dots, E_{13}$) in $N_\calE$. The clocks within each group are in free fall, comoving and synchronized, i.e. all clocks in one group give the same readings along the simultaneity 
hypersurfaces of their rest frames, respectively $u_\calO$ or $u_\calE$. The readings of the clocks will be denoted by $t_\calO$ and $t_\calE$.  The relative motion of the two groups may be arbitrary and is considered uncontrolled by the experimentator.
\begin{table}
\begin{tabular}{| l | l || l | l |}
\hline
 $1$ & $(0,0,0)$ &  $8$ &$ (L, L, 0)$  \\
 $2$ &$ (L, 0, 0)$ &  $9$ & $(-L ,-L,0)$   \\
 $3$ & $(-L,0,0)$ &  10 & $(L,0,L)$ \\
 $4$ & $(0,L,0)$ &  11 & $(-L,0,-L)$ \\
 $5$ & $(0,-L,0)$ & 12 & $(0,L,L)$ \\
 $6$ & $(0,0,L)$ & 13 & $(0,-L,-L)$ \\
 $7$ & $(0,0,-L)$ & \begin{centering} --- \end{centering} & \begin{centering} ---\end{centering} \\
 \hline
\end{tabular}
\caption{Arrangement of both the emitters and the receivers described in their local spatial coordinates. } \label{tab:pos}
\end{table}

\bfi
\includegraphics[width=0.4\textwidth]{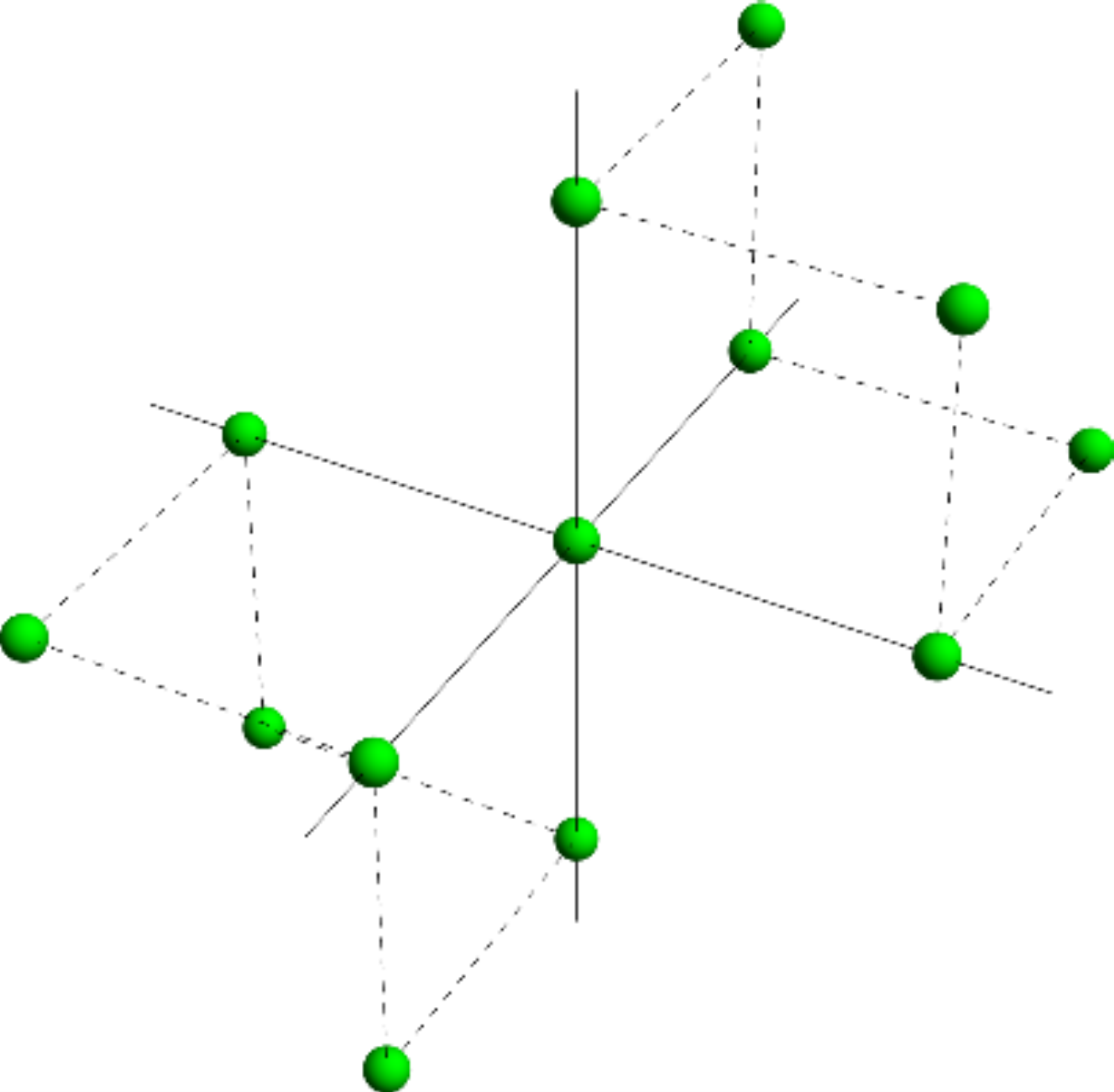}
\caption{Arrangement of the emitters  and the receivers within their respective groups, given in their locally flat, comoving frame.
We present here the receivers. The three solid lines are the $X$, $Y$ and $Z$ axes.
$O_1$ is the central receiver, $O_2$-$O_7$ are positioned pairwise at the opposite sides at the three axes at the distance of $L$ from the center. $O_{8}$-$O_{13}$ lie on the three normal planes of the axes.  }
\label{fig:geometricsetup}
\efi

 Let $L$ define the size of each group. We position 13 emitters and 13 receivers according to Table \ref{tab:pos},
see also Figure \ref{fig:geometricsetup}. Their positions are measured wrt to two OLF coordinate systems at $N_\calO$ and $N_\calE$ respectively, 
constructed from two orthonormal, oriented tetrads, $(u_\calO, e_i)$ at $\calO$ and $(u_\calE, f_i)$ and $\calE$. Therefore, the attitude of both groups of clocks with respect to each other, with respect to the line of sight, or with respect to any other external reference frame, is also arbitrary and will be treated as an uncontrolled variable.

All emitters now send time- and source-stamped signals at 3 equally spaced moments in time, corresponding to the readings $t_\calE - L$,  $t_\calE$ and $t_\calE + L$.  All signals are then received by all 13 receivers, their origins are recognized and their precise TOA's wrt to the receivers' time $t_\calO$ are recorded. This yields $13\times 13\times 3 = 507 $ measurements of TOA's between precisely localized pairs of events. After the measurement we extract the signal from the data using the procedure sketched below, effectively using  only 57 of them. 

In the first step we use the emission time $t_\calE$ of the second group of signals and the TOA $t_\calO$ the second signal sent from the central emitter $E_1$ and registered by the central receiver $O_1$ to define the respective offsets for all time measurements by the emitters and the receivers.  By subtracting these offsets from all readings  we obtain the time coordinates $\delta x_\calE^0$ and $\delta x_\calO^{0'}$ respectively from the times registered by the  clocks in each group, with the reference points $\calE$, $\calO$ defined by the emission and the reception of the second signal by the central clocks of the ensembles. This way we define also the fiducial null geodesic $\gamma_0$ through these points.

We now pick a set of 56 other times of arrival which we will further use to determine $\bf L$ and $\bf Q$. As we have seen in Sec \ref{sec:TOAs}, the TOA's are associated to 7-dimensional sub-vectors 
defining the event of emission in $N_\calE$ and the spatial position of the receivers $N_\calO$. Therefore the sub-vectors consisting of those components may be used for indexing 
the individual TOA measurements. We now need to define a set $V$ of sub-vectors we will use for the measurement. We first take the  TOA's related to the following 14 sub-vectors $\bf H^{\bf a}$ with only one non-vanishing component:
\bea
 {\bf H}_{k+}^{\bf a} &=& \left\{\begin{array}{l} 0 \qquad \textrm{for } {\bf a} \neq k \\
 L \qquad \textrm{for } {\bf a} = k \end{array} \right. \nonumber  \\ 
 {\bf H}_{k-}^{\bf a} &=& \left\{\begin{array}{l} 0 \qquad \textrm{for } {\bf a} \neq k \\
 -L \qquad \textrm{for } {\bf a} = k \end{array} \right. \nonumber .
\eea
They correspond to the second signal emitted by $E_1$ at $\calE$ and received by the receivers $O_2$-$O_7$, the second signals emitted by $E_2$-$E_7$ and received by $O_1$ and
the first and third signals emitted by $E_1$ and received by $O_1$.
We also define 42 sub-vectors with two non-vanishing components:
\bea
 {\bf H}_{k+,l+}^{\bf a} &=& \left\{\begin{array}{l} 0 \qquad \textrm{for } {\bf a} \neq k \textrm{ and } {\bf a} \neq l\\
 L \qquad \textrm{for } {\bf a} = k\textrm{ or } {\bf a} = l  \end{array} \right. \nonumber \\
 {\bf H}_{k-,l-}^{\bf a} &=& \left\{\begin{array}{l} 0 \qquad \textrm{for } {\bf a} \neq k \textrm{ and } {\bf a} \neq l\\
- L \qquad \textrm{for } {\bf a} = k\textrm{ or } {\bf a} = l  \end{array} \nonumber \right. ,  
\eea
%\bea
%{\bf H}_{k-,l+}^{\bf a} &=& \left\{\begin{array}{l} 0 \qquad \textrm{for } {\bf a} \neq k \textrm{ and } {\bf a} \neq l\\
%- L \qquad \textrm{for } {\bf a} = k\textrm{ or } {\bf a} = l \\
%L \qquad \textrm{for } {\bf a} = l  \end{array} \right. \\
%{\bf H}_{k+,l-}^{\bf a} &=& \left\{\begin{array}{l} 0 \qquad \textrm{for } {\bf a} \neq k \textrm{ and } {\bf a} \neq l\\
%L \qquad \textrm{for } {\bf a} = k\\
%-L \qquad \textrm{for } {\bf a} = l  
%\end{array} \right. .
%\eea
we assume here that $k \neq l$ and for the sake of unique indexing we follow the convention that $k < l$ in the subscripts when describing them. The subvectors correspond to the second signal
from $E_1$ received by $O_8$-$O_{13}$, second signals from $E_8$-$E_{13}$ received by $O_1$, some of the second signals form $O_2$-$O_7$ received by $E_2$-$E_7$, the some of the first and third signals from $E_2$-$E_7$ received by $O_1$ and the first and third signals from $E_1$, received by some of the receivers $O_2$-$O_7$.

 These sub-vectors, taken together with the
measured TOA of the signal $\tau( {\bf H}_{*}^{\bf a} )$,  constitute the full 8-dimensional vectors $ {\bf H}_{*} \in T_\calO M \oplus T_\calE M$, i.e.
\bea
  {\bf H}_{*} = \left(\begin{array}{l} \tau({\bf H}_{*}^{\bf a} )  \\ {\bf H}_{*}^{\bf a} \end{array} \right), \nonumber 
 \eea
or, equivalently, ${\bf H}_{*}^{\bf 0} =  \tau({\bf H}_{*}^{\bf a} )$, with $*$ denoting any of the subscripts for vectors in $V$, as defined above.

The set $V$ contains $2 \times 7 + 21 \times 2 = 56$ sub-vectors and this is the number of TOA's we will use. It turns out that with this particular placement of the emitters and receivers it is possible to obtain exact expressions for the components of $\bf L$ and $\bf Q$ in the pair of ON tetrads $(u_\calO,e_i)$ and $(u_\calE,f_j)$ from the TOA's.
We first note that the set $V$ is centrally symmetric with respect to the origin. Namely, for every ${\bf H^{\bf a}} \in V$ we also have $-{\bf H^{\bf a}} \in V$, i.e. all sub-vectors from $V$ can be arranged in pairs related by a point reflection through 0 \footnote{ On the other hand, note that this  does not have to apply to the set of corresponding full 8-dimensional vectors, simply because the TOA's $\tau({\bf H}_*^{\bf a} )$, constituting the $\bf 0$ component, will not obey the central symmetry in general.}.
Indeed, we have $-\bf H^{\bf a}_{k+} = \bf H^{\bf a}_{k-}$ and $-\bf H^{\bf a}_{k+,l+} = \bf H^{\bf a}_{k-,l-}$. 
This  feature is important, because it simplifies greatly the problem of solving the system  (\ref{eq:TOAfromQX}) separately for the components of $\bf L$ and $\bf Q$. Note that the linear and quadratic terms in (\ref{eq:TOAfromQX}) (or equivalently in (\ref{eq:TOAfromdeltax})) 
differ in the way they behave when we flip the signs of all components of the subvector $\bf X^{\mathbf{a}}$: ${\bf L}_{\bf a}\,{\bf X}^{\bf a} = -{\bf L}_{\bf a}\,(-{\bf X}^{\bf a} )$, but
${\bf Q}_{\bf ab}\,{\bf X}^{\bf a}\,{\bf X}^{\bf b} = {\bf Q}_{\bf ab}\,(-{\bf X}^{\bf a})\,(-{\bf X}^{\bf b} )$. The idea now is to make a direct use of this fact in order to decouple the equations 
governing the linear and quadratic terms.

We calculate from (\ref{eq:TOAfromQX}) the difference between the TOA's for the opposite subvectors $\bf H_{\bf k+}^{\bf a}$ and $-\bf H_{\bf k+}^{\bf a} =\bf H_{\bf k-}^{\bf a} $:
\bea
 \frac{1}{2}\left(\tau({\bf H}^{\bf a}_{k+}) - \tau({\bf H}^{\bf a}_{k-} )\right) = - {\bf L}_{\bf a}\,{\bf H}^{\bf a}_{k+} = -L\,{\bf L}_{\bf k}. \nonumber 
\eea
Thus we have obtained an exact expression for the components of $\bfL_{\bf a}$
\bea
 {\bf L}_{\bf k} = -\frac{L^{-1}}{2}\left(\tau({\bf H}^{\bf a}_{k+}) - \tau({\bf H}^{\bf a}_{k-} )\right),
\eea
with $\bf k = 1\dots 7$ (recall that the missing component ${\bf L}_{\bf 0} $ is equal to 1 due to the gauge condition  (\ref{eq:gauge}) we have imposed in order to fix the normalization of $\bf L$ and $\bf U$).

We then calculate in the same way the average of the TOA's for the opposite sub-vectors
\bea
\frac{1}{2}\left(\tau({\bf H}^{\bf a}_{k+}) + \tau({\bf H}^{\bf a}_{k-} )\right) = -\frac{1}{2}{\bf Q}_{\bf a b}\,{\bf H}_{k,+}^{\bf a}\,{\bf H}_{k,+}^{\bf b} =
-\frac{L^2}{2}\,{\bf Q}_{\bf k \bf k}, \nonumber 
\eea
obtaining the relation
\bea
{\bf Q}_{\bf k \bf k} = -L^{-2}\,\left(\tau({\bf H}^{\bf a}_{k+}) + \tau({\bf H}^{\bf a}_{k-} )\right).
\eea
Finally we obtain the mixed terms $\bf Q_{k l}$ with the help of the polarization identity
\bea {\bf Q(A,B)} = \frac{1}{2}\left({\bf Q(A+B,A+B) - Q(A,A) - Q(B,B)}\right) \nonumber 
\eea
applied to ${\bf A}^{\bf a} = {\bf H}_{k+}^{\bf a}$ and ${\bf B}^{\bf a} = {\bf H}_{l+}^{\bf a}$ with $ k <  l$:
\bea
{\bf Q}_{\bf k \bf l} &=& L^{-2}\, {\bf Q}_{\bf a \bf b}\,{\bf H}_{k+}^{\bf a}\,{\bf H}_{l+}^{\bf b} \nonumber \\
&=& \frac{ L^{-2}}{2} \left({\bf Q}_{\bf a \bf b}\,{\bf H}_{k+,l+}^{\bf a}\,{\bf H}_{k+,l+}^{\bf b} - {\bf Q}_{\bf a \bf b}\,{\bf H}_{k+}^{\bf a}\,{\bf H}_{k+}^{\bf b} -
{\bf Q}_{\bf a \bf b}\,\,{\bf H}_{l+}^{\bf a}\,{\bf H}_{l+}^{\bf b} \right) \nonumber 
\eea
The three quadratic terms in the brackets can then be related to the appropriate averages of  the TOA's for pairs of opposite sub-vectors, just like before. We obtain finally
\bea
{\bf Q}_{\bf k \bf l} = \frac{L^{-2}}{2}\,\left(-\tau({\bf H}_{k+,l+}) - \tau({\bf H}_{k-,l-}) + \tau({\bf H}_{k+}) + \tau({\bf H}_{k-})  + \tau({\bf H}_{l+}) + \tau({\bf H}_{l-})\right)
\eea

\ZZ{Z62 p120/1-6}

We have thus presented exact relations between the components of $\bf L$ and $\bf Q$ in the two internal ON tetrads and  combinations of TOA's between chosen receiver-emitter pairs, proving this way
the feasibility of recovering completely both objects from TOA's.  

After that we divide $\bfL = \left(\begin{array}{ll} 1 & \bfL_{\bf a} \end{array} \right)$ into $l_{\calO\,\mu'}$ and $l_{\calO\,\nu}$ according to (\ref{eq:Lll}) and $\bf Q$ into $Q_{\calO\calO\,i'j'}$, $Q_{\calO\calE\,i'\mu}$, $Q_{\calE\calE\,\mu\nu}$ according to (\ref{eq:Qqq}).
In the final step we apply relations (\ref{eq:mufromQepsilon})-(\ref{eq:epsilonE}) in order to obtain $\mu$ and $\nu$.

\section{Summary and conclusions}

We have provided a method of measuring the integrated curvature along the line of sight by measuring the variations of times of arrival of electromagnetic signals between two regions in spacetime. In particular, we have demonstrated how it is possible to extract the two first moments of the mass density distribution along the line of sight.  However, the method may require modifications when applied to real life physical situations or astronomical observations.

\subsection{Alternative setups and protocols}

The protocol described in Section \ref{sec:protocol} can be modified in many ways. In particular, there is no need to place the clocks exactly according to the Table \ref{tab:pos}, or to send the signals precisely at the moments we have prescribed there. The clock placement and the emission moments can in principle be arbitrary (the only restriction being the non-degeneracy in the sense defined in Section \ref{sec:inverse}), as long as they are all precisely measured within each ensemble. In particular, small deviations from the shape defined in Table \ref{tab:pos} should not affect the measurement, provided that we take them into account while solving the inverse problem from Section \ref{sec:inverse} via equation (\ref{eq:TOAfromdeltax}). 

In a broader context, note that it is also possible to drop the assumptions of the clocks being comoving within each group, at least as long as we are able to translate each clock's proper time to locally flat coordinates. This is because the solution of the inverse problem requires only the coordinates of all emission and reception events as input, not the details of the motions of each individual clock we use. Therefore, given sufficiently precise local tracking of clocks within each group, we can in principle perform the measurement in a purely passive mode, with no attempts to steer or place the clocks in any way in each ensemble. Even small, uncontrolled forces (drag, radiation pressure, tidal forces) acting on the clocks  can be in principle taken into account in this case.

Precise tracking and motion measurements of this kind can be achieved if each ensemble of clocks can also work as a clock compass, i.e. a device capable of determining the local inertial frame and measuring motions with respect to it by exchanging of electromagnetic signals between clocks. An example of such device, based on clock frequency comparisons, has been presented in \textcite{PhysRevD.98.024032, Obukhov2019, PhysRevD.102.044027}.  Therefore, two distant clock compasses, each additionally capable of clock tracking and long-range communication, can in principle probe the mass density distribution along their connecting line. More precise discussion of this kind of setup warrants a separate publication.

\subsection{Dark matter within the Solar System}

Two ensembles of clocks within the Solar System could in principle detect or provide upper bounds for the tiny background matter density within the Solar System, including the dark matter. The insensitivity of this method to the tidal perturbations by all bodies of the Solar System and the details of the motion of the clock ensembles makes it particularly attractive in this setting. However, as we will see, the  measurement requires enormous precision of pulse timing and huge distances in order to be competitive with  precise tracking of the Solar System bodies over long periods \cite{2013AstL...39..141P}.

Using the results of Section~\ref{sec:estimate} we may estimate the minimal mass contained within the connecting cylinder such setup could detect.
 Assuming the timing precision of $10\,\textrm{ns}$ for a single measurement, comparable to the timing precision of the Global Navigation Satellite Systems (GNSS) like GPS, or to the precision of the millisecond pulsar timing \cite{smits:2011}, 
 we get the  Schwarzschild radius of $3\,\textrm{m}$. This is a large value, corresponding to the mass of $2\cdot 10^{27} \textrm{kg}$, close to the Jupiter mass. 
 Assuming the dark matter mass density bounds of $\rho_{DM} < 10^{-19}\,\textrm{g}\,\textrm{cm}^{-3}$, taken from \textcite{2013AstL...39..141P}, we would need the connecting cylinder volume of at least $10^{10} \,\textrm{AU}^3$, far exceeding the size of the Solar System.   We see that without a significant improvement of the timing precision it seems unfeasible to achieve bounds comparable to  \cite{2013AstL...39..141P} with two clock ensembles. 

\subsection{Binary pulsars as sources}

Instead of using artificial sources of signals we may consider natural ones, provided by binary pulsars or the double pulsar \cite{Lorimer2008}. In this case the orbital motion of the pulsars provides a natural sampling of the TOA's over the emitters' region, while the Earth's orbital motion provides the sampling  over the receivers' region. The receiver-source distance   is also larger by many orders of magnitude than what could be achieved with artificial sources,  so  we may expect much larger integrated mass 
along the line of sight, and therefore also a significantly larger signal to measure. 

  Since we do not have independent measurements 
of the distances or orbital elements in the binary pulsar system, but rather we infer them from various relativistic effects via precise pulsar timing \cite{1992PhRvD..45.1840D}, the method of measurement would have to be modified.  The effects of propagation of signals through the curved spacetime between the binary pulsar and the Solar System, including the effects of $\mu, \nu \neq 0$, would need to be included into an extended model of the TOA's of the pulses, with a global fit of the observed data preformed over all unknown parameters \cite{smits:2011}. The feasibility of curvature measurements of this kind will be discussed in another publication.

\section*{Acknowledgements}
 The work was supported by the National Science Centre, Poland (NCN) via the SONATA BIS programme, Grant No.~2016/22/E/ST9/00578 for the project
\emph{``Local relativistic perturbative framework in hydrodynamics and general relativity and its application to cosmology''}.

\appendix

\section{Normal form of the quadric equation and the shape of the LSC} \label{app:normalform}

We begin by absorbing the linear term in the quadric equation \eqref{eq:sigmaexpansion}. Define the vector ${\bf Y} \in T_\calO M \oplus T_\calE M$ by ${\bf Y}  = (0 \quad \Delta \lambda\cdot l_\calE^\mu)$. From \eqref{eq:Uoowithl}-\eqref{eq:Ueewithl} we have ${\bf U}({\bf Y},\cdot) = \bf L$ and ${\bf U}({\bf Y},{\bf Y}) = 0$. It follows that  \eqref{eq:sigmaexpansion} is equivalent to
\bea
{\bf U}({\bf Z},{ \bf Z}) = 0, \label{eq:normalform1}
\eea
where the new variable ${\bf Z}$ is defined as ${\bf Z} = {\bf X} + {\bf Y}$. Note that this equation still contains no free term.

In the second step we need to diagonalize $\bf U$ in order to determine its invariant signature. It is easy to show that, irrespective of the spacetime geometry, the quadratic form $\bf U$ must be degenerate in at least one direction. Define the vector ${\bf L}^{\sharp} \in T_\calO M \oplus T_\calE M$ as the vector obtained by raising the index of $\bf L$ with the help of
the inverse metric ${\bf h}^{-1}$, i.e. ${\bf L}^{\sharp} = (l_\calO^{\mu'}\quad l_\calE^{\mu})$. 
This vector corresponds to a simultaneous displacement of both endpoints along $\gamma_0$, generated by the same infinitesimal variation of the affine parameter. Now, using again relations \eqref{eq:Uoowithl}-\eqref{eq:Ueewithl}, we can show that
${\bf U}({\bf L}^{\sharp},\cdot) = 0$.

The shape of the quadric depends on the signature of the form $\bf U$. However, apart from the single degenerate direction we have pointed out above, it is impossible to determine the signature without further assumptions regarding the Riemann tensor along the LOS. We can, however, calculate the signature in a flat spacetime and then use this result to gain some insight into the small curvature limit case 
as defined in Section \ref{sec:smallcurvature}.

Let $M$ be the Minkowski space in the standard coordinates $(x^\mu)$ and let $x_\calO^\mu$ and $x_\calE^\mu$ be two points connected by a null
line $\gamma_0$, given by $x^\mu(\lambda) = x_\calO^\mu + (\lambda - \lambda_\calO)\,l^\mu$.  The tangent vector $l^\mu$ satisfies $l^\mu =  \Delta \lambda^{-1}\,(x_\calE^\mu - x_\calO^\mu)$ and is null by assumption. 

The world function of the Minkowski spacetime reads \cite{SyngeBook}:
\bea
\sigma(x^{\mu},x'^\nu) = \frac{1}{2}\,\eta_{\mu\nu}\,(x^\mu - x'^\mu)\,(x^\nu - x'^\nu). \nonumber
\eea
From \eqref{eq:Lll} and \eqref{eq:Udef} we have 
\bea
{\bf U} &=& \frac{1}{\Delta \lambda}\left( \begin{matrix}  -\eta_{\mu\nu} && \eta_{\mu\nu}\\ \eta_{\mu\nu} && -\eta_{\mu\nu} \end{matrix} \right) \\
 \bfL &=& (\begin{matrix}l_{\mu} & -l_{\mu} \end{matrix}) \nonumber
\eea
for the null geodesic $\gamma_0$ between $x_\calO^\mu$ and $x_\calE^\mu$ \footnote{We do not distinguish the primed and unprimed indices in the Minkowski space since all tangent spaces in this case may be identified with each other.}.

The reduction of the quadric equation \eqref{eq:sigmaexpansion} to the normal form, together with the diagonalization of  $\bf U$, is fairly simple in the flat case. We first introduce the  vector 
$\xi^\mu =\delta  x_\calE^\mu + \Delta \lambda\,l^\mu - \delta x_\calO^\mu$. It is  equal to the position of one displaced point with respect to the other, i.e. $\xi^\mu = x_\calE^\mu + \delta x_\calE^\mu - x_\calO^\mu - \delta x_\calO^\mu$. We then express \eqref{eq:sigmaexpansion}  in the new variables $(\xi^\mu,\delta x_\calO^\mu)$ instead of the components of  $\bfX$. The quadric equation  takes then a very simple form
\bea
-\frac{1}{\Delta\lambda}\,\eta_{\mu\nu}\,\xi^\mu\,\xi^\nu = 0, \label{eq:normalform}
\eea 
with all $\delta x_\calO^\mu$ terms dropping out. This is, up to the prefactor,  the standard condition for the two displaced points to be connected by a null vector. It is  valid  for any pair of points, not only for small displacements of the two points around the geodesic endpoints, which means that in the Minkowski space the  quadric \eqref{eq:sigmaexpansion} is not an approximation, but rather it is the exact, ``global'' surface of communication. Note that in this form the quadric equation  contains no linear terms and that it is also diagonal  with respect
to the new pair of variables, which means that we have achieved the normal form.

Thus the LSC in the Minkowski space turns out to be  a highly degenerate quadric, with 4 degerate directions corresponding to $\delta x_\calO^\mu$. The reason of the four-fold degeneracy is the translational invariance of the Minkowski space, or, more precisely,  the fact that the shape of a light cone  does not depend on the position of its vertex. A pair of points $x_\calE^\mu + \delta x_\calE^\mu$ and $x_\calO^\mu + \delta x_\calO^\mu$ lies on the LSC iff the former is located on the light cone with its vertex located at the latter. Due to the lightcone shape invariance  the light cone equation depends only on the \emph{relative} position vector $\xi^\mu$ of first  endpoint  with respect to the vertex, with no dependence on the absolute position of the  vertex encoded in $\delta x_\calO^\mu$.

It follows from \eqref{eq:normalform} that the signature of $\bf U$ in the Minkowski space is  $(1,3,4)$, i.e. 1 plus sign, 3 minus signs, 4 zeros \footnote{Since there is no free term in equation \eqref{eq:normalform1}, the signs can actually be flipped to $(3,1,4)$ by multiplying both sides by -1, without affecting the normal form.}.
Moreover, a small perturbation of the quadratic form $\bf U$ of the type \eqref{eq:URintegral} (i.e. the small curvature limit) cannot alter the first four signs, but it can affect the zeros even at linear order. As we have seen, one zero must always remain in the signature, but it is easy to show that the three other zeros can be turned into  any other sign by a curvature perturbation. The resulting change of shape of the  quadric approximating the LSC is precisely the curvature effect our measurement is sensitive to. Summarizing, the invariant signature of the quadratic form in the small curvature limit contains at least one plus, at least 3 minuses and at least one 0, but otherwise depends on the spacetime geometry.

\paragraph{Remark.} We can extend these results to the quadratic form $\bf U^{\perp}$ defined on the 7-dimensional subspace $\bf L^{\perp}$, see  Section \ref{sec:TOAs}. Since it is equal to the extrinsic curvature of the LSC, we study this way the local shape of the LSC.
 
The vector $\bf L^{\sharp}$ lies in the subspace $\bf L^\perp$, so  from the results 
of Section \ref{sec:TOAs} we see that $\bf L^{\sharp}$ defines also a degenerate direction for $\bf U^{\perp}$, and thus also for the extrinsic curvature of the LSC. However, it is easy to show that $\bf U^{\perp}$ must always be
 degenerate in two different directions. Define two vectors ${\bf K}_\calO, {\bf K}_\calE \in T_\calO M \oplus T_\calE M$ via ${\bf K}_\calO = (l_\calO^{\mu'} \quad 0)$ and
${\bf K}_\calE = \left(0 \quad l_\calE^{\mu}\right)$. Both lie in the subspace $\bfL^{\perp}$ and it is easy to see from \eqref{eq:Uoowithl}-\eqref{eq:Ueewithl}  that 
$\bfU ({\bf K}_\calE,\cdot) =  -\bfU ({\bf K}_\calO,\cdot) = \frac{1}{\Delta\lambda}\,\bfL$. It follows that $\bfU^{\perp}$, defined as the restriction of $\bfU$ to the subspace $\bfL^{\perp}$, satisfies
$\bfU^\perp ({\bf K}_\calE,\cdot) =  \bfU^\perp ({\bf K}_\calO,\cdot) = 0$, so $\bf U^\perp$ is degenerate in the directions of  both vectors.

In analogy with $\bf U$ we also consider the flat case and the small curvature limit as special cases. In a flat spacetime we may diagonalize $\bf U^{\perp}$ by the same substitution we have used for  $\bfU$:  we introduce again $\xi^\mu$ and $\delta x_\calO^\mu$ as the new variables. The orthogonality
 condition  $\bfL(\bfX) = 0$ defining $\bfL^\perp$ becomes simply $l_{\mu}\,\xi^\mu = 0$, with no restrictions on $\delta x_\calO^\mu$. It follows that we can decompose $\xi^\mu$ according to
\bea
\xi^\mu = \xi^A\,e_{A}^\mu + \xi^l\,l^\mu,
\eea
where the two spatial vectors $e_A^\mu$, $A=1,2$, constitute an orthonormal basis of the transverse subspace orthogonal to $l^\mu$. It is straightforward to show then that
\bea
\bfU^{\perp}(\bfX,\bfX) = -\frac{1}{\Delta \lambda}\,\xi^A\,\xi^B\,\delta_{AB}.
\eea
Thus the extrinsic curvature is degenerate in 5 directions and negative in the two transverse directions of $\xi^\mu$, so the overall signature reads $(0,2,5)$. In the small curvature limit the two minus signs are stable against linear perturbations and two zeros are protected because of the identities proved above. Therfore up to 3 zeros out of  5 may turn into other signs due to linear curvature perturbations.

\section{$\bfU$ and the variations of the tangent vectors at $\lambda_\calO$ and $\lambda_\calE$} \label{app:Uvariations}

Let  $t_\calO^{\mu'}(x,x')$ and $t_\calE^{\mu}(x,x')$ denote the tangent vectors to a geodesic passing through $x$ and $x'$ and parametrized so that $\lambda - \lambda' = \Delta \lambda$
for a fixed $\Delta \lambda$. The vectors can be related to the derivatives of the world function via
\bea
 t_{\calO\,\mu'}(x,x') &=& -\frac{1}{\Delta \lambda}\,\sigma_{,\mu'} \label{eq:tangentO}\\
 t_{\calE\,\mu}(x,x') &=& \frac{1}{\Delta \lambda}\,\sigma_{,\mu}  \label{eq:tangentE}
\eea 
 see \cite{Poisson2011}. \ZZ{Z61 p. 66}
We consider now a LF coordinate system,  i.e. we have $\Gamma\UD{\mu'}{\nu'\alpha'} (\calO) = 
\Gamma\UD{\mu}{\nu\alpha}(\calE) = 0$. In these coordinates we calculate from (\ref{eq:tangentO})-(\ref{eq:tangentE}) the linear variation of the components of the tangent vectors under the variation of the geodesic endpoints, 
calculated around the fiducial geodesic corresponding with $x=\calE$, $x'=\calO$. It takes the form of
\bea
 \delta t_{\calO\,\mu'} &=&  -\frac{1}{\Delta \lambda}\left(\sigma_{,\mu'\nu'}\,\delta x_\calO^{\nu'} + \sigma_{,\mu'\nu}\,\delta x_\calE^\nu\right) \label{eq:vartO}\\
 \delta t_{\calE\,\mu} &=&  \frac{1}{\Delta \lambda}\left(\sigma_{,\mu\nu'}\,\delta x_\calO^{\nu'} + \sigma_{,\mu\nu}\,\delta x_\calE^\nu\right), \label{eq:vartE}\
\eea
with the partial derivatives of $\sigma(x,x')$ taken at $(\calE,\calO)$. But in the LF coordinates the standard second derivatives of $\sigma$ coincide with the 
covariant ones, i.e. $\sigma_{;\mu\nu} = \sigma_{,\mu\nu} $, $\sigma_{;\mu'\nu'} = \sigma_{,\mu'\nu'} $, $\sigma_{;\mu'\nu} = \sigma_{,\mu'\nu} $, 
$\sigma_{;\mu\nu'} = \sigma_{,\mu\nu'} $. Also, in these coordinates the coordinate-wise variations of $t_{\calO\,\mu'}$ and $t_{\calE\,\mu}$ can be identified with the 
covariant direction variations, i.e.
$ \Delta l_{\calO\,\mu'} = \delta t_{\calO\,\mu'} $ and $ \Delta l_{\calE\,\mu} = \delta t_{\calE\,\mu} $. In this case by comparing (\ref{eq:vartO})-(\ref{eq:vartE}) with the 
definition of $\bfU$ (\ref{eq:Udef}) we get (\ref{eq:Udef1})-(\ref{eq:Udef2}).

\section{Transformations between $\bfW$ and $\bfU$ in the  general case.} \label{app:UfromW}

By comparing (\ref{eq:Wdef1})-({\ref{eq:Wdef2}) to (\ref{eq:Udef1})-(\ref{eq:Udef2}) we see that the transformation between $\bfW$ and $\bfU$ corresponds to passing from the  initial value problem for the GDE, with 
the initial data for the perturbed geodesic given by the pair $(\delta x_\calO^{\mu'},\Delta l_\calE^{\nu'})$, to the boundary problem, with the displacement  vectors at both ends $(\delta x_\calO^{\mu'},\delta x_\calE^{\nu})$  defining the boundary values. Now, in a linear system of ODE's passing from one problem to the other is a fairly straightforward algebraic problem we will solve below.

 We  transform the relation (\ref{eq:Wdef1})-(\ref{eq:Wdef2})  in order to express $\Delta l_\calO^{\mu'}$ and $\Delta l_\calE^\nu$ by $\delta x_\calO^{\mu'}$ and $\delta x_\calE^\mu$: by left-multiplying both sides of (\ref{eq:Wdef1}) with $W_{XL}^{-1}$ we obtain $\Delta l_\calO^{\mu'}$, which we subsequently insert into (\ref{eq:Wdef2}) to get $\Delta l_\calE^\nu$. After collecting the like terms and small rearrangements  we get:
\bea
 \Delta l_\calO^{\mu'} &=& -\left(W_{XL}^{-1}\right)\UD{\mu'}{\nu} \,W_{XX}{}\UD{\nu}{\sigma'} \,\delta x_\calO^{\sigma'} + \left(W_{XL}^{-1}\right)\UD{\mu'}{\nu} \,\delta x_\calO^{\nu}\label{eq:Wtransformed1}\\
 \Delta l_\calE^{\mu} &=& \left(-W_{LL}{}\UD{\mu}{\nu'}\,\left(W_{XL}^{-1}\right)\UD{\nu'}{\kappa}\,W_{XX}{}\UD{\kappa}{\sigma'} + W_{LX}{}\UD{\mu}{\sigma'}\right)\,\delta x_\calO^{\sigma'} \nonumber \\
 & +& W_{LL}{}\UD{\mu}{\nu'}\,\left(W_{XL}^{-1}\right)\UD{\nu'}{\sigma}\,\delta x_\calE^\sigma\label{eq:Wtransformed2}
\eea
In the next step we need to lower the indices in both equations using $g_{\mu\nu}$ at $\calO$ in the first equation and $g_{\mu\nu}$ at $\calE$ in the second one.
By comparing (\ref{eq:Udef1})-(\ref{eq:Udef2}) with the result we obtain immediately the nonlinear relation
(\ref{eq:UfromW}).

The inverse relation can be derived similarly - we may reexpress $\delta x^{\mu}_{\calE}$ and $\Delta l^{\mu}_{\calE}$ in (\ref{eq:UfromW}) with $\delta x^{\mu'}_{\calO}$ and $\Delta l^{\mu '}_{\calO}$:
    \bea
        \delta x^{\mu}_{\calE} &=& -(U_{\calO\calE}^{-1})^{\mu\alpha'}U_{\calO\calO \ \alpha'\nu'}\  \delta x^{\nu'}_{\calO} + (U_{\calO\calE}^{-1})^{\mu\alpha'}g_{\alpha'\nu'} \ \Delta l_{\calO}^{\nu'}\label{eq:Utransformed1}\\
        \Delta l_{\calE}^{\mu} &=& g^{\mu\sigma}\bigg( U_{\calE\calE \ \sigma\omega} (U_{\calO\calE}^{-1})^{\omega\alpha'} U_{\calO\calO \ \alpha'\nu'} - U_{\calE\calO\ \sigma\nu'}  \bigg)\, \delta x_{\calO}^{\nu'} \nonumber\\
&-& g^{\mu\sigma}\, U_{\calE\calE \ \sigma\omega}(U_{\calO\calE}^{-1})^{\omega\gamma'}\,g_{\gamma'\nu'}\ \Delta l_{\calO}^{\nu'}\label{eq:Utransformed2}.  
    \eea
By comparing (\ref{eq:Wdef1})-(\ref{eq:Wdef2}) with (\ref{eq:Utransformed1})-(\ref{eq:Utransformed2}) we obtain the inverse relation (\ref{eq:WfromU}).

\section{Mapping $\bfW$ up to the linear order in curvature}\label{app:UW1}
In this appendix we  derive the formulas for 0th and 1st order of $\mw$, expressed in terms of curvature integrals along the null geodesic. This
has already been done for timelike geodesics \cite{PhysRevD.99.084044} and for null geodesic in the transverse subspace \cite{PhysRevD.83.083007, 2012AIPC.1471...82G, PhysRevD.97.084010},  albeit using different terminology and notation. For the sake of completeness we present a derivation in our framework.

Bilocal operators ($\wxx,\wxl$, etc.), constituting the  $4\times4$ submatrices of the matrix $\mw$, are solutions to the nonhomogeneous ODE (\ref{eq:ODEW}) defined along the null geodesic between the emitter and observer. From this ODE we get system of differential equations:
\begin{subequations}
    \begin{empheq}[left=\empheqlbrace]{align}
      & \dot{W}_{XX}=\wlx \notag\\
      & \dot{W}_{XL}=\wll \nonumber\\
      & \dot{W}_{LX}=R_{ll}\wxx \nonumber\\
      & \dot{W}_{LL}=R_{ll}\wxl, \nonumber
    \end{empheq}
\end{subequations}
with initial conditions:
\begin{subequations}
    \begin{empheq}[left=\empheqlbrace]{align}
      & \wxx(\lz)=I_{4} \nonumber\\
      & \wxl(\lz)=0 \nonumber\\
      & \wlx(\lz)=0 \nonumber\\
      & \wll(\lz)=I_{4}. \nonumber
    \end{empheq}
\end{subequations}
We may also rewrite this, by substituting relevant matrices, as system of second-order equations:
\begin{subequations}
\label{W1ord:eqs}
    \begin{empheq}[left=\empheqlbrace]{align}
      & \ddot{W}_{XX}=R_{ll}\wxx \label{wxx:eq}\\
      & \ddot{W}_{XL}=R_{ll}\wxl \\
      & \wlx=\dot{W}_{XX} \\
      & \wll=\dot{W}_{XL},
    \end{empheq}
\end{subequations}
with initial conditions:
\begin{subequations}
\label{W1ord_initial:eqs}
    \begin{empheq}[left=\empheqlbrace]{align}
      & \wxx(\lz) = I_4 \\
      &  \wxl(\lz) = 0 \\
      &  \wlx(\lz) = 0 \\
      & \wll(\lz) = I_4.
    \end{empheq}
\end{subequations}

The 2nd order equations for $W_{XX}$ and $W_{XL}$, together with the initial data  (\ref{W1ord_initial:eqs}), form two autonomous matrix ODE's. Once we solve them we can obtain $W_{LX}$ and $W_{LL}$  by subsequent differentiation of $W_{XX}$ and $W_{XL}$, see the two lower equations. This system of ODE's has been presented in \cite{Grasso:2018mei} and we will use it here to derive the perturbative expansion. 

In 0th order the solution of this system is the following:
\begin{subequations}
\label{W0th:sol}
    \begin{empheq}[left=\empheqlbrace]{align}
      & {W^{(0)}_{XX}}\UD{\bmu}{\bnu} = \delta\UD{\bmu}{\bnu} \\
      & {W^{(0)}_{XL}}\UD{\bmu}{\bnu} = (\lambda-\lz)\delta\UD{\bmu}{\bnu} \\
      & {W^{(0)}_{LX}}\UD{\bmu}{\bnu} = 0 \\
      & {W^{(0)}_{LL}}\UD{\bmu}{\bnu} = \delta\UD{\bmu}{\bnu}.
    \end{empheq}
\end{subequations}
In order to evaluate the 1st order we insert solutions (\ref{W0th:sol}) into relevant equations (\ref{W1ord:eqs}) containing curvature factor and integrate with respect to affine parameter. For $\wxx$ we have:
\begin{equation*}
\begin{split}
 {W^{(1)}_{XX}}\UD{\bmu}{\bnu} & = \int^{\lep}_{\lz}d\lambda \int^{\lambda}_{\lz} R^{\bmu}_{\ \bal\bbet\bnu}(\lambda')l^{\bal}l^{\bbet} \  \dd\lambda'.
\end{split}
\end{equation*}    
and $\wlx$, as its derivative:
\begin{equation*}
    {W^{(1)}_{LX}}\UD{\bmu}{\bnu} = \int^{\lep}_{\lz} \rll \  \dd\lambda.
\end{equation*}
Performing the same procedure for $\wxl$ and $\wll$, we get full 1st order $\mw$:
\begin{subequations}
    \begin{empheq}{align}
      & {W^{(1)}_{XX}}\UD{\bmu}{\bnu} = \int^{\lep}_{\lz} \dd\lambda \int^{\lambda}_{\lz}R^{\bmu}_{\ \bal\bbet\bnu}(\lambda')l^{\bal}l^{\bbet} \  \dd\lambda', \nonumber\\
      & {W^{(1)}_{XL}}\UD{\bmu}{\bnu} = \int^{\lep}_{\lz} \dd\lambda \int^{\lambda}_{\lz}R^{\bmu}_{\ \bal\bbet\bnu}(\lambda')l^{\bal}l^{\bbet} (\lambda'-\lz)  \ \dd\lambda',\nonumber\\
      & {W^{(1)}_{LX}}\UD{\bmu}{\bnu} = \int^{\lep}_{\lz}R^{\bmu}_{\ \bal\bbet\bnu}(\lambda)l^{\bal}l^{\bbet} \ \dd\lambda, \nonumber\\
      & {W^{(1)}_{LL}}\UD{\bmu}{\bnu} = \int^{\lep}_{\lz}R^{\bmu}_{\ \bal\bbet\bnu}(\lambda)l^{\bal}l^{\bbet} (\lambda-\lz) \ \dd\lambda. \nonumber
    \end{empheq}
\end{subequations}
In order to simplify the formulas for $\mathbf{W}^{(1)}$, we can rewrite them via integration by parts as integrals with kernel centered at $\lz$. For $\wxx$ we have: 
\begin{equation*}
\begin{split}
 {W^{(1)}_{XX}}\UD{\bmu}{\bnu} & = \int^{\lep}_{\lz}\dd\lambda \int^{\lambda}_{\lz} R^{\bmu}_{\ \bal\bbet\bnu}(\lambda')l^{\bal}l^{\bbet} \  \dd\lambda'=\int^{\lep}_{\lz} \rll(\lep-\lambda)\ \dd\lambda,
\end{split}
\end{equation*}    
where  $1$ is integrated to  \ $\lambda-\lz$ and  $\int^{\lambda}_{\lz} R^{\bmu}_{\ \bal\bbet\bnu}(\lambda')l^{\bal}l^{\bbet} \  d\lambda'$ is differentiated to $R^{\bmu}_{\ \bal\bbet\bnu}(\lambda)l^{\bal}l^{\bbet}$.
Applying the same to other operators we also get:
\begin{subequations}
\label{W1ord:final}
    \begin{empheq}{align}
      & {W^{(1)}_{XL}}\UD{\bmu}{\bnu} = \int^{\lep}_{\lz}\rll(\lep-\lambda)(\lambda-\lz)\ \dd\lambda,\\
      & {W^{(1)}_{LX}}\UD{\bmu}{\bnu} = \int^{\lep}_{\lz}\rll\ \dd\lambda, \\
      & {W^{(1)}_{LL}}\UD{\bmu}{\bnu} = \int^{\lep}_{\lz}\rll(\lambda-\lz)\ \dd\lambda.
    \end{empheq}
\end{subequations}

Finally, in order to make the formulas look more compact, we may rewrite $\mw^{(0)}$ and $\mw^{(1)}$ in the form of $8\times8$ matrices:
\begin{equation}
    \mw^{(0)} = \left(\begin{array}{l|l}
I_4 & \Delta \lambda \cdot I_4\\ \hline
0 & I_4
\end{array}\right)
\end{equation}
\begin{equation}
\label{W1rd:matrix}
    \mw^{(1)} = \int^{\lep}_{\lz} \left(\begin{array}{l|l} \rll\,(\lep-\lambda) & \rll\,(\lep-\lambda)(\lambda-\lz) \\ \hline  \rll & \rll\, (\lambda-\lz) \end{array}\right)\,\dd\lambda,
\end{equation}
where $\Delta\lambda=\lep-\lz$. This result is consistent with equation (46) from \textcite{ PhysRevD.83.083007} and equations (2.1)-(2.3) from \textcite{PhysRevD.99.084044}.

\begin{comment}
\begin{equation}
\label{W1rd:matrix}
    \mw^{(1)} = \int^{\lep}_{\lz}\rll \begin{pmatrix}
\lep-\lambda & (\lep-\lambda)(\lambda-\lz)\\
1 & \lambda-\lz
\end{pmatrix}\ d\lambda,
\end{equation}
where $\Delta\lambda=\lep-\lz$. 
For the sake of brevity we abuse here the norations slightly. The equation above should read as follows: the relevant bilocal operator is given by integral of Riemann tensor multiplied by relevant element of the matrix of polynomials, so it corresponds to the formulas (\ref{W1ord:final}).
\end{comment}

\section{Mapping $\bfU$ up to the linear order in curvature} \label{app:1PTR}
We are working in a parallel propagated tetrad, so the metric tensor satisfies $\eta_{\bar\mu\bar\nu} = \textrm{diag}(-1,1,1,1) = \eta^{\bar\mu\bar\nu}$.
$\uu$ and $\mw$ are related between each other with a nonlinear transformation (\ref{eq:UfromW}). Obviously the derivation of $\uu^{(0)}$ and $\uu^{(1)}$ requires evaluating $\wxlin$. Assuming that terms quadratic in curvature are small we have
\bea
    W_{XL}^{-1\ (0)}&=&(W_{XL}^{(0)})^{-1}, \nonumber \\
    W_{XL}^{-1\ (1)}&=&-(W_{XL}^{(0)})^{-1} W_{XL}^{(1)} (W_{XL}^{(0)})^{-1}. \nonumber
\eea
Substituting (\ref{W0th:sol}) and (\ref{W1ord:final}) we get
\begin{subequations}
\begin{align}
       &({W^{-1 \ (0)}_{XL}})\UD{\bmu}{\bnu} = -\frac{1}{\Delta\lambda} \delta\UD{\bmu}{\bnu} \nonumber \\
       &({W^{-1 \ (1)}_{XL}})\UD{\bmu}{\bnu} = -\int^{\lep}_{\lz}\rll(\lep-\lambda)(\lambda-\lz)\ \dd\lambda. \nonumber
\end{align}
\end{subequations}

In order to find $\uu^{(0)}$ we may use the relation (\ref{eq:UfromW}) between $\uu$ and $\mw$ at 0th order:
\begin{subequations}
\label{U:0rd}
\begin{align}
    & \uoo^{(0)}=-\eta W^{-1 \ (0)}_{XL}\wxx^{(0)}= -\frac{\eta}{\Delta\lambda}I_{4}\\
    & \uoe^{(0)}=\eta W^{-1 \ (0)}_{XL}=\frac{\eta}{\Delta\lambda}I_{4}\\
    & \ueo^{(0)}=\uoe^{(0)\ T}= \frac{\eta^{T}}{\Delta\lambda}I_{4}\\
    & \uee^{(0)}=-\eta\wll^{(0)}W^{-1 \ (0)}_{XL}=-\frac{\eta}{\Delta\lambda}I_{4},
\end{align}
\end{subequations}
$\eta$ denoting the metric. 
The linear order term $\uu^{(1)}$ requires linearizing the relation (\ref{eq:UfromW}) around $\bfW^{(0)}$: we assume
\begin{equation*}
    \uu=\uu^{(0)}+\uu^{(1)}.
\end{equation*}
We'll expand in this manner every bilocal operator contained in $\uu$, use the relevant relation (\ref{U:0rd}) and collect the products with respect to the right order, neglecting the quadratic and higher orders in curvature.
For $\uoo$ we have:
\begin{equation*}
\begin{split}
 \uoo^{(0)}+\uoo^{(1)} & = -\eta(\wxlinzer+\wxlion)\cdot(\wxx^{(0)}+\wxx^{(1)})
\end{split}
\end{equation*}   
or
\begin{equation*}
\begin{split}
 \uoo^{(0)}+\uoo^{(1)} & = \eta\left(
 -\wxlinzer\wxx^{(0)}-\wxlinzer\wxx^{(1)}-\wxlion\wxx^{(0)}-\wxlion\wxx^{(1)}\right).
\end{split}
\end{equation*} 
The first term on RHS in $\uoo^{(0)}$, while the last one is quadratic and can be neglected at first order. We then get
\begin{equation*}
    \begin{split}
        \uoo^{(1)} & =\eta\left(-\wxlinzer\wxx^{(1)}-\wxlion\wxx^{(0)}\right).
    \end{split}
\end{equation*}
Using now (\ref{W0th:sol}) and (\ref{W1ord:final}) we may rewrite the above result in the following form:
\begin{eqnarray*}
    U^{(1)}_{\mathcal{O}\mathcal{O}\,\bmu\bnu} &=& -\frac{1}{\dl^2}\int^{\lep}_{\lz}R_{\bmu \bal\bbet\bnu}l^{\bal}l^{\bbet} (\lep-\lambda)^2\ \dd\lambda,
\end{eqnarray*}
note the lowering of the first index  due to $\eta$ in front of the expressions.

We apply the same procedure to the other 3  blocks of  $\uu$ and obtain:
\begin{subequations}
    \begin{empheq}{align}
      U^{(1)}_{\mathcal{O}\mathcal{E}\bmu\bnu}&  = - \frac{1}{\dl^2}\int^{\lep}_{\lz}(\lep-\lambda)(\lambda-\lz)R_{\bmu \bal\bbet\bnu}l^{\bal}l^{\bbet}\ \dd\lambda,\nonumber\\
      U^{(1)}_{\mathcal{E}\mathcal{O}\bmu\bnu}& = -\frac{1}{\dl^2}\int^{\lep}_{\lz}(\lep-\lambda)(\lambda-\lz)R_{\bmu \bal\bbet\bnu}l^{\bal}l^{\bbet}\ \dd\lambda, \nonumber\\
      U^{(1)}_{\mathcal{E}\mathcal{E}\bmu\bnu}& = - \frac{1}{\dl^2}\int^{\lep}_{\lz}(\lambda-\lz)^2R_{\bmu \bal\bbet\bnu}l^{\bal}l^{\bbet}\ \dd\lambda. \nonumber
    \end{empheq}
\end{subequations}
The result is consistent with (\ref{eq:Usymmetric}). As in the case of $\mw$, we may rewrite $\uu^{(0)}$ and $\uu^{(1)}$ in more compact, matrix form:
\begin{equation}
    \uu^{(0)} = \frac{1}{\Delta\lambda}\begin{pmatrix}
-\eta_{\bmu\bnu} & \eta_{\bmu\bnu}\\
\eta_{\bmu\bnu} & -\eta_{\bmu\bnu}
\end{pmatrix}, \nonumber
\end{equation}
\begin{equation}
    \uu^{(1)} =  -\int^{\lep}_{\lz} \left(\begin{array}{l|l} R_{\bmu \bal\bbet\bnu}l^{\bal}l^{\bbet}\,\frac{(\lep-\lambda)^2}{\Delta\lambda^2} & R_{\bmu \bal\bbet\bnu}l^{\bal}l^{\bbet}\,\frac{(\lep-\lambda)(\lambda-\lz)}{\Delta\lambda^2} \\ \hline  R_{\bmu \bal\bbet\bnu}l^{\bal}l^{\bbet}\,\frac{(\lep-\lambda)(\lambda-\lz)}{\Delta\lambda^2} & R_{\bmu \bal\bbet\bnu}l^{\bal}l^{\bbet}\,\frac{(\lambda-\lz)^2}{\Delta\lambda^2} \end{array}\right)\ \dd\lambda. \nonumber
\end{equation}

\section{Proof of the invariance of $\mu$ and $\nu$.} \label{app:proof}

 %Proof:
%We  show that the submatrix $U_{\calO\calO}{}\UD{A'}{B'}$ transforms by a rigid rotation at most if we change the Sachs frame at $\calO$, and the same for
%$U_{\calO\calE}{}\UD{A'}{B}$ and  $U_{\calE\calE}{}\UD{A}{B}$. (Follows from the identity involving the action of $U$ on $l$.) 

We establish here the following Theorem:
\begin{theorem}
Let $(u_\calO, f_A, f_3)$ and $(u_\calE, g_A, g_3)$ be a pair of ON, adapted and properly oriented tetrads at $\calO$ and $\calE$ respectively.  Let $(\tilde u_\calO, \tilde f_A, \tilde f_3)$ and $(\tilde u_\calE, \tilde g_A, \tilde g_3)$ be another such pair. Then $\mu$ and $\nu$ given by (\ref{eq:muviaU}) and (\ref{eq:nuviaU}) have the same values when calculated in both pairs of adapted tetrads. \label{thm:frameindep}
\end{theorem}

Note that in this Appendix  we are not using primes for internal tetrad indices at $\calO$. 

\textbf{Proof. }
We first need the following Lemma, closely related to the shadow theorem by Sachs \cite{Korzynski:2017nas, sachs61}:
\begin{lemma} \label{lemma:trans}  In this setting we have the following relation between the transverse vectors
\bea
  \tilde f_A &=& R\UD{B}{A}\,f_B + a_A\,l_\calO \label{eq:ftransform} \\
  \tilde g_A &=& S\UD{B}{A}\,g_B + b_A\,l_\calE, \label{eq:gtransform}
\eea
where both $R\UD{A}{B}$ and $S\UD{A}{B}$ are 2D rotation matrices, i.e. $R^T\,R = S^T \,S = I_2$, $\det R = \det S = 1$, and $a_A$, $b_A$ are (irrelevant) 4 numbers.
\end{lemma}

{\bf Proof of Lemma \ref{lemma:trans}.} \ZZ{Z62 p. 42}
The proof proceeds in two steps: we first prove that relations (\ref{eq:ftransform})-(\ref{eq:gtransform}) indeed hold and that the matrices $R$ and $S$ need to be orthogonal. Then we show that their determinants must be positive as well.

{\bf Orthogonality of $R$ and $S$.} Let the dot denote the scalar product defined by the spacetime metric $g$. Since the $\tilde f_A$'s are orthogonal to both $\tilde f_3$ and $\tilde u_\calO$ and the $\tilde g_A$'s are orthogonal to both $\tilde g_3$ and $\tilde u_\calE$ we have $\tilde f_A \cdot l_\calO = 0$ and $\tilde g_A \cdot l_\calE = 0$, i.e. the transverse vectors are also orthogonal to the appropriate null tangents. Thus $\tilde f_A$'s and $\tilde e_A$'s lie in the subspaces orthogonal to the appropriate null tangent. 

Consider now the decomposition of 
$\tilde f_A$ in the $(u_\calO, f_A, f_3)$ frame, i.e. $\tilde f_A = C_A\, u_\calO + R\UD{B}{A}\, f_B + D_A\,f_3$, with undetermined so far coefficients $R\UD{B}{A}$, $C_A$ and $D_A$. Then the condition of orthogonality to $l_\calO = Q(- u_\calO + f_3)$ implies that $C_A = -D_A$, or
equivalently  $\tilde f_A = -D_A\,u_\calO + R\UD{B}{A}\,f_B + D_A\,f_3 = R\UD{B}{A}\,f_B + \frac{D_A}{Q}\,l_\calO$.
Applying the same reasoning to $g_A$ we prove the following relations:
\bea
\tilde f_A &=& R\UD{B}{A}\,f_B + a_A\, l_\calO  \label{eq:fA5}\\
 \tilde g_A &=& S\UD{B}{A}\,g_B + b_A\, l_\calE \label{eq:gA5}
\eea
with $R\UD{B}{A}$, $S\UD{B}{A}$, $a_A$ and $b_A$ arbitrary. 

We now impose the orthogonality and normalization conditions between the transverse vectors, $\tilde f_A\cdot \tilde f_B = \delta_{AB}$, $\tilde g_A \cdot \tilde g_B = \delta_{AB}$, on  relations (\ref{eq:fA5})-(\ref{eq:gA5}). They imply then that 
$R\UD{B}{A}\,R\UD{D}{C}\,\delta_{BD} = \delta_{AC}$ and $S\UD{B}{A}\,S\UD{D}{C}\,\delta_{BD} = \delta_{AC}$, i.e. both $R\UD{B}{A}$ and $S\UD{B}{A}$ must be orthogonal 2-by-2 matrices.
We now only need to prove that they are also special orthogonal, i.e. $\det S\UD{B}{A} = \det R\UD{B}{A} = 1$. 

{\bf The determinants of $R$ and $S$ are equal to +1.} Since both matrices are orthogonal, it suffices to show that $\det R\UD{B}{A} > 0$ and $\det S\UD{B}{A} > 0$.  We will do it for
$R\UD{B}{A}$, because the reasoning for $S\UD{B}{A}$ is identical.

Let $\kappa$ denote the exterior product of all base vectors, i.e. $\kappa =  u_\calO \wedge f_1 \wedge  f_2 \wedge f_3$. Since both tetrads $(u_\calO,f_A,f_3)$ and $(\tilde u_\calO,\tilde f_A,\tilde f_3)$ are assumed to be properly
oriented we must also have $\kappa = \tilde u_\calO \wedge \tilde f_1 \wedge \tilde f_2 \wedge \tilde f_3$ and an analogous relation for the other tetrad at $\calE$. 
We now show that this implies $\det R\UD{B}{A}>0$.

At $\calO$ we have $l_\calO=\tilde Q(-\tilde u_\calO+\tilde f_3)$, so 
\bea
\kappa = \frac{1}{\tilde Q}\,\tilde u_\calO\wedge \tilde f_1 \wedge \tilde f_2 \wedge l_\calO. \nonumber
\eea
Relation between the 4-velocities:
\bea
 u_\calO = {\gamma}\,\tilde u_\calO + \gamma\,\beta^i\,\tilde f_i = {\gamma}\,\tilde u_\calO + \gamma\,\beta^A\,\tilde f_A + \gamma\,\beta^3\,\tilde f_3 \nonumber
\eea
where $\beta^i\beta^j\delta_{ij} < 1$ and $\gamma = \left(1 - \beta^i\,\beta^j\,\delta_{ij}\right)^{-1/2}$. It follows that
\bea
 u_\calO = \gamma(1+\beta^3)\,\tilde u_\calO + \gamma\,\beta^A\,\tilde f_A + \frac{\gamma\beta^3}{\tilde Q}\,l_\calO, \nonumber
\eea
We can now substitute $\tilde u_\calO$ by $ u_\calO$ in $\kappa$:
\bea
\kappa = \frac{\gamma(1+\beta^3)}{\tilde Q} \, u_\calO \wedge \tilde f_1 \wedge \tilde f_2 \wedge l_\calO. \nonumber
\eea
We can now make use of the relation (\ref{eq:fA5}) between the two transverse vectors $\tilde f_A$ and $f_A$. Note that we have 
\bea
 \tilde f_1 \wedge \tilde f_2 = (\det R\UD{B}{A})\,f^1 \wedge f^2 + l_\calO \wedge \alpha, \nonumber
\eea
where $\alpha$ is an irrelevant 1-form. Then
\bea
\kappa = \frac{\gamma(1+\beta^3)\,\det R\UD{B}{A}}{\tilde Q} \, u_\calO \wedge f_1 \wedge f_2 \wedge l_\calO. \nonumber
\eea
In the last step we exchange $l_\calO$ for $f^3$ using $l_\calO = Q(- u_\calO + f_3)$:
\bea
 \kappa = \frac{Q\gamma(1+\beta^3)\,\det R\UD{B}{A}}{\tilde Q}  \,u_\calO \wedge f_1 \wedge f_2 \wedge f_3 \nonumber.
\eea
But we also have $\kappa = u_\calO \wedge f_1 \wedge f_2 \wedge f_3$, so 
\bea \frac{Q\gamma(1+\beta^3)\,\det R\UD{B}{A}}{\tilde Q} = 1. \nonumber
\eea
Now,  since $\gamma$, $\tilde Q$, $Q$ and $1 + \beta^3$ are all positive, it follows that $\det R\UD{B}{A} > 0$ as well.
The same argument can be applied to $\det S\UD{B}{A}$.

{\bf End of proof of Lemma \ref{lemma:trans}.}

We now evaluate the transverse components of (lowered-index) $U_{\calO\calO}$ in the new frame:
\bea
U_{\calO\calO}(\tilde f_A,\tilde f_B) &=&  U_{\calO\calO}(f_C,f_D)\,R\UD{C}{A}\,R\UD{D}{B} + U_{\calO\calO}(l_\calO, f_D)\,\,a_A\,R\UD{D}{B} \nonumber  \\
&+&  U_{\calO\calO}(f_C,l_\calO)\,R\UD{C}{A}\,a_B + U_{\calO\calO}(l_\calO, l_\calO)\,a_A\,a_B \label{eq:Utrans}.
\eea
From  (\ref{eq:Uoowithl}) we know that for every $X \in T_\calO M$ we have
\bea
 U_{\calO\calO}(X, l_\calO) = U_{\calO\calO}( l_\calO,X) = -\frac{1}{\Delta\lambda}\,l_\calO\cdot X, \nonumber
\eea
where the dot again denotes the spacetime metric scalar product. Now, we have
\bea
l_\calO\cdot f _A = l_\calO\cdot \tilde f_A = l_\calO\cdot l_\calO = 0 \nonumber
\eea
from the definition of an adapted base and from the null condition on $l_\calO$. Thus the last three terms in (\ref{eq:Utrans}) simply vanish and we have the transformation rule in
the form of 
\bea
U_{\calO\calO}(\tilde f_A,\tilde f_B) &=&  U_{\calO\calO}(f_C,f_D)\,R\UD{C}{A}\,R\UD{D}{B}.\nonumber
\eea
The same reasoning gives then the following transformation rules:
\bea
U_{\calE\calE}(\tilde g_A,\tilde g_B) &=&  U_{\calE\calE}(g_C,g_D)\,S\UD{C}{A}\,S\UD{D}{B} \nonumber\\
U_{\calO\calE}(\tilde f_A,\tilde g_B) &=&  U_{\calO\calE}(f_C,g_D)\,R\UD{C}{A}\,S\UD{D}{B} \nonumber\\
U_{\calE\calO}(\tilde g_A,\tilde f_B) &=&  U_{\calE\calO}(g_C,f_D)\,S\UD{C}{A}\,R\UD{D}{B} . \nonumber
\eea
Since the rotation matrices $S$ and $R$ have  a unit determinant, they have no impact on the value of the sub-determinants of $U_{\calO\calO}$ and others: 
\bea
\det U_{\calO\calO}(\tilde f_A,\tilde f_B) &=& \det  U_{\calO\calO}(f_A,f_B)\nonumber\\
\det U_{\calE\calE}(\tilde g_A,\tilde g_B) &=& \det U_{\calE\calE}(g_A,g_B)\nonumber\\
\det U_{\calO\calE}(\tilde f_A,\tilde g_B) &=& \det U_{\calO\calE}(f_A,g_B)\nonumber\\
\det U_{\calE\calO}(\tilde g_A,\tilde f_B) &=& \det U_{\calE\calO}(g_A,f_B).\nonumber
\eea
Therefore the values of $\mu$ and $\nu$  defined by the expressions  (\ref{eq:muviaU}) and (\ref{eq:nuviaU}) are the same in any pair of  adapted  tetrads at $\calO$ and $\calE$. 

 {\bf End of proof of Theorem \ref{thm:frameindep}.}

Thus $\mu$ and $\nu$ are independent of the choice of the adapted ON tetrad at $\calO$ and $\calE$, and thus $u_\calO$- and $u_\calE$-independent. Equation (\ref{eq:muviaU}) can be used with any pair of adapted tetrads at $\calO$ and $\calE$. \ZZ{Z62 p. 32.}

\section{$\mu$, $\nu$ and the cross-sectional area of infinitesimal bundles of rays} \label{app:cross}

For simplicity we derive only the expressions for $\mu$, noting here that those for $\nu$ have exactly the same form, but with the role of $\calO$ and $\calE$ reversed.
The problem of tracking the intersection of an infinitesimal bundle with a Sachs screen of an observer is an old one. It is well-known that the 
area of this intersection  is independent of the choice of the observer's frame \cite{perlick-lrr}.

 We begin by picking an orthonormal, adapted tetrad $(u, f_A, f_3)$ at $\calO$ and parallel-propagating it along $\gamma_0$. We
consider an infinitesimal bundle of light rays intersecting the area spanned by $f_1$ and $f_2$, defined as ${\cal A}_\calO = f_1 \wedge f_2$, such that the null geodesics are parallel at $\calO$. Its behaviour is determined by the geodesic deviation equation along $\gamma_0$. From \eqref{eq:Wdef1} we see that at $\calE$ this bundle crosses the area element spanned by $W_{XX}(f_1)$ and $W_{XX}(f_2)$. The cross-section of the bundle by the Sachs screen space $f_1$ and $f_2$ is given by the projection of these vectors to the Sachs screen. It can described by the products $f_B \cdot W_{XX}(f_A)$, with $\cdot$ defined by the spacetime metric.

In \cite{Grasso:2018mei} the following definition of $\mu$ was given:
\bea
\mu = 1 - \det {w_\perp} \UD A B = 1 - \det (\delta \UD A B + {m_\perp} \UD A B),
\eea
where $m_\perp : \calP_\calO \to \calP_\calE$ was the perpendicular part of the emitter-observer assymetry operator. It is easy to show that by construction, ${w_\perp} \UD A B = {\wxx} \UD A B$, i.e. it is   a 2-by-2 submatrix of ${\wxx} \UD {\bar \mu} {\bar \nu}$, corresponding to the projection onto the screen space spanned by $f_1$ and $f_2$. Namely,
we have
\bea
{w_\perp}\UD{A}{B} = [f_A] \cdot w_\perp ([f_B]) 
\eea
where $[f_A]$ is the equivalence class defined by the equivalence relation $X \sim X + C\, l$, where $C$ is a constant. But from the definition in \cite{Grasso:2018mei} we also have $[f_A] \cdot w_\perp ([f_B]) 
 = f_A\cdot W_{XX}(f_B)$, which proves the equality. Thus
 \bea
\mu = 1 - \det {W_{XX}} \UD A B .
\eea
The determinant $\det{W_{XX}} \UD A B$ plays the role of the  coefficient defining how an area element of the screen space $\calO$ appears rescaled on the screen space at $\calE$ when carried by the beam. More precisely, it is equal to the ratio of the area element $ W_{XX}(f_1) \wedge W_{XX}(f_2)$ projected to the screen space  and  $f_1 \wedge f_2$ at $\calE$. Since the latter defines the cross-sectional area of the infinitesimal bundle at $\calO$,  we can simply define $\mu$ as the ratio of (signed) cross-sectional areas of the bundle at $\calO$ and $\calE$:
 \bea
 \mu = 1 - \frac{\calA_\calE}{\calA_\calO}. \label{eq:mufromA}
 \eea
This means that $\mu$ measures the focusing power of the spacetime along $\gamma_0$ by a simple comparison of the cross-sectional areas of an initially parallel bundle of rays sent backwards in time towards $\calE$. Moreover, this formula leads immediately to a simple evolution equation for $\mu$. Let $g_A = W_{XX}(\lambda)\,f_A$ be the solutions of the GDE corresponding to $f_1$ and $f_2$ at $\calO$, defining the shape of the bundle. We start by differentiating $\calA(\lambda) = g_1 \wedge g_2$:
\begin{equation}
\label{eq:da}
\begin{split}
\dfrac{\dd \calA}{\dd \lambda} &= \nabla_l g_1 \wedge g_2 + g_1 \wedge \nabla_l g_2 = (g^C_1 \nabla_C l) \wedge g_2 + g_1 \wedge (g^C_2 \nabla_C l) \\&= (g_1 \wedge g_2) \nabla_C l^C = \calA \,\theta,
\end{split}
\end{equation}
where $\theta$ is the bundle expansion. We have  used here the properties of wedge product and the commutation relation $\nabla_l g=\nabla_g l$.
By differentiating  \eqref{eq:mufromA} we get
\bea
 \frac{\dd \mu}{\dd \lambda} = \theta\, (\mu - 1), \label{eq:mufromtheta}
\eea
with the initial data of the form $\mu(\lambda_\calO) = 0$.

Equation \eqref{eq:mufromtheta} requires the expansion $\theta (\lambda)$ of the infinitesimal bundle along $\gamma_0$ as input. It can be obtained, together with shear $\sigma_{AB}(\lambda)$, using the null Raychaudhuri equation along $\gamma_0$ and the shear equation \cite{perlick-lrr, Wald9, poisson_2004}, also known as the Sachs equations:
\bea
\label{eq:nray}
\dfrac{\dd \theta}{\dd \lambda} &=& - \frac {\theta^2}{ 2} - \sigma_{AB}\,\sigma^{AB} - R_{\mu\nu}\,l^\mu\,l^\nu \\
\dfrac{\dd \sigma_{AB} } {\dd\lambda} &=& -\theta\,\sigma_{AB}+C_{A\mu\nu B}\,l^\mu\,l^\nu.
\eea
As the initial data we take $\theta(\lambda_\calO) = 0$ and $\sigma_{AB}(\lambda_\calO) = 0$, corresponding to an initially parallel bundle. 
We do not include the twist, because  it must vanish at $\calO$ together with $\theta$ and $\sigma$, and thus also along the whole bundle. 

\begin{comment}\MK{Do we include that? Also: $\sqrt{A}$ is a bit problematic with the definition we used above.} Another type of second order ODE can be derived by  differentiating $\sqrt \calA$ twice and using (\ref{eq:da}) and (\ref{eq:nray}). One gets:
\bea
\dfrac{d^2 \sqrt \calA}{d \lambda^2} = -\left (\sigma^2 + \frac {R_{ll}} 2\right) \sqrt \calA
\eea
 Now an evolution equation for $\mu$ is straightforward:
\bea
\frac{d \mu}{d \lambda} = (\mu - 1)\theta
\eea
In the similar way:
\bea
\dfrac{d^2 \mu}{d \lambda^2} = \frac{\dot \mu^2}{2(\mu - 1)} - 2 (\mu - 1) \left (\sigma^2 + \frac{R_{ll}}{2} \right )
\eea
\end{comment}
\bibliography{bibbib}

\begin{thebibliography}{58}%
\makeatletter
\providecommand \@ifxundefined [1]{%
 \@ifx{#1\undefined}
}%
\providecommand \@ifnum [1]{%
 \ifnum #1\expandafter \@firstoftwo
 \else \expandafter \@secondoftwo
 \fi
}%
\providecommand \@ifx [1]{%
 \ifx #1\expandafter \@firstoftwo
 \else \expandafter \@secondoftwo
 \fi
}%
\providecommand \natexlab [1]{#1}%
\providecommand \enquote  [1]{``#1''}%
\providecommand \bibnamefont  [1]{#1}%
\providecommand \bibfnamefont [1]{#1}%
\providecommand \citenamefont [1]{#1}%
\providecommand \href@noop [0]{\@secondoftwo}%
\providecommand \href [0]{\begingroup \@sanitize@url \@href}%
\providecommand \@href[1]{\@@startlink{#1}\@@href}%
\providecommand \@@href[1]{\endgroup#1\@@endlink}%
\providecommand \@sanitize@url [0]{\catcode `\\12\catcode `\$12\catcode
  `\&12\catcode `\#12\catcode `\^12\catcode `\_12\catcode `\%12\relax}%
\providecommand \@@startlink[1]{}%
\providecommand \@@endlink[0]{}%
\providecommand \url  [0]{\begingroup\@sanitize@url \@url }%
\providecommand \@url [1]{\endgroup\@href {#1}{\urlprefix }}%
\providecommand \urlprefix  [0]{URL }%
\providecommand \Eprint [0]{\href }%
\providecommand \doibase [0]{http://dx.doi.org/}%
\providecommand \selectlanguage [0]{\@gobble}%
\providecommand \bibinfo  [0]{\@secondoftwo}%
\providecommand \bibfield  [0]{\@secondoftwo}%
\providecommand \translation [1]{[#1]}%
\providecommand \BibitemOpen [0]{}%
\providecommand \bibitemStop [0]{}%
\providecommand \bibitemNoStop [0]{.\EOS\space}%
\providecommand \EOS [0]{\spacefactor3000\relax}%
\providecommand \BibitemShut  [1]{\csname bibitem#1\endcsname}%
\let\auto@bib@innerbib\@empty
%</preamble>
\bibitem [{\citenamefont {Grasso}\ \emph {et~al.}(2019)\citenamefont {Grasso},
  \citenamefont {Korzy\'nski},\ and\ \citenamefont
  {Serbenta}}]{Grasso:2018mei}%
  \BibitemOpen
  \bibfield  {author} {\bibinfo {author} {\bibfnamefont {M.}~\bibnamefont
  {Grasso}}, \bibinfo {author} {\bibfnamefont {M.}~\bibnamefont {Korzy\'nski}},
  \ and\ \bibinfo {author} {\bibfnamefont {J.}~\bibnamefont {Serbenta}},\
  }\href {\doibase 10.1103/PhysRevD.99.064038} {\bibfield  {journal} {\bibinfo
  {journal} {Phys. Rev. D}\ }\textbf {\bibinfo {volume} {99}},\ \bibinfo
  {pages} {064038} (\bibinfo {year} {2019})},\ \Eprint
  {http://arxiv.org/abs/1811.10284} {arXiv:1811.10284 [gr-qc]} \BibitemShut
  {NoStop}%
\bibitem [{\citenamefont {Korzy\'nski}\ and\ \citenamefont
  {Kopi\'nski}(2018)}]{Korzynski:2017nas}%
  \BibitemOpen
  \bibfield  {author} {\bibinfo {author} {\bibfnamefont {M.}~\bibnamefont
  {Korzy\'nski}}\ and\ \bibinfo {author} {\bibfnamefont {J.}~\bibnamefont
  {Kopi\'nski}},\ }\href {\doibase 10.1088/1475-7516/2018/03/012} {\bibfield
  {journal} {\bibinfo  {journal} {JCAP}\ }\textbf {\bibinfo {volume} {1803}},\
  \bibinfo {pages} {012} (\bibinfo {year} {2018})},\ \Eprint
  {http://arxiv.org/abs/1711.00584} {arXiv:1711.00584 [gr-qc]} \BibitemShut
  {NoStop}%
%%CITATION = ARXIV:1711.00584;%%
\bibitem [{\citenamefont {Damour}(1992)}]{Damour1992}%
  \BibitemOpen
  \bibfield  {author} {\bibinfo {author} {\bibfnamefont {T.}~\bibnamefont
  {Damour}},\ }\href {http://www.jstor.org/stable/53916} {\bibfield  {journal}
  {\bibinfo  {journal} {Philosophical Transactions: Physical Sciences and
  Engineering}\ }\textbf {\bibinfo {volume} {341}},\ \bibinfo {pages} {135}
  (\bibinfo {year} {1992})}\BibitemShut {NoStop}%
\bibitem [{\citenamefont {{Lorimer}}(2008)}]{Lorimer2008}%
  \BibitemOpen
  \bibfield  {author} {\bibinfo {author} {\bibfnamefont {D.~R.}\ \bibnamefont
  {{Lorimer}}},\ }\href {\doibase 10.12942/lrr-2008-8} {\bibfield  {journal}
  {\bibinfo  {journal} {Living Reviews in Relativity}\ }\textbf {\bibinfo
  {volume} {11}},\ \bibinfo {eid} {8} (\bibinfo {year} {2008})},\ \Eprint
  {http://arxiv.org/abs/0811.0762} {arXiv:0811.0762 [astro-ph]} \BibitemShut
  {NoStop}%
\bibitem [{\citenamefont {{Dahal}}(2020)}]{Dahal2020}%
  \BibitemOpen
  \bibfield  {author} {\bibinfo {author} {\bibfnamefont {P.~K.}\ \bibnamefont
  {{Dahal}}},\ }\href {\doibase 10.1007/s12036-020-9625-y} {\bibfield
  {journal} {\bibinfo  {journal} {Journal of Astrophysics and Astronomy}\
  }\textbf {\bibinfo {volume} {41}},\ \bibinfo {eid} {8} (\bibinfo {year}
  {2020})},\ \Eprint {http://arxiv.org/abs/2002.01954} {arXiv:2002.01954
  [astro-ph.IM]} \BibitemShut {NoStop}%
\bibitem [{\citenamefont {{Refsdal}}(1964)}]{Refsdahl1964}%
  \BibitemOpen
  \bibfield  {author} {\bibinfo {author} {\bibfnamefont {S.}~\bibnamefont
  {{Refsdal}}},\ }\href {\doibase 10.1093/mnras/128.4.307} {\bibfield
  {journal} {\bibinfo  {journal} {\mnras}\ }\textbf {\bibinfo {volume} {128}},\
  \bibinfo {pages} {307} (\bibinfo {year} {1964})}\BibitemShut {NoStop}%
\bibitem [{\citenamefont {{Suyu}}\ \emph {et~al.}(2017)\citenamefont {{Suyu}},
  \citenamefont {{Bonvin}}, \citenamefont {{Courbin}}, \citenamefont
  {{Fassnacht}}, \citenamefont {{Rusu}}, \citenamefont {{Sluse}}, \citenamefont
  {{Treu}}, \citenamefont {{Wong}}, \citenamefont {{Auger}}, \citenamefont
  {{Ding}}, \citenamefont {{Hilbert}}, \citenamefont {{Marshall}},
  \citenamefont {{Rumbaugh}}, \citenamefont {{Sonnenfeld}}, \citenamefont
  {{Tewes}}, \citenamefont {{Tihhonova}}, \citenamefont {{Agnello}},
  \citenamefont {{Blandford}}, \citenamefont {{Chen}}, \citenamefont
  {{Collett}}, \citenamefont {{Koopmans}}, \citenamefont {{Liao}},
  \citenamefont {{Meylan}},\ and\ \citenamefont {{Spiniello}}}]{Suyu2017}%
  \BibitemOpen
  \bibfield  {author} {\bibinfo {author} {\bibfnamefont {S.~H.}\ \bibnamefont
  {{Suyu}}}, \bibinfo {author} {\bibfnamefont {V.}~\bibnamefont {{Bonvin}}},
  \bibinfo {author} {\bibfnamefont {F.}~\bibnamefont {{Courbin}}}, \bibinfo
  {author} {\bibfnamefont {C.~D.}\ \bibnamefont {{Fassnacht}}}, \bibinfo
  {author} {\bibfnamefont {C.~E.}\ \bibnamefont {{Rusu}}}, \bibinfo {author}
  {\bibfnamefont {D.}~\bibnamefont {{Sluse}}}, \bibinfo {author} {\bibfnamefont
  {T.}~\bibnamefont {{Treu}}}, \bibinfo {author} {\bibfnamefont {K.~C.}\
  \bibnamefont {{Wong}}}, \bibinfo {author} {\bibfnamefont {M.~W.}\
  \bibnamefont {{Auger}}}, \bibinfo {author} {\bibfnamefont {X.}~\bibnamefont
  {{Ding}}}, \bibinfo {author} {\bibfnamefont {S.}~\bibnamefont {{Hilbert}}},
  \bibinfo {author} {\bibfnamefont {P.~J.}\ \bibnamefont {{Marshall}}},
  \bibinfo {author} {\bibfnamefont {N.}~\bibnamefont {{Rumbaugh}}}, \bibinfo
  {author} {\bibfnamefont {A.}~\bibnamefont {{Sonnenfeld}}}, \bibinfo {author}
  {\bibfnamefont {M.}~\bibnamefont {{Tewes}}}, \bibinfo {author} {\bibfnamefont
  {O.}~\bibnamefont {{Tihhonova}}}, \bibinfo {author} {\bibfnamefont
  {A.}~\bibnamefont {{Agnello}}}, \bibinfo {author} {\bibfnamefont {R.~D.}\
  \bibnamefont {{Blandford}}}, \bibinfo {author} {\bibfnamefont {G.~C.~F.}\
  \bibnamefont {{Chen}}}, \bibinfo {author} {\bibfnamefont {T.}~\bibnamefont
  {{Collett}}}, \bibinfo {author} {\bibfnamefont {L.~V.~E.}\ \bibnamefont
  {{Koopmans}}}, \bibinfo {author} {\bibfnamefont {K.}~\bibnamefont {{Liao}}},
  \bibinfo {author} {\bibfnamefont {G.}~\bibnamefont {{Meylan}}}, \ and\
  \bibinfo {author} {\bibfnamefont {C.}~\bibnamefont {{Spiniello}}},\ }\href
  {\doibase 10.1093/mnras/stx483} {\bibfield  {journal} {\bibinfo  {journal}
  {\mnras}\ }\textbf {\bibinfo {volume} {468}},\ \bibinfo {pages} {2590}
  (\bibinfo {year} {2017})},\ \Eprint {http://arxiv.org/abs/1607.00017}
  {arXiv:1607.00017 [astro-ph.CO]} \BibitemShut {NoStop}%
\bibitem [{\citenamefont {Bauch}(2019)}]{Bauch}%
  \BibitemOpen
  \bibfield  {author} {\bibinfo {author} {\bibfnamefont {A.}~\bibnamefont
  {Bauch}},\ }\enquote {\bibinfo {title} {Time and frequency metrology in the
  context of relativistic geodesy},}\ in\ \href {\doibase
  10.1007/978-3-030-11500-5_1} {\emph {\bibinfo {booktitle} {Relativistic
  Geodesy: Foundations and Applications}}},\ \bibinfo {editor} {edited by\
  \bibinfo {editor} {\bibfnamefont {D.}~\bibnamefont {Puetzfeld}}\ and\
  \bibinfo {editor} {\bibfnamefont {C.}~\bibnamefont {L{\"a}mmerzahl}}}\
  (\bibinfo  {publisher} {Springer International Publishing},\ \bibinfo
  {address} {Cham},\ \bibinfo {year} {2019})\ pp.\ \bibinfo {pages}
  {1--24}\BibitemShut {NoStop}%
\bibitem [{\citenamefont {Puetzfeld}\ and\ \citenamefont
  {L{\"a}mmerzahl}(2019)}]{puetzfeldbooke}%
  \BibitemOpen
  \bibinfo {editor} {\bibfnamefont {D.}~\bibnamefont {Puetzfeld}}\ and\
  \bibinfo {editor} {\bibfnamefont {C.}~\bibnamefont {L{\"a}mmerzahl}},\ eds.,\
  \href {\doibase 10.1007/978-3-030-11500-5_1} {\emph {\bibinfo {title}
  {Relativistic Geodesy: Foundations and Applications}}}\ (\bibinfo
  {publisher} {Springer International Publishing},\ \bibinfo {address} {Cham},\
  \bibinfo {year} {2019})\BibitemShut {NoStop}%
\bibitem [{\citenamefont {{Bahder}}(2003)}]{Bahder2003}%
  \BibitemOpen
  \bibfield  {author} {\bibinfo {author} {\bibfnamefont {T.~B.}\ \bibnamefont
  {{Bahder}}},\ }\href {\doibase 10.1103/PhysRevD.68.063005} {\bibfield
  {journal} {\bibinfo  {journal} {\prd}\ }\textbf {\bibinfo {volume} {68}},\
  \bibinfo {eid} {063005} (\bibinfo {year} {2003})},\ \Eprint
  {http://arxiv.org/abs/gr-qc/0306076} {arXiv:gr-qc/0306076 [gr-qc]}
  \BibitemShut {NoStop}%
\bibitem [{\citenamefont {{Ashby}}(2003)}]{Ashby2003}%
  \BibitemOpen
  \bibfield  {author} {\bibinfo {author} {\bibfnamefont {N.}~\bibnamefont
  {{Ashby}}},\ }\href {\doibase 10.12942/lrr-2003-1} {\bibfield  {journal}
  {\bibinfo  {journal} {Living Reviews in Relativity}\ }\textbf {\bibinfo
  {volume} {6}},\ \bibinfo {eid} {1} (\bibinfo {year} {2003})}\BibitemShut
  {NoStop}%
\bibitem [{\citenamefont {Shapiro}(1964)}]{shapiro:1964}%
  \BibitemOpen
  \bibfield  {author} {\bibinfo {author} {\bibfnamefont {I.~I.}\ \bibnamefont
  {Shapiro}},\ }\href {\doibase 10.1103/PhysRevLett.13.789} {\bibfield
  {journal} {\bibinfo  {journal} {Phys. Rev. Lett.}\ }\textbf {\bibinfo
  {volume} {13}},\ \bibinfo {pages} {789} (\bibinfo {year} {1964})}\BibitemShut
  {NoStop}%
\bibitem [{\citenamefont {{Synge}}(1960)}]{SyngeBook}%
  \BibitemOpen
  \bibfield  {author} {\bibinfo {author} {\bibfnamefont {J.~L.}\ \bibnamefont
  {{Synge}}},\ }\href@noop {} {\emph {\bibinfo {title} {{Relativity: The
  general theory}}}}\ (\bibinfo  {publisher} {North-Holland Publishing
  Company},\ \bibinfo {address} {Amsterdam},\ \bibinfo {year}
  {1960})\BibitemShut {NoStop}%
\bibitem [{\citenamefont {{Teyssandier}}\ \emph {et~al.}()\citenamefont
  {{Teyssandier}}, \citenamefont {{Le Poncin-Lafitte}},\ and\ \citenamefont
  {{Linet}}}]{Teyssandier-book}%
  \BibitemOpen
  \bibfield  {author} {\bibinfo {author} {\bibfnamefont {P.}~\bibnamefont
  {{Teyssandier}}}, \bibinfo {author} {\bibfnamefont {C.}~\bibnamefont {{Le
  Poncin-Lafitte}}}, \ and\ \bibinfo {author} {\bibfnamefont {B.}~\bibnamefont
  {{Linet}}},\ }\enquote {\bibinfo {title} {{A Universal Tool for Determining
  the Time Delay and the Frequency Shift of Light: Synge's World Function}},}\
  in\ \href {\doibase 10.1007/978-3-540-34377-6_6} {\emph {\bibinfo {booktitle}
  {Lasers, Clocks and Drag-Free Control: Exploration of Relativistic Gravity in
  Space}}},\ \bibinfo {series} {Astrophysics and Space Science Library}, Vol.\
  \bibinfo {volume} {349},\ \bibinfo {editor} {edited by\ \bibinfo {editor}
  {\bibfnamefont {H.}~\bibnamefont {{Dittus}}}, \bibinfo {editor}
  {\bibfnamefont {C.}~\bibnamefont {{Lammerzahl}}}, \ and\ \bibinfo {editor}
  {\bibfnamefont {S.~G.}\ \bibnamefont {{Turyshev}}}},\ p.\ \bibinfo {pages}
  {153}\BibitemShut {NoStop}%
\bibitem [{\citenamefont {{Teyssandier}}\ and\ \citenamefont {{Le
  Poncin-Lafitte}}(2008)}]{Teyssandier:2008}%
  \BibitemOpen
  \bibfield  {author} {\bibinfo {author} {\bibfnamefont {P.}~\bibnamefont
  {{Teyssandier}}}\ and\ \bibinfo {author} {\bibfnamefont {C.}~\bibnamefont
  {{Le Poncin-Lafitte}}},\ }\href {\doibase 10.1088/0264-9381/25/14/145020}
  {\bibfield  {journal} {\bibinfo  {journal} {Classical and Quantum Gravity}\
  }\textbf {\bibinfo {volume} {25}},\ \bibinfo {eid} {145020} (\bibinfo {year}
  {2008})},\ \Eprint {http://arxiv.org/abs/0803.0277} {arXiv:0803.0277 [gr-qc]}
  \BibitemShut {NoStop}%
\bibitem [{\citenamefont {Linet}\ and\ \citenamefont
  {Teyssandier}(2002)}]{LinetTeyssandier}%
  \BibitemOpen
  \bibfield  {author} {\bibinfo {author} {\bibfnamefont {B.}~\bibnamefont
  {Linet}}\ and\ \bibinfo {author} {\bibfnamefont {P.}~\bibnamefont
  {Teyssandier}},\ }\href {\doibase 10.1103/PhysRevD.66.024045} {\bibfield
  {journal} {\bibinfo  {journal} {Phys. Rev. D}\ }\textbf {\bibinfo {volume}
  {66}},\ \bibinfo {pages} {024045} (\bibinfo {year} {2002})}\BibitemShut
  {NoStop}%
\bibitem [{Note1()}]{Note1}%
  \BibitemOpen
  \bibinfo {note} {This point is important, because in certain cases two or
  more signals may arrive almost at the same moment, leading to ambiguities in
  the data interpretation. Distinguishing of the incoming signals can be
  achieved, for example, by including a data stamp containing an identifier
  encoded in the signal (for example, see the method used to transmit the
  P-code in the Global Positioning System \cite {Bahder2003}), or by using
  different frequencies for each signal according to a prescribed
  sequence.}\BibitemShut {Stop}%
\bibitem [{\citenamefont {Neumann}\ \emph {et~al.}(2020)\citenamefont
  {Neumann}, \citenamefont {Puetzfeld},\ and\ \citenamefont
  {Rubilar}}]{PhysRevD.102.044027}%
  \BibitemOpen
  \bibfield  {author} {\bibinfo {author} {\bibfnamefont {G.}~\bibnamefont
  {Neumann}}, \bibinfo {author} {\bibfnamefont {D.}~\bibnamefont {Puetzfeld}},
  \ and\ \bibinfo {author} {\bibfnamefont {G.~F.}\ \bibnamefont {Rubilar}},\
  }\href {\doibase 10.1103/PhysRevD.102.044027} {\bibfield  {journal} {\bibinfo
   {journal} {Phys. Rev. D}\ }\textbf {\bibinfo {volume} {102}},\ \bibinfo
  {pages} {044027} (\bibinfo {year} {2020})}\BibitemShut {NoStop}%
\bibitem [{\citenamefont {Puetzfeld}\ \emph {et~al.}(2018)\citenamefont
  {Puetzfeld}, \citenamefont {Obukhov},\ and\ \citenamefont
  {L\"ammerzahl}}]{PhysRevD.98.024032}%
  \BibitemOpen
  \bibfield  {author} {\bibinfo {author} {\bibfnamefont {D.}~\bibnamefont
  {Puetzfeld}}, \bibinfo {author} {\bibfnamefont {Y.~N.}\ \bibnamefont
  {Obukhov}}, \ and\ \bibinfo {author} {\bibfnamefont {C.}~\bibnamefont
  {L\"ammerzahl}},\ }\href {\doibase 10.1103/PhysRevD.98.024032} {\bibfield
  {journal} {\bibinfo  {journal} {Phys. Rev. D}\ }\textbf {\bibinfo {volume}
  {98}},\ \bibinfo {pages} {024032} (\bibinfo {year} {2018})}\BibitemShut
  {NoStop}%
\bibitem [{\citenamefont {Puetzfeld}\ and\ \citenamefont
  {Obukhov}(2016)}]{PhysRevD.93.044073}%
  \BibitemOpen
  \bibfield  {author} {\bibinfo {author} {\bibfnamefont {D.}~\bibnamefont
  {Puetzfeld}}\ and\ \bibinfo {author} {\bibfnamefont {Y.~N.}\ \bibnamefont
  {Obukhov}},\ }\href {\doibase 10.1103/PhysRevD.93.044073} {\bibfield
  {journal} {\bibinfo  {journal} {Phys. Rev. D}\ }\textbf {\bibinfo {volume}
  {93}},\ \bibinfo {pages} {044073} (\bibinfo {year} {2016})}\BibitemShut
  {NoStop}%
\bibitem [{\citenamefont {{Szekeres}}(1965)}]{Szekeres1965}%
  \BibitemOpen
  \bibfield  {author} {\bibinfo {author} {\bibfnamefont {P.}~\bibnamefont
  {{Szekeres}}},\ }\href {\doibase 10.1063/1.1704788} {\bibfield  {journal}
  {\bibinfo  {journal} {Journal of Mathematical Physics}\ }\textbf {\bibinfo
  {volume} {6}},\ \bibinfo {pages} {1387} (\bibinfo {year} {1965})}\BibitemShut
  {NoStop}%
\bibitem [{\citenamefont {Ciufolini}\ and\ \citenamefont
  {Demia\'nski}(1986)}]{PhysRevD.34.1018}%
  \BibitemOpen
  \bibfield  {author} {\bibinfo {author} {\bibfnamefont {I.}~\bibnamefont
  {Ciufolini}}\ and\ \bibinfo {author} {\bibfnamefont {M.}~\bibnamefont
  {Demia\'nski}},\ }\href {\doibase 10.1103/PhysRevD.34.1018} {\bibfield
  {journal} {\bibinfo  {journal} {Phys. Rev. D}\ }\textbf {\bibinfo {volume}
  {34}},\ \bibinfo {pages} {1018} (\bibinfo {year} {1986})}\BibitemShut
  {NoStop}%
\bibitem [{\citenamefont {Ba\.za\'nski}(1977{\natexlab{a}})}]{Bazanski1}%
  \BibitemOpen
  \bibfield  {author} {\bibinfo {author} {\bibfnamefont {S.~L.}\ \bibnamefont
  {Ba\.za\'nski}},\ }\href {http://www.numdam.org/item/AIHPA_1977__27_2_115_0}
  {\bibfield  {journal} {\bibinfo  {journal} {Annales de l'I.H.P. Physique
  th\'eorique}\ }\textbf {\bibinfo {volume} {27}},\ \bibinfo {pages} {115}
  (\bibinfo {year} {1977}{\natexlab{a}})}\BibitemShut {NoStop}%
\bibitem [{\citenamefont {Ba\.za\'nski}(1977{\natexlab{b}})}]{Bazanski2}%
  \BibitemOpen
  \bibfield  {author} {\bibinfo {author} {\bibfnamefont {S.~L.}\ \bibnamefont
  {Ba\.za\'nski}},\ }\href {http://www.numdam.org/item/AIHPA_1977__27_2_145_0}
  {\bibfield  {journal} {\bibinfo  {journal} {Annales de l'I.H.P. Physique
  th\'eorique}\ }\textbf {\bibinfo {volume} {27}},\ \bibinfo {pages} {145}
  (\bibinfo {year} {1977}{\natexlab{b}})}\BibitemShut {NoStop}%
\bibitem [{\citenamefont {Pirani}(1956)}]{Pirani1956}%
  \BibitemOpen
  \bibfield  {author} {\bibinfo {author} {\bibfnamefont {F.~A.~E.}\
  \bibnamefont {Pirani}},\ }\href@noop {} {\bibfield  {journal} {\bibinfo
  {journal} {Acta Phys. Polon.}\ }\textbf {\bibinfo {volume} {15}},\ \bibinfo
  {pages} {389} (\bibinfo {year} {1956})}\BibitemShut {NoStop}%
\bibitem [{\citenamefont {Korzy\ifmmode~\acute{n}\else \'{n}\fi{}ski}\ and\
  \citenamefont {Villa}(2020)}]{korzynskivilla}%
  \BibitemOpen
  \bibfield  {author} {\bibinfo {author} {\bibfnamefont {M.}~\bibnamefont
  {Korzy\ifmmode~\acute{n}\else \'{n}\fi{}ski}}\ and\ \bibinfo {author}
  {\bibfnamefont {E.}~\bibnamefont {Villa}},\ }\href {\doibase
  10.1103/PhysRevD.101.063506} {\bibfield  {journal} {\bibinfo  {journal}
  {Phys. Rev. D}\ }\textbf {\bibinfo {volume} {101}},\ \bibinfo {pages}
  {063506} (\bibinfo {year} {2020})}\BibitemShut {NoStop}%
\bibitem [{\citenamefont {Klioner}(2003)}]{Klioner_2003}%
  \BibitemOpen
  \bibfield  {author} {\bibinfo {author} {\bibfnamefont {S.~A.}\ \bibnamefont
  {Klioner}},\ }\href {\doibase 10.1086/367593} {\bibfield  {journal} {\bibinfo
   {journal} {The Astronomical Journal}\ }\textbf {\bibinfo {volume} {125}},\
  \bibinfo {pages} {1580} (\bibinfo {year} {2003})}\BibitemShut {NoStop}%
\bibitem [{\citenamefont {{Butkevich}}\ and\ \citenamefont
  {{Lindegren}}(2014)}]{2014A&A...570A..62B}%
  \BibitemOpen
  \bibfield  {author} {\bibinfo {author} {\bibfnamefont {A.~G.}\ \bibnamefont
  {{Butkevich}}}\ and\ \bibinfo {author} {\bibfnamefont {L.}~\bibnamefont
  {{Lindegren}}},\ }\href {\doibase 10.1051/0004-6361/201424483} {\bibfield
  {journal} {\bibinfo  {journal} {A\& A}\ }\textbf {\bibinfo {volume} {570}},\
  \bibinfo {eid} {A62} (\bibinfo {year} {2014})},\ \Eprint
  {http://arxiv.org/abs/1407.4664} {arXiv:1407.4664 [astro-ph.IM]} \BibitemShut
  {NoStop}%
\bibitem [{\citenamefont {\v{C}erven\'y}(2001)}]{cerveny_2001}%
  \BibitemOpen
  \bibfield  {author} {\bibinfo {author} {\bibfnamefont {V.}~\bibnamefont
  {\v{C}erven\'y}},\ }\href {\doibase 10.1017/CBO9780511529399} {\emph
  {\bibinfo {title} {Seismic Ray Theory}}}\ (\bibinfo  {publisher} {Cambridge
  University Press},\ \bibinfo {year} {2001})\BibitemShut {NoStop}%
\bibitem [{\citenamefont {Farra}(1999)}]{farra2002}%
  \BibitemOpen
  \bibfield  {author} {\bibinfo {author} {\bibfnamefont {V.}~\bibnamefont
  {Farra}},\ }\href {\doibase 10.1046/j.1365-246X.1999.00733.x} {\bibfield
  {journal} {\bibinfo  {journal} {Geophysical Journal International}\ }\textbf
  {\bibinfo {volume} {136}},\ \bibinfo {pages} {205 } (\bibinfo {year}
  {1999})}\BibitemShut {NoStop}%
\bibitem [{\citenamefont {\v{C}erven\'y}\ \emph {et~al.}(2012)\citenamefont
  {\v{C}erven\'y}, \citenamefont {Iversen},\ and\ \citenamefont
  {P\v{s}en\v{c}\'ik}}]{cerveny2012}%
  \BibitemOpen
  \bibfield  {author} {\bibinfo {author} {\bibfnamefont {V.}~\bibnamefont
  {\v{C}erven\'y}}, \bibinfo {author} {\bibfnamefont {E.}~\bibnamefont
  {Iversen}}, \ and\ \bibinfo {author} {\bibfnamefont {I.}~\bibnamefont
  {P\v{s}en\v{c}\'ik}},\ }\href {\doibase 10.1111/j.1365-246X.2012.05430.x}
  {\bibfield  {journal} {\bibinfo  {journal} {Geophysical Journal
  International}\ }\textbf {\bibinfo {volume} {189}},\ \bibinfo {pages} {1597}
  (\bibinfo {year} {2012})}\BibitemShut {NoStop}%
\bibitem [{\citenamefont {bin Waheed}\ \emph {et~al.}(2013)\citenamefont {bin
  Waheed}, \citenamefont {P\v{s}en\v{c}\'ik}, \citenamefont {\v{C}erven\'y},
  \citenamefont {Iversen},\ and\ \citenamefont {Alkhalifah}}]{waheed2013}%
  \BibitemOpen
  \bibfield  {author} {\bibinfo {author} {\bibfnamefont {U.}~\bibnamefont {bin
  Waheed}}, \bibinfo {author} {\bibfnamefont {I.}~\bibnamefont
  {P\v{s}en\v{c}\'ik}}, \bibinfo {author} {\bibfnamefont {V.}~\bibnamefont
  {\v{C}erven\'y}}, \bibinfo {author} {\bibfnamefont {E.}~\bibnamefont
  {Iversen}}, \ and\ \bibinfo {author} {\bibfnamefont {T.}~\bibnamefont
  {Alkhalifah}},\ }\href {\doibase 10.1190/geo2012-0406.1} {\bibfield
  {journal} {\bibinfo  {journal} {Geophysics}\ }\textbf {\bibinfo {volume}
  {78}},\ \bibinfo {pages} {WC65} (\bibinfo {year} {2013})}\BibitemShut
  {NoStop}%
\bibitem [{\citenamefont {Isaacson}(1968{\natexlab{a}})}]{PhysRev.166.1263}%
  \BibitemOpen
  \bibfield  {author} {\bibinfo {author} {\bibfnamefont {R.~A.}\ \bibnamefont
  {Isaacson}},\ }\href {\doibase 10.1103/PhysRev.166.1263} {\bibfield
  {journal} {\bibinfo  {journal} {Phys. Rev.}\ }\textbf {\bibinfo {volume}
  {166}},\ \bibinfo {pages} {1263} (\bibinfo {year}
  {1968}{\natexlab{a}})}\BibitemShut {NoStop}%
\bibitem [{\citenamefont {Isaacson}(1968{\natexlab{b}})}]{PhysRev.166.1272}%
  \BibitemOpen
  \bibfield  {author} {\bibinfo {author} {\bibfnamefont {R.~A.}\ \bibnamefont
  {Isaacson}},\ }\href {\doibase 10.1103/PhysRev.166.1272} {\bibfield
  {journal} {\bibinfo  {journal} {Phys. Rev.}\ }\textbf {\bibinfo {volume}
  {166}},\ \bibinfo {pages} {1272} (\bibinfo {year}
  {1968}{\natexlab{b}})}\BibitemShut {NoStop}%
\bibitem [{\citenamefont {Poisson}\ \emph {et~al.}(2011)\citenamefont
  {Poisson}, \citenamefont {Pound},\ and\ \citenamefont {Vega}}]{Poisson2011}%
  \BibitemOpen
  \bibfield  {author} {\bibinfo {author} {\bibfnamefont {E.}~\bibnamefont
  {Poisson}}, \bibinfo {author} {\bibfnamefont {A.}~\bibnamefont {Pound}}, \
  and\ \bibinfo {author} {\bibfnamefont {I.}~\bibnamefont {Vega}},\ }\href
  {\doibase 10.12942/lrr-2011-7} {\bibfield  {journal} {\bibinfo  {journal}
  {Living Reviews in Relativity}\ }\textbf {\bibinfo {volume} {14}},\ \bibinfo
  {pages} {7} (\bibinfo {year} {2011})}\BibitemShut {NoStop}%
\bibitem [{\citenamefont {Low}(1998)}]{low:1998}%
  \BibitemOpen
  \bibfield  {author} {\bibinfo {author} {\bibfnamefont {R.}~\bibnamefont
  {Low}},\ }\href {\doibase 10.1063/1.532257} {\bibfield  {journal} {\bibinfo
  {journal} {Journal of Mathematical Physics}\ }\textbf {\bibinfo {volume}
  {39}},\ \bibinfo {pages} {3332} (\bibinfo {year} {1998})}\BibitemShut
  {NoStop}%
\bibitem [{\citenamefont {Perlick}(2004)}]{perlick-lrr}%
  \BibitemOpen
  \bibfield  {author} {\bibinfo {author} {\bibfnamefont {V.}~\bibnamefont
  {Perlick}},\ }\href {\doibase 10.12942/lrr-2004-9} {\bibfield  {journal}
  {\bibinfo  {journal} {Living Reviews in Relativity}\ }\textbf {\bibinfo
  {volume} {7}},\ \bibinfo {pages} {9} (\bibinfo {year} {2004})}\BibitemShut
  {NoStop}%
\bibitem [{\citenamefont {Vines}(2015)}]{Vines:2014oba}%
  \BibitemOpen
  \bibfield  {author} {\bibinfo {author} {\bibfnamefont {J.}~\bibnamefont
  {Vines}},\ }\href {\doibase 10.1007/s10714-015-1901-9} {\bibfield  {journal}
  {\bibinfo  {journal} {Gen. Rel. Grav.}\ }\textbf {\bibinfo {volume} {47}},\
  \bibinfo {pages} {59} (\bibinfo {year} {2015})},\ \Eprint
  {http://arxiv.org/abs/1407.6992} {arXiv:1407.6992 [gr-qc]} \BibitemShut
  {NoStop}%
%%CITATION = ARXIV:1407.6992;%%
\bibitem [{\citenamefont {Bahder}(2009)}]{BahderBook}%
  \BibitemOpen
  \bibfield  {author} {\bibinfo {author} {\bibfnamefont {T.~B.}\ \bibnamefont
  {Bahder}},\ }\href@noop {} {\emph {\bibinfo {title} {Clock Synchronization
  and Navigation in the Vicinity of the Earth}}}\ (\bibinfo  {publisher} {Nova
  Science Publishers, Inc.},\ \bibinfo {address} {New York},\ \bibinfo {year}
  {2009})\BibitemShut {NoStop}%
\bibitem [{Note2()}]{Note2}%
  \BibitemOpen
  \bibinfo {note} {Note that in the definitions of $M\times M$, $\protect
  \ensuremath {\protect \mathbf {L}}$ and $\protect \ensuremath {\protect
  \mathbf {X}}$ we have swapped the order of coordinates at $\protect
  \ensuremath {\protect \mathcal {O}}$ and $\protect \ensuremath {\protect
  \mathcal {E}}$ in comparison to the definition of the world function: we
  first take the four components related to the reception region $N_\protect
  \ensuremath {\protect \mathcal {O}}$, and then the four components related to
  the emission region $N_\protect \ensuremath {\protect \mathcal {E}}$, the
  opposite of the definition of the world function. Hopefully this should not
  lead to confusion.}\BibitemShut {Stop}%
\bibitem [{\citenamefont {Uzun}(2020)}]{Uzun_2020}%
  \BibitemOpen
  \bibfield  {author} {\bibinfo {author} {\bibfnamefont {N.}~\bibnamefont
  {Uzun}},\ }\href {\doibase 10.1088/1361-6382/ab60b5} {\bibfield  {journal}
  {\bibinfo  {journal} {Classical and Quantum Gravity}\ }\textbf {\bibinfo
  {volume} {37}},\ \bibinfo {pages} {045002} (\bibinfo {year}
  {2020})}\BibitemShut {NoStop}%
\bibitem [{\citenamefont {DeWitt}\ and\ \citenamefont
  {Brehme}(1960)}]{DeWittBrehme}%
  \BibitemOpen
  \bibfield  {author} {\bibinfo {author} {\bibfnamefont {B.~S.}\ \bibnamefont
  {DeWitt}}\ and\ \bibinfo {author} {\bibfnamefont {R.~W.}\ \bibnamefont
  {Brehme}},\ }\href {\doibase https://doi.org/10.1016/0003-4916(60)90030-0}
  {\bibfield  {journal} {\bibinfo  {journal} {Annals of Physics}\ }\textbf
  {\bibinfo {volume} {9}},\ \bibinfo {pages} {220 } (\bibinfo {year}
  {1960})}\BibitemShut {NoStop}%
\bibitem [{\citenamefont {Dixon}(1970)}]{Dixon2}%
  \BibitemOpen
  \bibfield  {author} {\bibinfo {author} {\bibfnamefont {W.~G.}\ \bibnamefont
  {Dixon}},\ }\href {\doibase 10.1098/rspa.1970.0020} {\bibfield  {journal}
  {\bibinfo  {journal} {Proceedings of the Royal Society of London. A.
  Mathematical and Physical Sciences}\ }\textbf {\bibinfo {volume} {314}},\
  \bibinfo {pages} {499} (\bibinfo {year} {1970})}\BibitemShut {NoStop}%
\bibitem [{\citenamefont {Gallo}\ and\ \citenamefont
  {Moreschi}(2011)}]{PhysRevD.83.083007}%
  \BibitemOpen
  \bibfield  {author} {\bibinfo {author} {\bibfnamefont {E.}~\bibnamefont
  {Gallo}}\ and\ \bibinfo {author} {\bibfnamefont {O.~M.}\ \bibnamefont
  {Moreschi}},\ }\href {\doibase 10.1103/PhysRevD.83.083007} {\bibfield
  {journal} {\bibinfo  {journal} {Phys. Rev. D}\ }\textbf {\bibinfo {volume}
  {83}},\ \bibinfo {pages} {083007} (\bibinfo {year} {2011})}\BibitemShut
  {NoStop}%
\bibitem [{\citenamefont {Crisnejo}\ and\ \citenamefont
  {Gallo}(2018)}]{PhysRevD.97.084010}%
  \BibitemOpen
  \bibfield  {author} {\bibinfo {author} {\bibfnamefont {G.}~\bibnamefont
  {Crisnejo}}\ and\ \bibinfo {author} {\bibfnamefont {E.}~\bibnamefont
  {Gallo}},\ }\href {\doibase 10.1103/PhysRevD.97.084010} {\bibfield  {journal}
  {\bibinfo  {journal} {Phys. Rev. D}\ }\textbf {\bibinfo {volume} {97}},\
  \bibinfo {pages} {084010} (\bibinfo {year} {2018})}\BibitemShut {NoStop}%
\bibitem [{\citenamefont {Flanagan}\ \emph {et~al.}(2019)\citenamefont
  {Flanagan}, \citenamefont {Grant}, \citenamefont {Harte},\ and\ \citenamefont
  {Nichols}}]{PhysRevD.99.084044}%
  \BibitemOpen
  \bibfield  {author} {\bibinfo {author} {\bibfnamefont {E.~E.}\ \bibnamefont
  {Flanagan}}, \bibinfo {author} {\bibfnamefont {A.~M.}\ \bibnamefont {Grant}},
  \bibinfo {author} {\bibfnamefont {A.~I.}\ \bibnamefont {Harte}}, \ and\
  \bibinfo {author} {\bibfnamefont {D.~A.}\ \bibnamefont {Nichols}},\ }\href
  {\doibase 10.1103/PhysRevD.99.084044} {\bibfield  {journal} {\bibinfo
  {journal} {Phys. Rev. D}\ }\textbf {\bibinfo {volume} {99}},\ \bibinfo
  {pages} {084044} (\bibinfo {year} {2019})}\BibitemShut {NoStop}%
\bibitem [{Note3()}]{Note3}%
  \BibitemOpen
  \bibinfo {note} {In practice the number of measurements can be made lower
  because the components of $\protect \ensuremath {\protect \mathbf {L}}$ and
  $\protect \ensuremath {\protect \mathbf {U}}$ are not independent, see (\ref
  {eq:Uoowithl})-(\ref {eq:Ueewithl}). However, we do not make any use of this
  possibility in this paper.}\BibitemShut {Stop}%
\bibitem [{Note4()}]{Note4}%
  \BibitemOpen
  \bibinfo {note} {On the other hand, note that this does not have to apply to
  the set of corresponding full 8-dimensional vectors, simply because the TOA's
  $\tau ({\protect \bf H}_*^{\protect \bf a} )$, constituting the $\protect \bf
  0$ component, will not obey the central symmetry in general.}\BibitemShut
  {Stop}%
\bibitem [{\citenamefont {Obukhov}\ and\ \citenamefont
  {Puetzfeld}(2019)}]{Obukhov2019}%
  \BibitemOpen
  \bibfield  {author} {\bibinfo {author} {\bibfnamefont {Y.~N.}\ \bibnamefont
  {Obukhov}}\ and\ \bibinfo {author} {\bibfnamefont {D.}~\bibnamefont
  {Puetzfeld}},\ }\enquote {\bibinfo {title} {Measuring the gravitational field
  in general relativity: From deviation equations and the gravitational compass
  to relativistic clock gradiometry},}\ in\ \href {\doibase
  10.1007/978-3-030-11500-5_3} {\emph {\bibinfo {booktitle} {Relativistic
  Geodesy: Foundations and Applications}}},\ \bibinfo {editor} {edited by\
  \bibinfo {editor} {\bibfnamefont {D.}~\bibnamefont {Puetzfeld}}\ and\
  \bibinfo {editor} {\bibfnamefont {C.}~\bibnamefont {L{\"a}mmerzahl}}}\
  (\bibinfo  {publisher} {Springer International Publishing},\ \bibinfo
  {address} {Cham},\ \bibinfo {year} {2019})\ pp.\ \bibinfo {pages}
  {87--130}\BibitemShut {NoStop}%
\bibitem [{\citenamefont {{Pitjev}}\ and\ \citenamefont
  {{Pitjeva}}(2013)}]{2013AstL...39..141P}%
  \BibitemOpen
  \bibfield  {author} {\bibinfo {author} {\bibfnamefont {N.~P.}\ \bibnamefont
  {{Pitjev}}}\ and\ \bibinfo {author} {\bibfnamefont {E.~V.}\ \bibnamefont
  {{Pitjeva}}},\ }\href {\doibase 10.1134/S1063773713020060} {\bibfield
  {journal} {\bibinfo  {journal} {Astronomy Letters}\ }\textbf {\bibinfo
  {volume} {39}},\ \bibinfo {pages} {141} (\bibinfo {year} {2013})},\ \Eprint
  {http://arxiv.org/abs/1306.5534} {arXiv:1306.5534 [astro-ph.EP]} \BibitemShut
  {NoStop}%
\bibitem [{\citenamefont {{Smits}}\ \emph {et~al.}(2011)\citenamefont
  {{Smits}}, \citenamefont {{Tingay}}, \citenamefont {{Wex}}, \citenamefont
  {{Kramer}},\ and\ \citenamefont {{Stappers}}}]{smits:2011}%
  \BibitemOpen
  \bibfield  {author} {\bibinfo {author} {\bibfnamefont {R.}~\bibnamefont
  {{Smits}}}, \bibinfo {author} {\bibfnamefont {S.~J.}\ \bibnamefont
  {{Tingay}}}, \bibinfo {author} {\bibfnamefont {N.}~\bibnamefont {{Wex}}},
  \bibinfo {author} {\bibfnamefont {M.}~\bibnamefont {{Kramer}}}, \ and\
  \bibinfo {author} {\bibfnamefont {B.}~\bibnamefont {{Stappers}}},\ }\href
  {\doibase 10.1051/0004-6361/201016141} {\bibfield  {journal} {\bibinfo
  {journal} {A\& A}\ }\textbf {\bibinfo {volume} {528}},\ \bibinfo {eid} {A108}
  (\bibinfo {year} {2011})},\ \Eprint {http://arxiv.org/abs/1101.5971}
  {arXiv:1101.5971 [astro-ph.IM]} \BibitemShut {NoStop}%
\bibitem [{\citenamefont {{Damour}}\ and\ \citenamefont
  {{Taylor}}(1992)}]{1992PhRvD..45.1840D}%
  \BibitemOpen
  \bibfield  {author} {\bibinfo {author} {\bibfnamefont {T.}~\bibnamefont
  {{Damour}}}\ and\ \bibinfo {author} {\bibfnamefont {J.~H.}\ \bibnamefont
  {{Taylor}}},\ }\href {\doibase 10.1103/PhysRevD.45.1840} {\bibfield
  {journal} {\bibinfo  {journal} {\prd}\ }\textbf {\bibinfo {volume} {45}},\
  \bibinfo {pages} {1840} (\bibinfo {year} {1992})}\BibitemShut {NoStop}%
\bibitem [{Note5()}]{Note5}%
  \BibitemOpen
  \bibinfo {note} {We do not distinguish the primed and unprimed indices in the
  Minkowski space since all tangent spaces in this case may be identified with
  each other.}\BibitemShut {Stop}%
\bibitem [{Note6()}]{Note6}%
  \BibitemOpen
  \bibinfo {note} {Since there is no free term in equation \protect \textup
  {\hbox {\mathsurround \z@ \protect \normalfont (\ignorespaces \ref
  {eq:normalform1}\unskip \@@italiccorr )}}, the signs can actually be flipped
  to $(3,1,4)$ by multiplying both sides by -1, without affecting the normal
  form.}\BibitemShut {Stop}%
\bibitem [{\citenamefont {{Gallo}}\ and\ \citenamefont
  {{Moreschi}}(2012)}]{2012AIPC.1471...82G}%
  \BibitemOpen
  \bibfield  {author} {\bibinfo {author} {\bibfnamefont {E.}~\bibnamefont
  {{Gallo}}}\ and\ \bibinfo {author} {\bibfnamefont {O.~M.}\ \bibnamefont
  {{Moreschi}}},\ }in\ \href {\doibase 10.1063/1.4756817} {\emph {\bibinfo
  {booktitle} {I Cosmosul: Cosmology and Gravitation in the Southern Cone}}},\
  \bibinfo {series} {American Institute of Physics Conference Series}, Vol.\
  \bibinfo {volume} {1471},\ \bibinfo {editor} {edited by\ \bibinfo {editor}
  {\bibfnamefont {J.}~\bibnamefont {{Alcaniz}}}, \bibinfo {editor}
  {\bibfnamefont {S.}~\bibnamefont {{Carneiro}}}, \bibinfo {editor}
  {\bibfnamefont {L.~P.}\ \bibnamefont {{Chimento}}}, \bibinfo {editor}
  {\bibfnamefont {S.}~\bibnamefont {{Del Campo}}}, \bibinfo {editor}
  {\bibfnamefont {J.~C.}\ \bibnamefont {{Fabris}}}, \bibinfo {editor}
  {\bibfnamefont {J.~A.~S.}\ \bibnamefont {{Lima}}}, \ and\ \bibinfo {editor}
  {\bibfnamefont {W.}~\bibnamefont {{Zimdahl}}}}\ (\bibinfo {year} {2012})\
  pp.\ \bibinfo {pages} {82--87},\ \Eprint {http://arxiv.org/abs/1212.2434}
  {arXiv:1212.2434 [gr-qc]} \BibitemShut {NoStop}%
\bibitem [{\citenamefont {Sachs}(1961)}]{sachs61}%
  \BibitemOpen
  \bibfield  {author} {\bibinfo {author} {\bibfnamefont {R.}~\bibnamefont
  {Sachs}},\ }\href {http://www.jstor.org/stable/2414993} {\bibfield  {journal}
  {\bibinfo  {journal} {Proceedings of the Royal Society of London. Series A,
  Mathematical and Physical Sciences}\ }\textbf {\bibinfo {volume} {264}},\
  \bibinfo {pages} {309} (\bibinfo {year} {1961})}\BibitemShut {NoStop}%
\bibitem [{\citenamefont {Wald}(1984)}]{Wald9}%
  \BibitemOpen
  \bibfield  {author} {\bibinfo {author} {\bibfnamefont {R.~M.}\ \bibnamefont
  {Wald}},\ }\enquote {\bibinfo {title} {{General Relativity}},}\ \ (\bibinfo
  {publisher} {Chicago University Press},\ \bibinfo {address} {Chicago, USA},\
  \bibinfo {year} {1984})\ Chap.~\bibinfo {chapter} {9}, p.\ \bibinfo {pages}
  {222}\BibitemShut {NoStop}%
\bibitem [{\citenamefont {Poisson}(2004)}]{poisson_2004}%
  \BibitemOpen
  \bibfield  {author} {\bibinfo {author} {\bibfnamefont {E.}~\bibnamefont
  {Poisson}},\ }\href {\doibase 10.1017/CBO9780511606601} {\emph {\bibinfo
  {title} {A Relativist's Toolkit: The Mathematics of Black-Hole Mechanics}}}\
  (\bibinfo  {publisher} {Cambridge University Press},\ \bibinfo {year}
  {2004})\BibitemShut {NoStop}%
\end{thebibliography}%
\end{document}